\documentclass[journal]{IEEEtran}

\usepackage[T1]{fontenc}
\usepackage{textcomp}
\usepackage{blindtext}
\usepackage[utf8]{inputenc}
\usepackage[noadjust]{cite}
\usepackage[font=footnotesize]{caption}
\usepackage{dblfloatfix}
\usepackage{graphicx}
\usepackage{psfrag}
\usepackage{subfig}
\usepackage{amsmath,amssymb,amsthm,mathrsfs,amsfonts,dsfont}
\usepackage{color}
\usepackage[algoruled]{algorithm2e}
\usepackage{fixltx2e}
\usepackage{multirow}
\usepackage{multicol}
\usepackage[hyphens]{url}
\usepackage[hidelinks]{hyperref}
\usepackage[anythingbreaks]{breakurl}
\usepackage[normalem]{ulem}
\usepackage{acronym}
\usepackage[table,xcdraw]{xcolor}
\usepackage{accents}
\usepackage{mwe}
\usepackage{hhline}
\usepackage{trfsigns}
\usepackage{siunitx}
\usepackage{lscape}
\usepackage{wasysym}
\usepackage{bm}

\theoremstyle{definition}

\newacro{3GPP}{3rd Generation Partnership Project}
\newacro{5G}{fifth generation}
\newacro{5G NR}{5G New Radio}
\newacro{6G}{sixth generation}
\newacro{A/D}{analog-to-digital}
\newacro{ADC}{analog-to-digital converter}
\newacro{AFE}{analog front-end}
\newacro{AGV}{automatic guided vehicle}
\newacro{AWGN}{additive white Gaussian noise}
\newacro{B5G}{beyond \ac{5G}}
\newacro{BB}{baseband}
\newacro{BER}{bit error ratio}
\newacro{BPSK}{binary phase-shift keying}
\newacro{BP}{band-pass}
\newacro{BS}{base station}
\newacro{CDM}{code-division multiplexing}
\newacro{CFO}{carrier frequency offset}
\newacro{CFR}{channel frequency response}
\newacro{CIR}{channel impulse response}
\newacro{CoMP}{coordinated multipoint}
\newacro{CP}{cyclic prefix}
\newacro{CPE}{common phase error}
\newacro{CPO}{carrier phase offset}
\newacro{CRLB}{Cram\'er–Rao lower bound}
\newacro{CS}{chirp sequence}
\newacro{CW}{continuous wave}
\newacro{CZT}{chirp Z-transform}
\newacro{D/A}{digital-to-analog}
\newacro{DAC}{digital-to-analog converter}
\newacro{DDS}{direct digital synthesis}
\newacro{DFRC}{dual-function radar-cunication or dual-functional radar-cunication}
\newacro{DFnT}{discrete Fresnel transform}
\newacro{DFT}{discrete Fourier transform}
\newacro{EVM}{error vector magnitude}
\newacro{FDE}{frequency-domain equalization}
\newacro{FDM}{frequency-division multiplexing}
\newacro{FMCW}{frequency-modulated continuous wave}
\newacro{FO}{frequency offset}
\newacro{FR2}{Frequency Range 2}
\newacro{gNB}{gNodeB}
\newacro{HP}{high-pass}
\newacro{IBFD}{in-band full duplex}
\newacro{ICI}{intercarrier interference}
\newacro{IDFT}{inverse discrete Fourier transform}
\newacro{IDFnT}{inverse discrete Fresnel transform}
\newacro{IF}{intermediate frequency}
\newacro{IHE}{Institute of Radio Frequency Engineering and Electronics}
\newacro{ISAC}{integrated sensing and communication}
\newacro{ISI}{intersymbol interference}
\newacro{ISLR}{integrated sidelobe level ratio}
\newacro{I/Q}{in-phase/quadrature}
\newacro{JCAS}{joint communication and sensing}
\newacro{KIT}{Karlsruhe Institute of Technology}
\newacro{LDPC}{low-density parity-check}
\newacro{LFSR}{linear-feedback shift register}
\newacro{LNA}{low-noise amplifier}
\newacro{LO}{local oscillator}
\newacro{LoS}{line-of-sight}
\newacro{LP}{low-pass}
\newacro{LS}{least squares}
\newacro{mmWave}{milimeter wave}
\newacro{MIMO}{multiple-input multiple-output}
\newacro{MLE}{maximum likelihood estimator}
\newacro{MLS}{maximum-length sequence}
\newacro{MUSIC}{multiple signal classification}
\newacro{NLoS}{non-line-of-sight}
\newacro{OCDM}{orthogonal chirp-division multiplexing}
\newacro{OFDM}{orthogonal frequency-division multiplexing}
\newacro{OOB}{out-of-band}
\newacro{OTA}{over-the-air}
\newacro{P/S}{parallel-to-serial}
\newacro{PA}{power amplifier}
\newacro{PACF}{periodic autocorrelation function}
\newacro{PCCF}{periodic cross-correlation function}
\newacro{PDF}{probability density function}
\newacro{PLC}{powerline cunication}
\newacro{PLL}{phase-locked loop}
\newacro{PMCW}{phase-modulated continuous wave}
\newacro{PMN}{perceptive mobile network}
\newacro{PN}{oscillator phase noise}
\newacro{PoC}{proof-of-concept}
\newacro{PPLR}{peak power loss ratio}
\newacro{PRBS}{pseudorandom binary sequence}
\newacro{PRS}{positioning reference signal}
\newacro{PSD}{power spectral density}
\newacro{PSLR}{peak sidelobe level ratio}
\newacro{PT-RS}{phase tracking reference signal}
\newacro{QAM}{quadrature amplitude modulation}
\newacro{QPSK}{quadrature phase-shift keying}
\newacro{RadCom}{radar-cunication}
\newacro{RCS}{radar cross section}
\newacro{RF}{radio-frequency}
\newacro{RIS}{reflective intelligent surface}
\newacro{RMSE}{root mean squared error}
\newacro{SC}[S\&C]{Schmidl \& Cox}
\newacro{SFO}{sampling frequency offset}
\newacro{SIC}{self-interference cancellation}
\newacro{SINR}{signal-to-interference-plus-noise ratio}
\newacro{SISO}{single-input single-output}
\newacro{SIR}{signal-to-interference ratio}
\newacro{SNR}{signal-to-noise ratio}
\newacro{SoC}{system-on-a-chip}
\newacro{STO}{symbol time offset}
\newacro{S/P}{serial-to-parallel}
\newacro{TDE}{time-domain equalization}
\newacro{TDM}{time-division multiplexing}
\newacro{TDR}{time-domain reflectometry}
\newacro{TITO}{tilt inference of time offset}
\newacro{TO}{time offset}
\newacro{UE}{user equipment}
\newacro{UWAC}{underwater acoustic cunication}
\newacro{V2V}{vehicle-to-vehicle}
\newacro{ZF}{zero forcing}
\newacro{ZP}{zero padding}
\usepackage{pgfplots}
\pgfplotsset{compat=newest}
\usetikzlibrary{plotmarks}
\usetikzlibrary{arrows.meta}
\usepgfplotslibrary{patchplots}
\usepackage{grffile}
\usepackage{amsmath}

\usepackage{calrsfs}

\usepackage{slashbox}

\usepackage{fancyhdr}
\fancypagestyle{empty}{fancy}

\begin{document}
	
	\title{On the Sensing Performance of OFDM-based ISAC under the Influence of Oscillator Phase Noise}
	
	\author{Lucas Giroto de Oliveira,~\IEEEmembership{Graduate Student Member,~IEEE}, Yueheng Li,\\ Benedikt Geiger,~\IEEEmembership{Graduate Student Member,~IEEE}, Laurent Schmalen,~\IEEEmembership{Fellow,~IEEE},\\ Thomas Zwick,~\IEEEmembership{Fellow,~IEEE}, and Benjamin Nuss,~\IEEEmembership{Senior Member,~IEEE}
		\thanks{Manuscript received DD MM, 2024. The authors acknowledge the financial support by the Federal Ministry of Education and Research of Germany in the projects ``KOMSENS-6G'' (grant number: 16KISK123) and ``Open6GHub'' (grant number: 16KISK010). \textit{(Corresponding author: Lucas Giroto de Oliveira.)}}
		\thanks{L. Giroto de Oliveira, Y. Li, T. Zwick, and B. Nuss are with the Institute of Radio Frequency Engineering and Electronics (IHE), Karlsruhe Institute of Technology (KIT), 76131 Karlsruhe, Germany (e-mail: {lucas.oliveira@kit.edu}, {yueheng.li@kit.edu}, {thomas.zwick@kit.edu}, {benjamin.nuss@kit.edu}).}
		\thanks{B. Geiger and L. Schmalen are with the Communications Engineering Laboratory (CEL), Karlsruhe Institute of Technology (KIT), 76187 Karlsruhe, Germany (e-mail: {benedikt.geiger@kit.edu}, {laurent.schmalen@kit.edu}).}
	}
	
	
	\maketitle
	
	\begin{abstract}
		Integrated sensing and communication (ISAC) is a novel capability expected for sixth generation (6G) cellular networks. To that end, several challenges must be addressed to enable both mono- and bistatic sensing in existing deployments. A common impairment in both architectures is oscillator phase noise (PN), which not only degrades communication performance, but also severely impairs radar sensing. To enable a broader understanding of orthogonal-frequency division multiplexing (OFDM)-based sensing impaired by PN, this article presents an analysis of sensing peformance in OFDM-based ISAC for different waveform parameter choices and settings in both mono- and bistatic architectures. In this context, the distortion of the adopted digital constellation modulation is analyzed and the resulting PN-induced effects in range-Doppler radar images are investigated both without and with PN compensation. These effects include peak power loss of target reflections and higher sidelobe levels, especially in the Doppler shift direction. In the conducted analysis, these effects are measured by the peak power loss ratio, peak-to-sidelobe level ratio, and integrated sidelobe level ratio parameters, the two latter being evaluated in both range and Doppler shift directions. In addition, the signal-to-interference ratio is analyzed to allow not only quantifying the distortion of a target reflection, but also measuring the interference floor level in a radar image. The achieved results allow to quantify not only the PN-induced impairments to a single target, but also how the induced degradation may impair the sensing performance of OFDM-based ISAC systems in multi-target scenarios.
	\end{abstract}
	
	
	\IEEEpeerreviewmaketitle
	
	
	\section{Introduction}\label{sec:introduction}
	
	\IEEEPARstart{O}{ne} of the most anticipated features of future \ac{6G} cellular networks is \ac{ISAC} \cite{chafii2023,liu2022}, which is expected to be addressed already in the early releases of \ac{3GPP} specifications for \ac{6G}. Supported by robust connectivity at high data rates \cite{viswanathan2020,rajatheva2020}, \ac{ISAC} is expected to enable a wide range of environment sensing use cases \cite{kadelka2023,shatov2024}. An example is performing radar measurements with existing cellular network deployments to contribute to the maintenance of real-time digital twins of the physical wireless environment. These can, in turn, be used to coordinate and optimize the performance of both communication and sensing subsystems \cite{alkhateeb2023,imran2024}.
	
	Enabling radar sensing as an additional functionality in cellular networks comes with a series of challenges and requirements \cite{thomae2021,wild2023}. In quasi-monostatic sensing architectures, henceforth referred to as monostatic for conciseness, \ac{IBFD} operation \cite{roberts2024} is required. This results in the need of high isolation between the closely located transmitter and receiver antenna arrays, besides \ac{SIC} at both \ac{RF} and \ac{BB} stages to avoid degrading the communication performance and reducing the sensing dynamic range of the \ac{ISAC} system \cite{barneto2019,barneto2021}. In bi- and multistatic architectures \cite{thomae2019,kanhere2021_multistatic,mollen2023,thomae2023,yajnanarayana2024}, inherent challenges from radar networks \cite{geng2020} such as time, frequency, and sampling frequency offsets are observed \cite{giroto2023_EuMW,pegoraro2024,brunner2024}. Although synchronization is possible with already existing \ac{5G} cellular resources \cite{omri2019}, much finer accuracy is demanded for radar sensing than for communication \cite{giroto2024}.
	
	Although the three aforementioned synchronization mismatches are only relevant when performing sensing between two distinct nodes, a fourth impairment is present in both monostatic and bistatic deployments, namely the \ac{PN}. \ac{PN} becomes particularly relevant at the \ac{mmWave} frequency range and above, and it has been widely investigated for communication systems \cite{bookSFO,khanzadi2014}. In addition, it is also known to degrade sensing performance. For example, \ac{PN} imposes challenges to important \ac{ISAC} tasks such as clutter removal for effective sensing, e.g., in indoor environments \cite{henninger2023,henninger2024}. In \ac{FMCW} radar systems, \ac{PN} introduces random frequency deviations to the chirp signals, which ultimately lead to \ac{SNR} loss and distort the \ac{IF} signal spectrum \cite{ayhan2016}. In the context of digital waveforms for radar sensing, the influence of \ac{PN} on \ac{PMCW} was also analyzed for radar networks in \cite{kou2024} and \cite{werbunat2024}. The latter study demonstrated that \ac{PN} leads to an \ac{SNR} reduction due to degradation of the correlation properties of the adopted \ac{PRBS} and a stripe along the Doppler shift direction for each target in the radar image. In \cite{schweizer2018,aguilar2024}, these effects were also briefly discussed for \ac{OFDM} radar and their causes attributed to a \ac{PN}-induced \ac{CPE} and \ac{ICI}. Further studies such as \cite{keskin2023,keskin2023_2,koivunen2024} derive \acp{CRLB} on the achievable accuracy of range and velocity, which is related to the Doppler shift, of a single radar target under \ac{PN} influence. In \cite{keskin2023}, a strategy to estimate the \ac{PN} parameters and use them to improve the accuracy of range and velocity estimates is also proposed. A detailed analysis of the \ac{PN} influence on the \ac{SNR} loss and distortion of radar images in both range and Doppler shift directions for different choices of \ac{OFDM} waveform parameters and scenarios has, however, not been reported in the literature so far. This article aims to fill this gap by performing an analysis of sensing performance degradation due to \ac{PN} in \ac{mmWave} \ac{OFDM}-based \ac{ISAC}. For that purpose, the distortion of digital modulation constellation and the resulting degradation of target reflection main lobe and sidelobes levels in both range and Doppler shift directions of the radar image are adopted as performance metrics. Furthermore, the \ac{PN} effect when changing individual \ac{OFDM} signal parameters is analyzed, and both single- and multi-target scenarios with moving and static targets are addressed.
	
	The contributions of this article can be summarized as follows:
	\begin{itemize}
		
		\item A thorough description and mathematical formulation of the effects of \ac{PN} on mono- and bistatic \ac{OFDM}-based \ac{ISAC} architectures is given. This includes closed-form expressions showing the effects of \ac{PN}-induced \ac{CPE} and \ac{ICI} on \ac{OFDM} symbols in the frequency domain.
		
		\item  An explanation of the relationship between \ac{PN} and \ac{OFDM} signal parameters, including the presentation of results to illustrate how digital modulation constellations are distorted is provided. In this sense, a comprehensive discussion on when \ac{CPE} or \ac{ICI} become the dominant \ac{PN}-induced degradation to the \ac{OFDM}-based \ac{ISAC} system is made. To numerically quantify the constellation distortion, the \ac{EVM} is adopted as a performance parameter and \ac{QPSK}, \mbox{16-\ac{QAM}}, \mbox{64-\ac{QAM}}, and \mbox{256-\ac{QAM}} modulations are considered, besides different \ac{OFDM} signal parameter choices and \ac{PN} levels.
		
		\item A comprehensive analysis of the \ac{PN}-induced sensing performance degradation in \ac{mmWave} \ac{OFDM}-based \ac{ISAC} systems with both flexible and \ac{3GPP}-compliant \ac{OFDM} signal parameterization is conducted. For that purpose, the distortion of main lobe and sidelobes of targets in the range-Doppler shift radar images are evaluated by means of the \ac{PPLR}, \ac{PSLR}, and \ac{ISLR} parameters. The \ac{PPLR} serves as a measure of power loss at the target reflection mainlobe, while \ac{PSLR} and \ac{ISLR} quantify the increase in the sidelobe level, which may eventually prevent the detection of the actual target or result in ghost targets. While these parameters are evaluated for a single target, they allow predicting the interaction between multiple targets in a radar image. In addition, the image \ac{SIR} was also considered to analyze not only the distortion of the target reflection itself, but also the increase in the interference floor in the radar image caused by \ac{PN}. The aforementioned parameters are also analyzed after \ac{CPE} estimation and correction, which allows to partly compensate the \ac{PN} effects at reasonable computational complexity costs.
				
	\end{itemize}
	
	The remainder of this article is organized as follows. Section~\ref{sec:sysModel} formulates the system model of an \ac{OFDM}-based \ac{ISAC} system under influence of \ac{PN}. In Section~\ref{sec:perfAnalysis}, a communication and radar sensing performance analysis of the aforementioned \ac{ISAC} system under \ac{PN} influence is performed. Finally, concluding remarks are presented in Section~\ref{sec:conclusion}.	
	
	\section{System Model}\label{sec:sysModel}
	
	\begin{figure}[!t]
		\centering
		
		\psfrag{A}[c][c]{\scriptsize $\SI{-10}{dB/dec}$}
		\psfrag{B}[c][c]{\scriptsize $\SI{-20}{dB/dec}$}
		
		\psfrag{CC}[c][c]{\scriptsize $10\log_{10}(L_0)$}
		\psfrag{DD}[c][c]{\scriptsize $10\log_{10}(L_\mathrm{floor})$}
		\psfrag{EE}[c][c]{\scriptsize $\log_{10}(f_\mathrm{corner})$}
		\psfrag{FF}[c][c]{\scriptsize $\log_{10}(B_\mathrm{PLL})$}
		
		\psfrag{G}[c][c]{\scriptsize $10\log_{10}\left(S_{\theta^\mathrm{PN}_\mathrm{Tx}}(f)\right)$}
		\psfrag{H}[c][c]{\scriptsize $\log_{10}(f)$}
		
		\includegraphics[width=8.5cm]{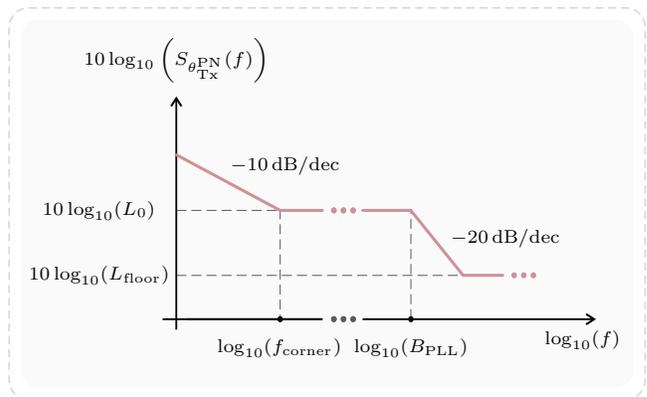}
		\captionsetup{justification=raggedright,labelsep=period,singlelinecheck=false}
		\caption{\ Double-sided PSD profile of the assumed PN model for an PLL-based LO \cite{bookSFO,khanzadi2014,schweizer2018}.}\label{fig:pnPSD_model}	
		
	\end{figure}
	
	In an \ac{OFDM}-based \ac{ISAC} system, constellation symbols \mbox{$\mathbf{d}_\text{Tx}$} belonging to a modulation alphabet \mbox{$\mathcal{M}$} such that \mbox{$\mathbf{d}_\text{Tx}^\text{OFDM}\in\mathcal{M}^{NM\times1}\subset\mathbb{C}^{NM\times1}$} undergo \ac{S/P} conversion and are reshaped into a discrete-frequency domain frame containing \mbox{$N\in\mathbb{N}_{>0}$} subcarriers and \mbox{$M\in\mathbb{N}_{>0}$} \ac{OFDM} symbols. This frame is then transformed into the discrete-time domain via \acp{IDFT} along the subcarriers at each \ac{OFDM} symbol. A \ac{CP} of length \mbox{$N_\mathrm{CP}\in\mathbb{N}_{\geq0}$} is then prepended to each of the aforementioned \ac{OFDM} symbols.  After \ac{P/S} conversion on the resulting discrete-time domain frame, \ac{D/A} conversion with sampling rate $F_\mathrm{s}$ is performed on the resulting sample stream and the generated signal undergoes \ac{LP} filtering, being then denoted as $x_\mathrm{BB}(t)\in\mathbb{C}$. It is considered that the \ac{BB} analog signal $x_\mathrm{BB}(t)$ has bandwidth $B\leq F_\mathrm{s}$, which results in an \ac{OFDM} subcarrier spacing \mbox{$\Delta f=B/N$}. After up-converting $x_\mathrm{BB}(t)$ to the carrier frequency $f_\text{c}\gg B$ with an \ac{I/Q} mixer, the \ac{RF} analog signal $x_\mathrm{RF}(t)\in\mathbb{C}$ expressed as
	\begin{figure*}[!b]
		\hrulefill
		\vspace*{4pt}
		\setcounter{equation}{3}
		\begin{equation}\label{eq:y_RF_2}
			y_\mathrm{RF}(t) = \sum_{p=0}^{P-1}\alpha_p~\left[x_\mathrm{BB}(t-\tau_p)~\e^{\im[2\pi f_\mathrm{c}(t-\tau_p)+\psi_\mathrm{Tx}+\theta^\mathrm{PN}_\mathrm{Tx}(t-\tau_p)]}\right]~\e^{\im 2\pi f_{\mathrm{D},p}t}	
		\end{equation}
	\end{figure*}
	\begin{figure*}[!b]
		\setcounter{equation}{5}
		\hrulefill
		\vspace*{4pt}
		\begin{align}\label{eq:y_BB_extended}
			y_\mathrm{BB}(t) &= \left\{\sum_{p=0}^{P-1}\alpha_p~\left[x_\mathrm{BB}(t-\tau_p)~\e^{\im[2\pi f_\mathrm{c}(t-\tau_p)+\psi_\mathrm{Tx}+\theta^\mathrm{PN}_\mathrm{Tx}(t-\tau_p)]}\right]~\e^{\im 2\pi f_{\mathrm{D},p}t}\right\}~\e^{-\im[2\pi f_\mathrm{c}t+\psi_\mathrm{Rx}+\theta^\mathrm{PN}_\mathrm{Rx}(t)]}\nonumber\\
			&= \sum_{p=0}^{P-1}\alpha_p~ x_\mathrm{BB}(t-\tau_p)~\e^{\im[\psi_\mathrm{Tx}-\psi_\mathrm{Rx}-2\pi f_\mathrm{c}\tau_p]}~\e^{\im 2\pi f_{\mathrm{D},p}t}~\e^{\im\left[\theta^\mathrm{PN}_\mathrm{Tx}(t-\tau_p)-\theta^\mathrm{PN}_\mathrm{Rx}(t)\right]}
		\end{align}
	\end{figure*}
	\begin{figure*}[!b]
		\setcounter{equation}{9}
		\hrulefill
		\vspace*{4pt}
		\begin{equation}\label{eq:y_BB_simplifTheta}
			y_\mathrm{BB}(t) = \left[\sum_{p=0}^{P-1}\alpha_p~ x_\mathrm{BB}(t-\tau_p)~\e^{\im\psi_p}~\e^{\im 2\pi f_{\mathrm{D},p}t}\right]\left\{1+\im[\theta^\mathrm{PN}_\mathrm{Tx}(t-\tau_p)-\theta^\mathrm{PN}_\mathrm{Rx}(t)]\right\}
		\end{equation}
	\end{figure*}
	\begin{figure*}[!t]
		\setcounter{equation}{11}
		\begin{align}\label{eq:y_samp_time}
			y_{n,m} =& \left(\sum_{p=0}^{P-1}\alpha_p~ x_{\left<n-n_{\Delta,p}\right>_N,m}~\e^{\im\psi_p}~\e^{\im 2\pi f_{\mathrm{D},p}[m(N+N_\mathrm{CP})+N_\mathrm{CP}+n]/B}\right)\nonumber\\
			&\left[1+\im\left(\theta^\mathrm{PN}_{\mathrm{Tx}}\left([m(N+N_\mathrm{CP})+N_\mathrm{CP}+n-n_{\Delta,p}]/B\right)-\theta^\mathrm{PN}_{\mathrm{Rx}}\left([m(N+N_\mathrm{CP})+N_\mathrm{CP}+n]/B\right)\right)\right]
		\end{align}
		\vspace*{4pt}
		\hrulefill
		\setcounter{equation}{12}
		\begin{align}\label{eq:Y_samp_freq}
			Y_{k,m} =& \left[\sum_{p=0}^{P-1}\alpha_{p}~X_{k,m}~\e^{\im\psi_p}~\e^{-\im 2\pi k\tau_pB/N}~\e^{\im 2\pi f_{\mathrm{D},p}[m(N+N_\mathrm{CP})+N_\mathrm{CP}]/B}\right]\nonumber\\
			&+ \frac{\im}{N}\sum_{\kappa=-N/2}^{N/2}\left[\sum_{p=0}^{P-1}\alpha_{p}~X_{\kappa,m}~\e^{\im\psi_p}~\e^{-\im 2\pi \kappa\tau_pB/N}~\e^{\im 2\pi f_{\mathrm{D},p}[m(N+N_\mathrm{CP})+N_\mathrm{CP}]/B}\right]\nonumber\\
			&\phantom{+}\times\sum_{n=0}^{N-1}\left(\theta^\mathrm{PN}_{\mathrm{Tx}}\left([m(N+N_\mathrm{CP})+N_\mathrm{CP}+n-n_{\Delta,p}]/B\right)-\theta^\mathrm{PN}_{\mathrm{Rx}}\left([m(N+N_\mathrm{CP})+N_\mathrm{CP}+n]/B\right)\right)\e^{\im 2\pi(\kappa-k)n/N}
		\end{align}
		\vspace*{4pt}
		\hrulefill
	\end{figure*}
	\setcounter{equation}{0}
	\begin{equation}\label{eq:x_RF}
		x_\mathrm{RF}(t) = x_\mathrm{BB}(t)~\e^{\im\left[2\pi f_\mathrm{c}t+\psi_\mathrm{Tx}+\theta^\mathrm{PN}_\mathrm{Tx}(t)\right]}
	\end{equation}
	is obtained. It is worth highlighting that $x_\mathrm{RF}(t)$ is only modeled as a complex-valued signal for simplicity. In practice, it should be real-valued and both the frequency band of interest centered at $f_\mathrm{c}$ and its mirror image centered at $-f_\mathrm{c}$ should be considered when describing $x_\mathrm{RF}(t)$ and the \ac{RF} signals derived from it. Furthermore, the term \mbox{$\psi_\mathrm{Tx}\in\mathbb{R}$} in \eqref{eq:x_RF} denotes the transmitter carrier phase offset, and \mbox{$\theta^\mathrm{PN}_\mathrm{Tx}(t)\in\mathbb{R}$} represents the transmitter \ac{PN} that has a double-sided \ac{PSD} \mbox{$S_{\theta^\mathrm{PN}_\mathrm{Tx}}(f)\in\mathbb{R}_{\geq0}$} \cite{walt2023}. The profile of the aforementioned \ac{PSD} is shown in Fig.~\ref{fig:pnPSD_model} assuming that the \ac{LO} is implemented by using a \ac{PLL} \cite{bookSFO,khanzadi2014,schweizer2018}. In this figure, $f_\mathrm{corner}$ and $B_\mathrm{PLL}$ denote the flicker corner frequency and the \ac{PLL} loop bandwidth, respectively. It is assumed that the \ac{PN} \ac{PSD} has a slope of $\SI{-10}{dB/dec}$ until $f_\mathrm{corner}$. Between $f_\mathrm{corner}$ and $B_\mathrm{PLL}$, the \ac{PSD} assumes the value $L_0$, which is defined as the in-band \ac{PN} level. After that interval, the \ac{PSD} decays at a rate of $\SI{-20}{dB/dec}$ until it reaches the \ac{PN} floor $L_\mathrm{floor}$. Both $L_0$ and $L_\mathrm{floor}$ are measured in $\mathrm{rad}^2/\mathrm{Hz}$. When the base-10 logarithm of the \ac{PSD} of the \ac{PN} is taken, the resulting unit becomes $\mathrm{dBc}/\mathrm{Hz}$. Based on these parameters, the transmit \ac{PN} double-sided \ac{PSD} can be mathematically described by \cite{bookSFO,schweizer2018}
	\setcounter{equation}{1}
	\begin{equation}\label{eq:pnPSD}
		S_{\theta^\mathrm{PN}_\mathrm{Tx}}(f) = \frac{B_\mathrm{PLL}^2L_0}{B_\mathrm{PLL}^2+f^2}\left(1+\frac{f_\mathrm{corner}}{f}\right)+L_\mathrm{floor}.
	\end{equation}
	
	After being amplified by a \ac{PA} and radiated by the transmit antenna of the considered \ac{ISAC} system, $x_\mathrm{RF}(t)$ propagates through \mbox{$P\in\mathbb{N}_{>0}$} labeled as $p\in\{0,1,\dots,P-1\}$. Assuming that $p\mathrm{th}$ path is associated with attenuation $\alpha_p$, delay $\tau_p$, and Doppler shift $f_{\mathrm{D},p}$, the receive captured signal by the receive antenna of the considered \ac{ISAC} system can be denoted as $y_\mathrm{RF}(t)\in\mathbb{C}$, which is expressed as
	\begin{equation}\label{eq:y_RF_1}
		y_\mathrm{RF}(t) = \sum_{p=0}^{P-1}\alpha_p~x_\mathrm{RF}(t-\tau_p)~\e^{\im 2\pi f_{\mathrm{D},p}t}.
	\end{equation}
	By substituting \eqref{eq:x_RF} into \eqref{eq:y_RF_1}, \eqref{eq:y_RF_2} is obtained. After $y_\mathrm{RF}(t)$ is amplified by a \ac{LNA}, it undergoes \ac{BP} filtering and is down-converted with an \ac{I/Q} mixer. The resulting signal from the aforementioned down-conversion, $y_\mathrm{BB}(t)\in\mathbb{C}$, can be expressed as	
	\setcounter{equation}{4}	
	\begin{equation}\label{eq:y_BB}
		y_\mathrm{BB}(t) = y_\mathrm{RF}(t)~\e^{-\im\left[2\pi f_\mathrm{c}t+\psi_\mathrm{Rx}+\theta^\mathrm{PN}_\mathrm{Rx}(t)\right]},
	\end{equation}
	where \mbox{$\psi_\mathrm{Rx}\in\mathbb{R}$} and \mbox{$\theta^\mathrm{PN}_\mathrm{Rx}(t)\in\mathbb{R}$}  denote the receiver carrier phase offset and the receiver \ac{PN} time series in radians, respectively. In this article, it is assumed that the receiver \ac{LO} is also derived from a \ac{PLL} and that its \ac{PN} has a double-sided \ac{PSD} \mbox{$S_{\theta^\mathrm{PN}_\mathrm{Rx}}(f)\in\mathbb{R}_{\geq0}$} with the same profile as its counterpart at the transmitter side given by \eqref{eq:pnPSD} and shown in Fig.~\ref{fig:pnPSD_model}. By substituting \eqref{eq:y_RF_2} into \eqref{eq:y_BB}, the latter equation can be expanded as in \eqref{eq:y_BB_extended}.
	It is next assumed that \mbox{$\theta^\mathrm{PN}_\mathrm{Tx}(t-\tau_p)-\theta^\mathrm{PN}_\mathrm{Rx}(t)\ll1$}, which is reasonable for typical \ac{PN} levels in practice. Therefore, it holds that \mbox{$\cos\left[\theta^\mathrm{PN}_\mathrm{Tx}(t-\tau_p)-\theta^\mathrm{PN}_\mathrm{Rx}(t)\right]\approx 1$} and \mbox{$\sin\left[\theta^\mathrm{PN}_\mathrm{Tx}(t-\tau_p)-\theta^\mathrm{PN}_\mathrm{Rx}(t)\right]\approx \theta^\mathrm{PN}_\mathrm{Tx}(t-\tau_p)-\theta^\mathrm{PN}_\mathrm{Rx}(t)$}, which allows further simplifying
	\setcounter{equation}{6}
	\begin{align}\label{eq:PNsimplif_aux}
		e^{\im[\theta^\mathrm{PN}_\mathrm{Tx}(t-\tau_p)-\theta^\mathrm{PN}_\mathrm{Rx}(t)]}=&\cos\left[\theta^\mathrm{PN}_\mathrm{Tx}(t-\tau_p)-\theta^\mathrm{PN}_\mathrm{Rx}(t)\right]\nonumber\\
		&+ \im\sin\left[\theta^\mathrm{PN}_\mathrm{Tx}(t-\tau_p)-\theta^\mathrm{PN}_\mathrm{Rx}(t)\right]
	\end{align}
	as
	\begin{equation}\label{eq:PNsimplif}
		\e^{\im[\theta^\mathrm{PN}_\mathrm{Tx}(t-\tau_p)-\theta^\mathrm{PN}_\mathrm{Rx}(t)]} \approx 1 + \im\left[\theta^\mathrm{PN}_\mathrm{Tx}(t-\tau_p)-\theta^\mathrm{PN}_\mathrm{Rx}(t)\right].
	\end{equation}
	If, in addition, the phase terms associated with transmitter, receiver and the delay of the $p\mathrm{th}$ path are jointly represented by $\psi_p$ such that
	\begin{equation}\label{eq:psi_p}
		\psi_p = \psi_\mathrm{Tx}-\psi_\mathrm{Rx}-2\pi f_\mathrm{c}\tau_p,
	\end{equation}
	then \eqref{eq:y_BB_extended} can be rewritten as in \eqref{eq:y_BB_simplifTheta}.
	\begin{figure*}[!b]
		\setcounter{equation}{13}
		\hrulefill
		\vspace*{4pt}
		\begin{equation}\label{eq:Y_samp_freq_woPN}
			\tilde{Y}_{k,m} = \sum_{p=0}^{P-1}\alpha_{p}~X_{k,m}~\e^{\im\psi_p}~\e^{-\im 2\pi k\tau_pB/N}~\e^{\im 2\pi f_{\mathrm{D},p}[m(N+N_\mathrm{CP})+N_\mathrm{CP}]/B}
		\end{equation}
	\end{figure*}
	\begin{figure*}[!b]
		\setcounter{equation}{14}
		\hrulefill
		\vspace*{4pt}
		\begin{align}\label{eq:PN_samp_freq}
			\eta^\mathrm{PN}_{k,m} =&\frac{\im}{N}\sum_{\kappa=-N/2}^{N/2}\alpha~X_{\kappa,m}~\e^{\im\psi}~\e^{-\im 2\pi \kappa\tau B/N}~\e^{\im 2\pi f_\mathrm{D}[m(N+N_\mathrm{CP})+N_\mathrm{CP}]/B}\nonumber\\
			&\times\sum_{n=0}^{N-1}\left(\theta^\mathrm{PN}_{\mathrm{Tx}}\left([m(N+N_\mathrm{CP})+N_\mathrm{CP}+n-n_{\Delta,p}]/B\right)-\theta^\mathrm{PN}_{\mathrm{Rx}}\left([m(N+N_\mathrm{CP})+N_\mathrm{CP}+n]/B\right)\right)\e^{\im 2\pi(\kappa-k)n/N}
		\end{align}
	\end{figure*}
	After \ac{LP} filtering, the receive \ac{BB} signal $y_\mathrm{BB}(t)$ undergoes \ac{A/D} conversion with sampling rate $F_\mathrm{s}$, which is associated with the sampling period \mbox{$T_\mathrm{s}=1/F_\mathrm{s}$}. If \mbox{$F_\mathrm{s}>B$}, it is assumed that digital low-pass filtering and downsampling to $B$ is performed. Disregarding \ac{AWGN}, this yields the discrete-time domain sequence $y[s]\in\mathbb{C}$ that can expressed as
	\setcounter{equation}{10}
	\begin{equation}\label{eq:rxSampling}
		y[s] = y_\mathrm{BB}(t)\Big|_{t=sT_\mathrm{s}},
	\end{equation}
	with \mbox{$s\in\{0,1,\dots,M(N+N_\mathrm{CP})-1\}$}. By performing \ac{S/P} conversion on $y[s]$ and discarding the \ac{CP} at each \ac{OFDM} symbol, the discrete-time domain frame $\mathbf{y}\in\mathbb{C}^{N\times M}$ is obtained. Based on \eqref{eq:y_BB_simplifTheta} and \eqref{eq:rxSampling}, the sample at the $n\mathrm{th}$ row and $m\mathrm{th}$ column of $\mathbf{y}$ can be expressed as in \eqref{eq:y_samp_time} assuming that the Doppler-shift induced \ac{ICI} is negligible for all $P$ targets for simplicity. Without loss of generality, \mbox{$n_{\Delta,p}=\tau_pB$} in this equation is assumed to be an integer to simplify the notation. In addition, $\left<\cdot\right>_N$ is the modulo $N$ operator. Next, the \ac{OFDM} symbols contained in $\mathbf{y}$ undergo \acp{DFT}. This results in the discrete-frequency domain \ac{OFDM} frame $\mathbf{Y}\in\mathbb{C}^{N\times M}$, whose sample at $k\mathrm{th}$ subcarrier, \mbox{$k\in\{-N/2,1,\dots,N/2-1\}$}, at the $m\mathrm{th}$ \ac{OFDM} symbol is expressed as in \eqref{eq:Y_samp_freq}.
	
	Considering a single path, i.e., \mbox{$P=1$}, and omitting the subscript $p$ for the sake of simplicity, the \ac{OFDM} frame $\mathbf{Y}$ is split into the \ac{PN}-free \ac{OFDM} frame $\tilde{\mathbf{Y}}\in\mathbb{C}^{N\times M}$ and the \ac{PN}-induced distortion to the \ac{OFDM} frame $\bm{\eta}^\mathrm{PN}\in\mathbb{C}^{N\times M}$. Consequently, the sample at the $k\mathrm{th}$ row and $m\mathrm{th}$ column of the \ac{PN}-free \ac{OFDM} frame $\tilde{\mathbf{Y}}$ is expressed as in \eqref{eq:Y_samp_freq_woPN}, while the distortion at the same position of $\bm{\eta}^\mathrm{PN}$ is expressed as in \eqref{eq:PN_samp_freq}. Splitting \eqref{eq:PN_samp_freq} into the cases where the sum index $\kappa$ is equal to or different than the analyzed subcarrier index $k$, i.e., $\kappa=k$ or $\kappa\neq k$, \mbox{$\eta^\mathrm{PN,CPE}_{k,m}\in\mathbb{C}$} and \mbox{$\eta^\mathrm{PN,ICI}_{k,m}$} are respectively obtained. These variables represent the \ac{PN}-induced \ac{CPE} and the \ac{PN}-induced \ac{ICI}, which are expressed as in \eqref{eq:pnCPE} and \eqref{eq:pnICI}, respectively \cite{armada2001}.
	\begin{figure*}[!t]
		\setcounter{equation}{15}
		\begin{equation}\label{eq:pnCPE}
			\eta^\mathrm{PN,CPE}_{k,m} =\frac{\im}{N}\tilde{Y}_{k,m}~\sum_{n=0}^{N-1}\left(\theta^\mathrm{PN}_{\mathrm{Tx}}\left([m(N+N_\mathrm{CP})+N_\mathrm{CP}+n-n_{\Delta,p}]/B\right)-\theta^\mathrm{PN}_{\mathrm{Rx}}\left([m(N+N_\mathrm{CP})+N_\mathrm{CP}+n]/B\right)\right)
		\end{equation}
		\vspace*{4pt}
		\hrulefill
	\end{figure*}
	\begin{figure*}[!t]
		\begin{align}\label{eq:pnICI}
			\eta^\mathrm{PN,ICI}_{k,m}=&\frac{\im}{N}\sum_{\kappa=-N/2,\kappa\neq k}^{N/2}~\tilde{Y}_{\kappa,m}\nonumber\\
			&\times\sum_{n=0}^{N-1} \left(\theta^\mathrm{PN}_{\mathrm{Tx}}\left([m(N+N_\mathrm{CP})+N_\mathrm{CP}+n-n_{\Delta,p}]/B\right)-\theta^\mathrm{PN}_{\mathrm{Rx}}\left([m(N+N_\mathrm{CP})+N_\mathrm{CP}+n]/B\right)\right)\e^{\im 2\pi(\kappa-k)n/N}
		\end{align}
		\vspace*{4pt}
		\hrulefill
	\end{figure*}
	\begin{figure*}[!t]
		\begin{equation}\label{eq:thetaMean}
			\Phi^\mathrm{PN,CPE}_m = \frac{1}{N}\sum_{n=0}^{N-1}\left(\theta^\mathrm{PN}_{\mathrm{Tx}}\left([m(N+N_\mathrm{CP})+N_\mathrm{CP}+n-n_{\Delta,p}]/B\right)-\theta^\mathrm{PN}_{\mathrm{Rx}}\left([m(N+N_\mathrm{CP})+N_\mathrm{CP}+n]/B\right)\right)
		\end{equation}
		\vspace*{4pt}
		\hrulefill
	\end{figure*}
	\begin{figure*}[!t]
		\centering
		\subfloat[ ]{
			
			\psfrag{A}[c][c]{\tiny $\tau_0,R_0$}
			\psfrag{B}[c][c]{\tiny $\theta^\mathrm{PN}_\mathrm{Tx}(t-\tau_0)$}
			\psfrag{C}[c][c]{\tiny $\theta^\mathrm{PN}_\mathrm{Tx}(t)$}
			
			\includegraphics[height=5cm]{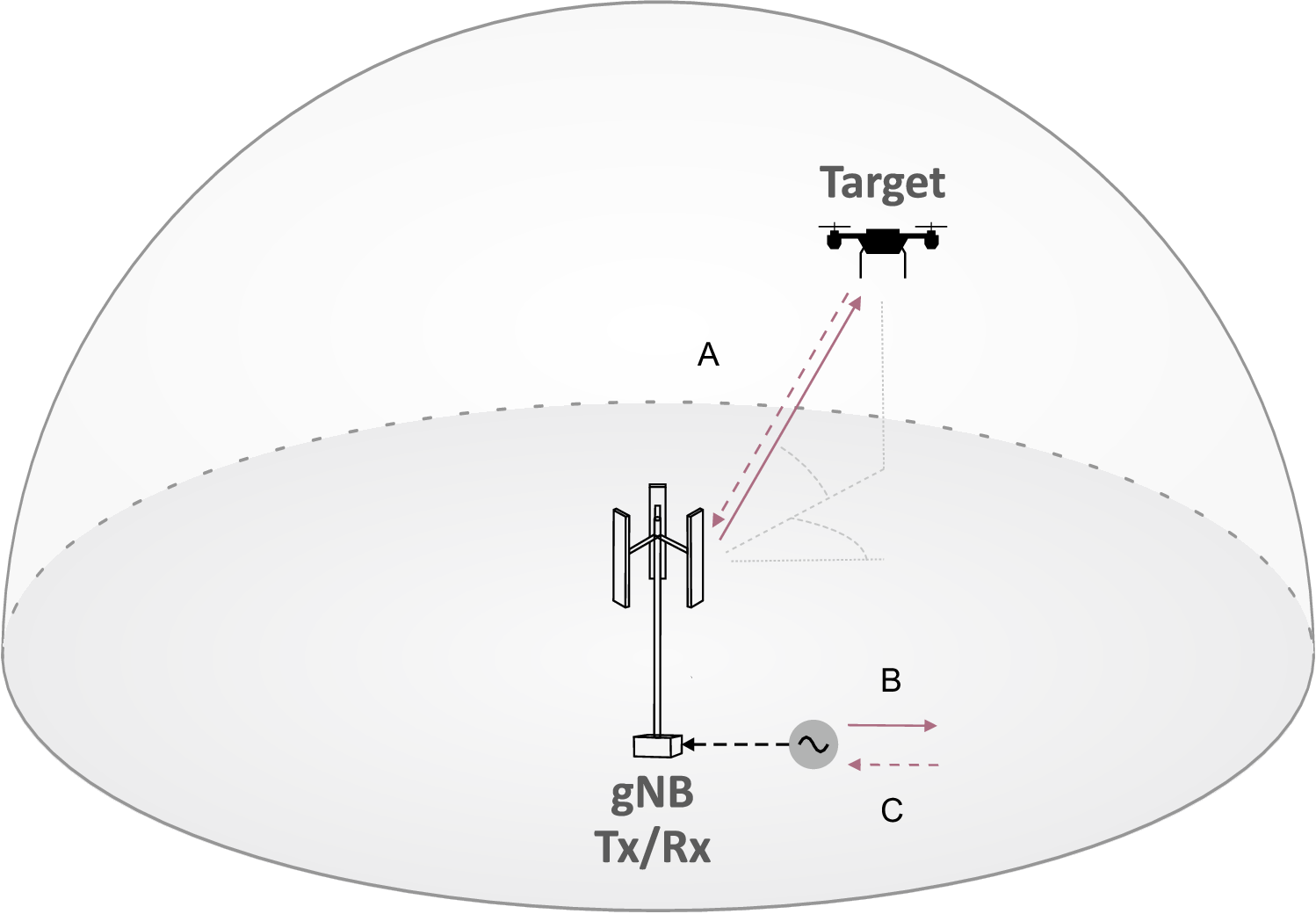}
		}\hspace{0.25cm}
		\subfloat[ ]{
			
			\psfrag{A}[c][c]{\tiny $\tau^\mathrm{Tx-T}_{1},R^\mathrm{Tx-T}_1$}
			\psfrag{B}[c][c]{\tiny $\tau^\mathrm{T-Rx}_{1},R^\mathrm{T-Rx}_1$}
			\psfrag{C}[c][c]{\tiny $\tau_{0},R_0$}
			
			\psfrag{D}[c][c]{\tiny $\theta^\mathrm{PN}_\mathrm{Tx}(t-\tau_1)$}
			\psfrag{E}[c][c]{\tiny $\theta^\mathrm{PN}_\mathrm{Tx}(t-\tau_0)$}
			
			\psfrag{F}[c][c]{\tiny $\theta^\mathrm{PN}_\mathrm{Rx}(t)$}
			\psfrag{G}[c][c]{\tiny $\theta^\mathrm{PN}_\mathrm{Rx}(t)$}
			
			\includegraphics[height=5cm]{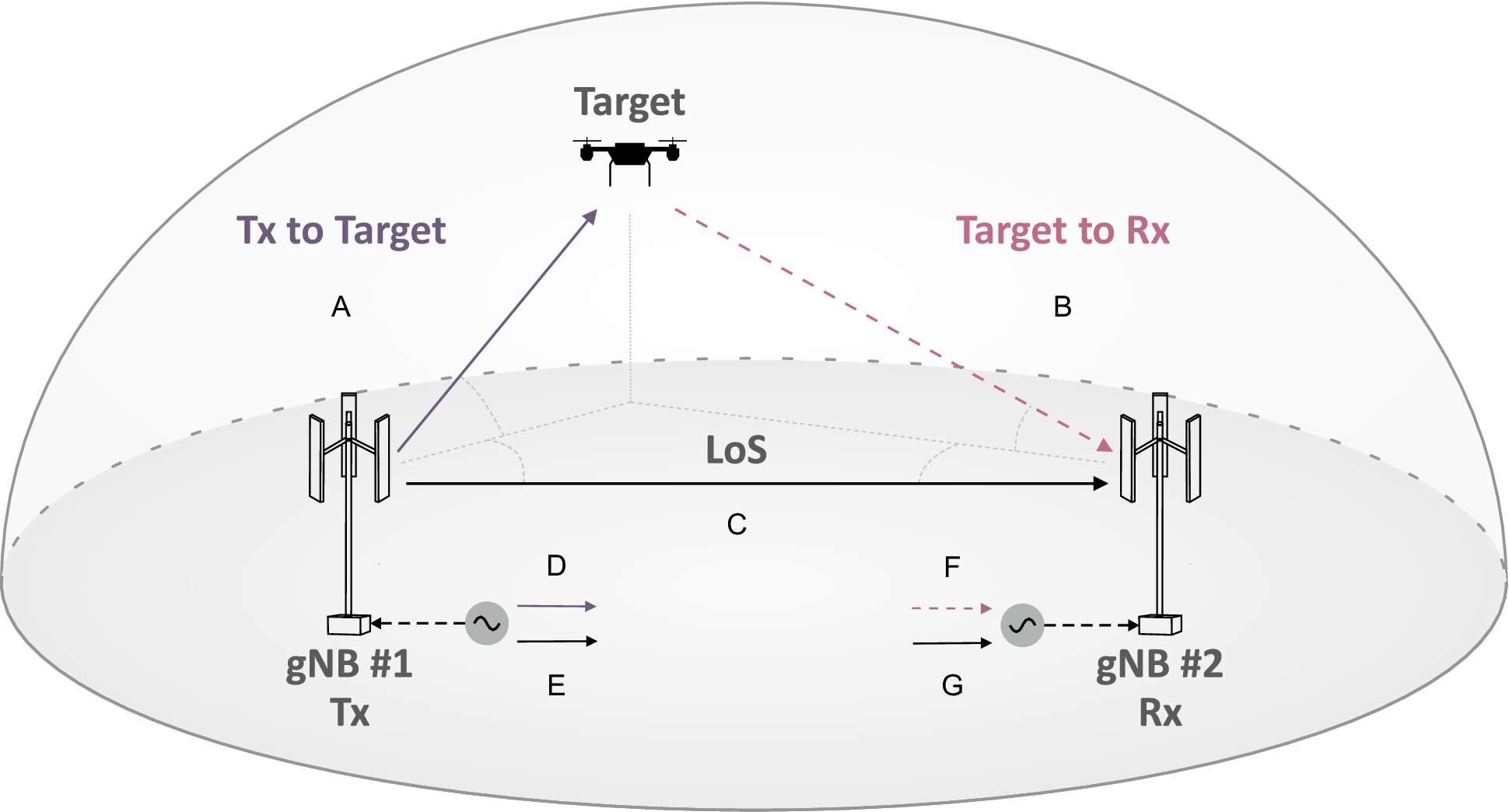}
		}
		
		\captionsetup{justification=raggedright,labelsep=period,singlelinecheck=false}
		\caption{\ Monostatic (a) and bistatic (b) ISAC scenarios. In the monostatic scenario in (a), a gNB performs monostatic sensing to detect a single target of index \mbox{$p=0$} at a range $R_0$, which results in a single propagation path (\mbox{$P=1$}) with propagation delay \mbox{$\tau_0=R_0/c_0$}, where $c_0$ is the speed of light in vacuum. In this scenario, the effective transmit PN contribution is \mbox{$\theta^\mathrm{PN}_\mathrm{Tx}(t-\tau_0)$}, while the receive PN is \mbox{$\theta^\mathrm{PN}_\mathrm{Rx}(t)=\theta^\mathrm{PN}_\mathrm{Tx}(t)$}. In the bistatic ISAC scenario in (b), two gNBs communicate with one another through an LoS path of index \mbox{$p=0$}, physical length $R_0$, and propagation delay \mbox{$\tau_0=R_0/c_0$}. In addition, the aforementioned gNBs also perfom bistatic sensing of a single target of index \mbox{$p=1$} at a range $R_1$, and with corresponding propagation delay \mbox{$\tau_1=R_1/c_0$}. It holds that \mbox{$\tau_1=\tau^\mathrm{Tx-T}_{1}+\tau^\mathrm{T-Rx}_{1}$}, where \mbox{$\tau^\mathrm{Tx-T}_{1}=R^\mathrm{Tx-T}_{1}/c_0$}, \mbox{$\tau^\mathrm{T-Rx}_{1}=R^\mathrm{T-Rx}_{1}/c_0$}, and $R^\mathrm{Tx-T}_{1}$ and $R^\mathrm{T-Rx}_{1}$ are respectively the ranges between transmitting gNB and radar target, and radar target and receiving gNB. In the bistatic case, the transmit PN contributions for the LoS path ($p=0$) and the path associated with the radar target are \mbox{$\theta^\mathrm{PN}_\mathrm{Tx}(t-\tau_0)$} and \mbox{$\theta^\mathrm{PN}_\mathrm{Tx}(t-\tau_1)$}, respectively. Unlike in the monostatic case, the receive PN contribution for both paths, \mbox{$\theta^\mathrm{PN}_\mathrm{Rx}(t)$}, has no correlation with \mbox{$\theta^\mathrm{PN}_\mathrm{Tx}(t)$} in bistatic ISAC as different oscillators are used by the transmitting and receiving gNBs.}\label{fig:isacArchitectures}
		
	\end{figure*}
	
	According to \eqref{eq:pnCPE}, the \ac{CPE} causes a phase rotation due to the multiplication of $\tilde{Y}^p_{k,m}$ by $\im\Phi^\mathrm{PN,CPE}_m$, where $\Phi^\mathrm{PN,CPE}_m\in\mathbb{R}$ is expressed as in \eqref{eq:thetaMean} and does not depend on the subcarrier index $k$. Consequently, all subcarriers within the same \ac{OFDM} symbol suffer the same phase rotation. While this can be corrected to a certain extent for communication by performing channel estimation and equalization with pilot subcarriers strategically placed in the \ac{OFDM} frame, it may become an issue for radar sensing. This is due to the fact that the goal of radar sensing is to perform highly accurate channel estimation over consecutive \ac{OFDM} symbols to allow estimating range and Doppler shift of targets \cite{giroto2021_tmtt}. While a phase rotation does not degrade the range profiles obtained at each \ac{OFDM} symbol, it significantly impairs the Doppler shift estimation. This happens since the aforementioned process consists of converting the Doppler-shift induced phases into a Doppler shift estimate by performing one \ac{DFT} over multiple \ac{OFDM} symbols for. Since \ac{CPE} alters the phase of each \ac{OFDM} symbol, the obtained Doppler shift profiles will be distorted, presenting possible loss of the target peak power and increase of the sidelobe levels along the Doppler shift direction at the target range bin of the ultimately obtained range-Doppler shift  radar image. The \ac{PN}-induced \ac{ICI}, in turn, already degrades each \ac{OFDM} symbol individually and not only when analyzing the whole frame as in the \ac{CPE} case. The \ac{ICI} causes loss of orthogonality between the subcarriers, reducing their effective \ac{SINR} and degrading the performance of both \ac{OFDM}-based communication and also sensing, resulting in peak power loss and sidelobe level increase in the obtained range profiles at each \ac{OFDM} symbol and ultimately degrading the obtained range-Doppler shift radar image. According to \cite{armada2001,bookSFO,schweizer2018}, it is assumed that the \ac{CPE} is the dominant \ac{PN}-induced effect if \mbox{$\Delta f/2>B_\mathrm{PLL}$}, while \ac{ICI} becomes the most relevant effect for \mbox{$\Delta f/2<B_\mathrm{PLL}$}. Since $B_\mathrm{PLL}$ values of $\SI{50}{\kilo\hertz}$ \cite{dhar2017}, through around $\SI{150}{\kilo\hertz}$ \cite{keysight2024}, and even up to $\SI{200}{\kilo\hertz}$ \cite{schweizer2018} are reported for \ac{mmWave}, both aforementioned cases can happen in practice, e.g., when subcarrier spacings ranging from $\SI{15}{\kilo\hertz}$ to $\SI{120}{\kilo\hertz}$ are used according to \ac{3GPP} specifications \cite{3GPPTS38211}.	
	
	The influence of \ac{PN} on the \ac{OFDM}-based \ac{ISAC} system performance will depend on whether $\theta^\mathrm{PN}_{\mathrm{Tx}}(t)$ and $\theta^\mathrm{PN}_{\mathrm{Rx}}(t)$ are correlated. While correlated \ac{PN} is observed in monostatic radar sensing due to the use of the same oscillator by both transmitter and receiver, the use of distinct oscillators results in non-correlated \ac{PN} for bistatic radar sensing and communication. The aforementioned scenarios are depicted in Fig.~\ref{fig:isacArchitectures}, where a \ac{gNB} with full-duplex capability \cite{barneto2019,barneto2021} performs monostatic sensing in Fig.~\ref{fig:isacArchitectures}a and two \acp{gNB} perform simultaneous communication and bistatic sensing following the system concept from \cite{giroto2023_EuMW,brunner2024} in Fig.~\ref{fig:isacArchitectures}b.
	
	Besides quantifying the \ac{PN}-induced impairments, there is also a growing interest in the literature regarding \ac{PN} estimation and compensation. In \ac{OFDM}-based optical communication, for instance, a dedicated \ac{CW} tone is simultaneously transmitted at  the \ac{OFDM} signal to fully estimate \ac{PN} fully, i.e., both \ac{CPE} and \ac{ICI} \cite{hussin2012}. A \ac{CW}-based approach has also been proposed in \cite{werbunat2024}, but it is only applicable to \ac{PMCW}-based radar sensing or \ac{ISAC}. Further \ac{PN} compensation strategies were proposed in \cite{casas2002,zou2007,syrjala2009,suyama2009,syrjala2010,wang2016,leshem2017}. Although these methods allow fully estimating the \ac{PN}, they either require knowledge of the channel or perform joint \ac{PN} and channel estimation. In \ac{ISAC} applications, this would lead to rather high computational complexity as a dedicated estimate for each \ac{OFDM} symbol would be needed, especially in scenarios with multiple targets with different Doppler shifts. In addition, the \ac{PN} estimation would be biased towards estimating the \ac{PN} of a dominant, reference path. This is, e.g., the case in bistatic sensing or communication with a \ac{LoS} component between transmitter and receiver, as depicted in Fig.~\ref{fig:isacArchitectures}b. In monostatic sensing, a reference path is usually present due to coupling between transmit and receive antennas. Nevertheless, since this path has virtually zero range, the correlation between $\theta^\mathrm{PN}_{\mathrm{Tx}}(t)$ and $\theta^\mathrm{PN}_{\mathrm{Rx}}(t)$ will be maximum for it, and therefore the \ac{PN} influence on this path will be significantly suppressed. Consequently, the transmit-receive coupled path cannot be used as a reference for \ac{CPE} estimation. Using other targets at higher ranges also not a reliable strategy, as it cannot be ensured that they are static and their Doppler-shift-induced phases may also potentially bias the \ac{PN} estimate.
	
	While the \ac{PN}-induced \ac{ICI} cannot be efficiently combated, \ac{CPE} can be estimated and corrected to a certain extent \cite{oh2020}. This partial \ac{PN} compensation method is particularly efficient when the \ac{PN} varies slowly within an \ac{OFDM} symbol duration, which leads to dominant \ac{CPE} over \ac{PN}-induced \ac{ICI} \cite{armada1998}. In \ac{OFDM}-based communication, \ac{CPE} estimation can be performed by using pilot subcarriers \cite{robertson1995}, at which the observed phase of the modulation symbol at the receiver side is compared with the one of the originally transmitted symbol. The aforementioned procedure can be mathematically described as
	\begin{equation}\label{eq:CPE_est}
		\widehat{\Theta}^\mathrm{PN,CPE}_m=\arg\left\{\sum_{k=0}^{N-1}Y_{k,m}X^*_{k,m}\right\},
	\end{equation}
	where \mbox{$\widehat{\Theta}^\mathrm{PN,CPE}_m\in\mathbb{R}$} is the \ac{CPE} estimate for the \mbox{$m\mathrm{th}$} \ac{OFDM} symbol, \mbox{$(\cdot)^*$}, and \mbox{$\arg\{\cdot\}$} returns the argument of a complex number. This estimate is then used for \ac{CPE} correction on the receive frame $\mathbf{Y}$, producing the frame \mbox{$\mathbf{Y}^\mathrm{corr}\in\mathbb{C}^{N\times M}$} whose element at the \mbox{$k\mathrm{th}$} and \mbox{$m\mathrm{th}$} column is expressed as
	\begin{equation}\label{eq:CPE_corr}
		Y_{k,m}^\mathrm{corr} = Y_{k,m}\e^{-\im \widehat{\Phi}^\mathrm{PN,CPE}_m}.
	\end{equation}	
	In \ac{5G NR}, for instance, the \ac{PT-RS} is used to estimate, track, and allow correcting \ac{PN}-induced \ac{CPE} \cite{qi2018}. The discussed \ac{CPE} estimation and correction can improve the constellation shape and therefore communication and sensing performances. It is, however, is only suitable when there is a dominant, reference path, and therefore only suitable for bistatic \ac{ISAC} architectures as previously discussed. As for the discussed full \ac{PN} estimation strategies, the effectiveness of the \ac{CPE} correction for sensing may also be limited if the \ac{PN} changes rapidly, as multiple targets with rather low \ac{RCS} and at different ranges other than the one of the reference path are expected \cite{zzhang2023,jzhang2024}. These targets will experience significantly different \ac{PN} realizations and consequently \acp{CPE}. In such cases, \ac{CPE} estimation and correction is still desirable as it will improve the communication performance of the bistatic \ac{ISAC} system. To decouple the \ac{CPE} of the reference path and other targets, the \ac{CPE} at a given \ac{OFDM} symbol can be estimated based on the phase of the peak associated with the reference path after obtaining a \ac{CIR} estimate from the pilot subcarriers.
	
	Based on the presented \ac{OFDM}-based \ac{ISAC} system model under \ac{PN} influence, the implications of the \ac{PN}-induced degradation of communication and radar sensing performance in mono- and bistatic \ac{ISAC} architectures as well as the effectiveness of the described \ac{CPE} estimation and correction for communication and bistatic sensing are analyzed in Section~\ref{sec:perfAnalysis}.
	
	\begin{table}[!t]
		\renewcommand{\arraystretch}{1.5}
		\arrayrulecolor[HTML]{708090}
		\setlength{\arrayrulewidth}{.1mm}
		\setlength{\tabcolsep}{4pt}
		
		\centering
		\captionsetup{width=43pc,justification=centering,labelsep=newline}
		\caption{\textsc{Adopted OFDM signal parameters}}
		\label{tab:ofdmParameters}
		\resizebox{\columnwidth}{!}{
			\begin{tabular}{|cc|}
				\hhline{|==|}
				\multicolumn{1}{|c|}{\textbf{Carrier frequency} ($f_\text{c}$)}      & $\SI{26.2}{\giga\hertz}$ \\ \hline
				\multicolumn{1}{|c|}{\textbf{Bandwidth} ($B$)}      & $\SI{1}{\giga\hertz}$ \\ \hline
				\multicolumn{1}{|c|}{\textbf{No. of subcarriers} ($N$)}      & $\{256, 512, 1024, 2048, 4096, 8192, 16384\}$ \\ \hline
				\multicolumn{1}{|c|}{\textbf{CP length} ($N_\mathrm{CP}$)}      & $\{0, N/4, N\}$ \\ \hline
				\multicolumn{1}{|c|}{\textbf{No. of OFDM symbols} ($M$)}      & $\{32, 128, 512, 2048\}$ \\ \hline
				\multicolumn{1}{|c|}{\textbf{Modulation alphabet}}      & \{QPSK, 16-QAM, 64-QAM, 256-QAM\} \\  \hhline{|==|}	
				
			\end{tabular}
		}
	\end{table}	
	\begin{figure}[!t]
		\centering
		\subfloat[ ]{

			\psfrag{22}[c][c]{\scriptsize $10^2$}
			\psfrag{44}[c][c]{\scriptsize $10^4$}
			\psfrag{66}[c][c]{\scriptsize $10^6$}
			\psfrag{88}[c][c]{\scriptsize $10^8$}
			
			\psfrag{-160}[c][c]{\scriptsize -$160$}
			\psfrag{-140}[c][c]{\scriptsize -$140$}
			\psfrag{-120}[c][c]{\scriptsize -$120$}
			\psfrag{-100}[c][c]{\scriptsize -$100$}
			\psfrag{-80}[c][c]{\scriptsize -$80$}
			
			\psfrag{Frequency offset (Hz)}[c][c]{\footnotesize Frequency offset (Hz)}
			\psfrag{Phase noise PSD (dBc/Hz)}[c][c]{\vspace{-0.1cm}\footnotesize PN PSD (dBc/Hz)}
			
			\includegraphics[width=3.75cm]{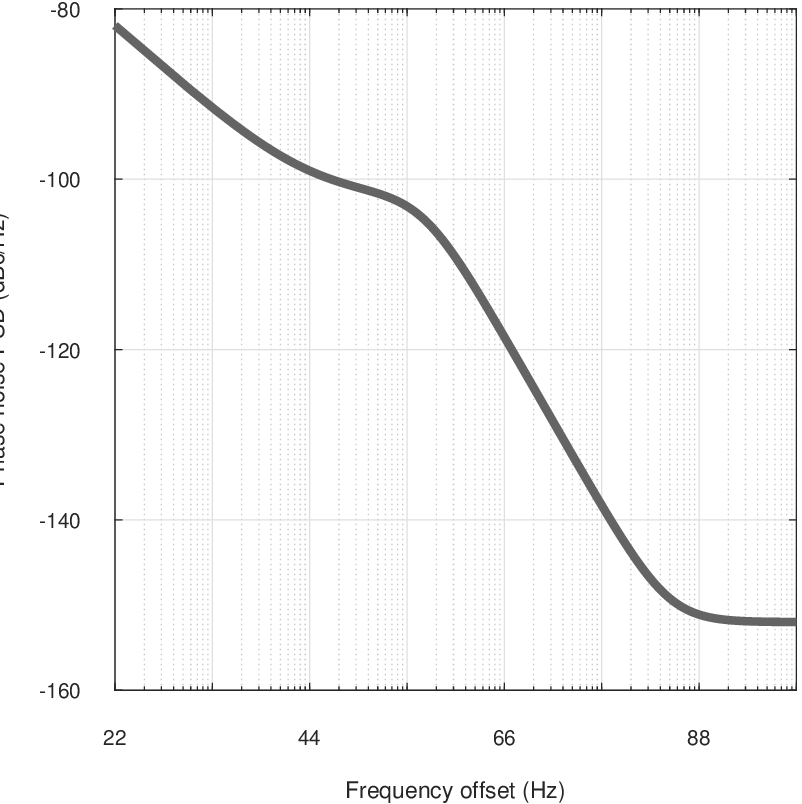}\label{fig:PNpsd}			
		}\hspace{0.1cm}
		\subfloat[ ]{
			
			\psfrag{-0.02}[c][c]{\scriptsize -$0.02$}
			\psfrag{-0.01}[c][c]{\scriptsize -$0.01$}
			\psfrag{0}[c][c]{\scriptsize $0$}
			\psfrag{0.01}[c][c]{\scriptsize $0.01$}
			\psfrag{0.02}[c][c]{\scriptsize $0.02$}
			
			\psfrag{0}[c][c]{\scriptsize $0$}
			\psfrag{50}[c][c]{\scriptsize $50$}
			\psfrag{100}[c][c]{\scriptsize $100$}
			\psfrag{150}[c][c]{\scriptsize $150$}
			
			\psfrag{Phase noise (rad)}[c][c]{\footnotesize Phase offset (rad)}
			\psfrag{Probability density}[c][c]{\footnotesize Probability density}
			
			\includegraphics[width=3.75cm]{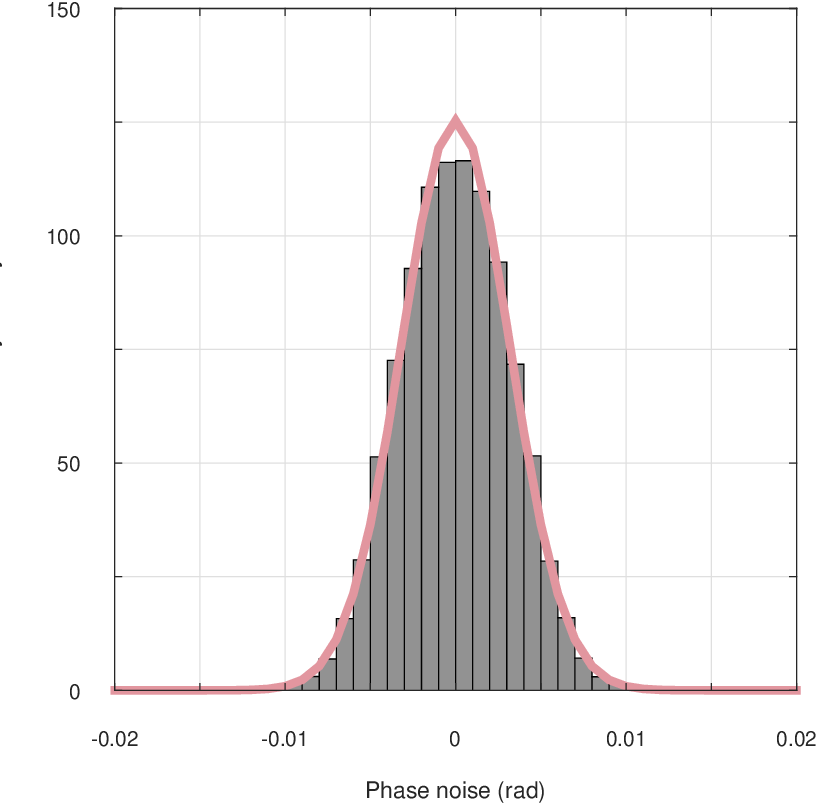}\label{fig:PNdist}			
		}
		
		\captionsetup{justification=raggedright,labelsep=period,singlelinecheck=false}
		\caption{\ Adopted PN model: (a) double-sided PN PSD and (b) PDF of the PN in the time domain.}\label{fig:PN_psd_dist}
		
	\end{figure}
	
	\section{OFDM-based ISAC Performance Analysis}\label{sec:perfAnalysis}
	
	In this section, the performance of a \ac{mmWave} \ac{OFDM}-based \ac{ISAC} system under \ac{PN} is analyzed. Both monostatic and bistatic architectures are considered, and the \ac{PN}-induced degradation of both communication and radar sensing performances is investigated. For that purpose, a corner frequency of \mbox{$f_\mathrm{corner}=\SI{10}{\kilo\hertz}$} and a \ac{PLL} loop bandwidth of \mbox{$B_\mathrm{PLL}=\SI{150}{\kilo\hertz}$} were assumed, besides the in-band \ac{PN} level of \mbox{$L_0=\SI{-105}{dBc/Hz}$} and the \ac{PN} floor \mbox{$L_\mathrm{floor}=\SI{-155}{dBc/Hz}$}. The aforementioned parameters were chosen in order to simulate a \ac{PN} \ac{PSD} that is in between the specified ones of the E8257D PSG Microwave Analog Signal Generator from Keysight Technologies \cite{keysight2024} for $\SI{20}{\giga\hertz}$ and $\SI{40}{\giga\hertz}$. Those frequencies cover the initial part of the \ac{5G NR} \ac{FR2}, which was used in the setups for \ac{ISAC} measurements in previous works \cite{li2024,giroto2024}. In addition, the simulations in this sections consider the listed \ac{OFDM} signal parameters in Table~\ref{tab:ofdmParameters}, with the specific values used mentioned in each individual analysis.
	
	\begin{figure}[!t]
		\centering
		\subfloat[ ]{

			\psfrag{44}[c][c]{\scriptsize $10^4$}
			\psfrag{55}[c][c]{\scriptsize $10^5$}
			\psfrag{66}[c][c]{\scriptsize $10^6$}
			\psfrag{77}[c][c]{\scriptsize $10^7$}
			\psfrag{88}[c][c]{\scriptsize $10^8$}
			\psfrag{99}[c][c]{\scriptsize $10^9$}
			
			\psfrag{-57}[c][c]{\scriptsize -$57$}
			\psfrag{-54}[c][c]{\scriptsize -$54$}
			\psfrag{-51}[c][c]{\scriptsize -$51$}
			\psfrag{-48}[c][c]{\scriptsize -$48$}
			\psfrag{-45}[c][c]{\scriptsize -$45$}
			
			\psfrag{Frequency offset (Hz)}[c][c]{\footnotesize Frequency offset (Hz)}
			\psfrag{Int. phase noise level (dBc)}[c][c]{\footnotesize Int. PN level (dBc)}
			
			\includegraphics[width=3.75cm]{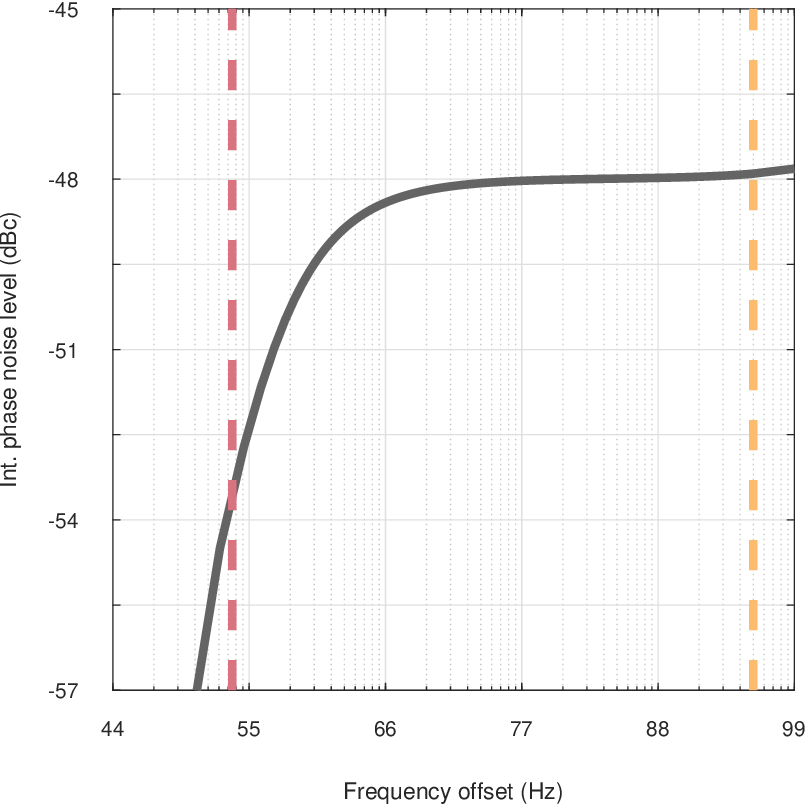}\label{fig:intPN}			
		}\hspace{0.1cm}
		\subfloat[ ]{
			\psfrag{8}[c][c]{\scriptsize $8$}
			\psfrag{9}[c][c]{\scriptsize $9$}
			\psfrag{10}[c][c]{\scriptsize $10$}
			\psfrag{11}[c][c]{\scriptsize $11$}
			\psfrag{12}[c][c]{\scriptsize $12$}
			\psfrag{13}[c][c]{\scriptsize $13$}
			\psfrag{14}[c][c]{\scriptsize $14$}
			
			\psfrag{44}[c][c]{\scriptsize $10^4$}
			\psfrag{55}[c][c]{\scriptsize $10^5$}
			\psfrag{66}[c][c]{\scriptsize $10^6$}
			\psfrag{77}[c][c]{\scriptsize $10^7$}
			
			\psfrag{log2N}[c][c]{\footnotesize $\log_2(N)$}
			\psfrag{Delta}[c][c]{\footnotesize $\Delta f$ (Hz)}
			
			\includegraphics[width=3.75cm]{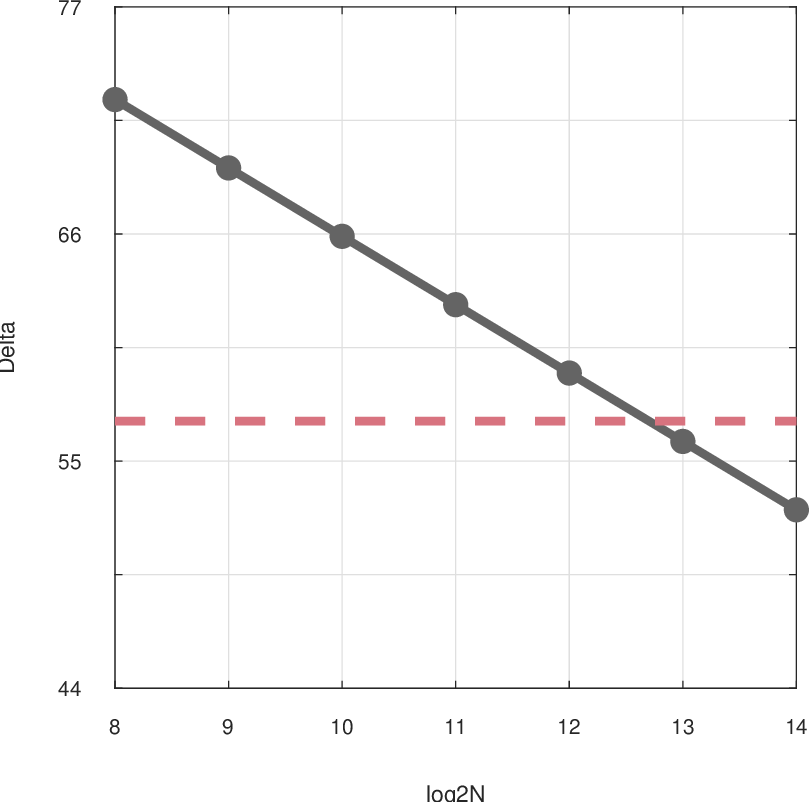}\label{fig:Delta_f}
		}\\
		\subfloat[ ]{
			\psfrag{8}[c][c]{\scriptsize $8$}
			\psfrag{9}[c][c]{\scriptsize $9$}
			\psfrag{10}[c][c]{\scriptsize $10$}
			\psfrag{11}[c][c]{\scriptsize $11$}
			\psfrag{12}[c][c]{\scriptsize $12$}
			\psfrag{13}[c][c]{\scriptsize $13$}
			\psfrag{14}[c][c]{\scriptsize $14$}
			
			\psfrag{0}[c][c]{\scriptsize $0$}
			\psfrag{5}[c][c]{\scriptsize $5$}
			\psfrag{10}[c][c]{\scriptsize $10$}
			\psfrag{15}[c][c]{\scriptsize $15$}
			
			\psfrag{log2N}[c][c]{\footnotesize $\log_2(N)$}
			\psfrag{Delta2Bpll}[c][c]{\footnotesize $(\Delta f/2)/B_\mathrm{PLL}$}
			
			\includegraphics[width=3.75cm]{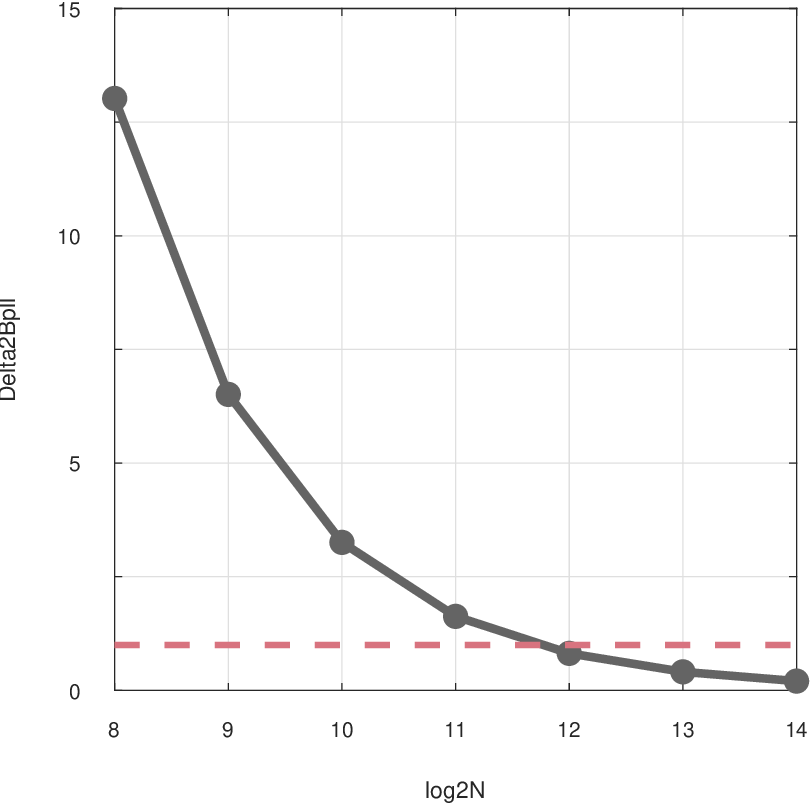}\label{fig:Delta2Bpll}
		}

		\captionsetup{justification=raggedright,labelsep=period,singlelinecheck=false}
		\caption{\ Transition from CPE to ICI as dominant PN-induced effect: (a) integrated \ac{PN} level as a function of frequency offset (double-sided) for the frequency range of interest, (b) subcarrier spacing $\Delta f$ as a function of the number of subcarriers $N$, and (c) ratio between half of $\Delta f$ and the PLL loop bandwidth $B_\mathrm{PLL}$ as a function of $N$ ({\color[rgb]{0.3922,0.3922,0.3922}$\CIRCLE$}). In (a) and (b), $B_\mathrm{PLL}$ is highlighted ({\color[rgb]{0.8471,0.4510,0.4980}\textbf{\textendash~\textendash}}), whereas in (b), the ratio \mbox{$(\Delta f/2)/B_\mathrm{PLL}=1$} below which ICI becomes more relevant than CPE is highlighted with the same line type. In addition, the offset of \mbox{$B/2=\SI{500}{\mega\hertz}$} is also highlighted in (a) ({\color[rgb]{0.9882,0.7333,0.4275}\textbf{\textendash~\textendash}}), as this figure shows integrated double-sided PSD and therefore this offset corresponds to the bandwidth \mbox{$B=\SI{1}{\giga\hertz}$} listed in Table~\ref{tab:ofdmParameters}.}\label{fig:intPNDelta2Bpll}
		
	\end{figure}
	
	Based on the assumed parameters, the resulting double-sided \ac{PN} \ac{PSD}, which is assumed to be the same for both $S_{\theta^\mathrm{PN}_\mathrm{Tx}}(f)$ and $S_{\theta^\mathrm{PN}_\mathrm{Rx}}(f)$, and the \ac{PDF} of the \ac{PN} time series are shown in Figs.~\ref{fig:PNpsd} and \ref{fig:PNdist}, respectively. It is worth highlighting that this model is assumed for both transmitter and receiver. However, the interaction of both aforementioned \ac{PN} terms will depend on whether a monostatic or bistatic \ac{ISAC} architecture is assumed. In addition, the results in Fig.~\ref{fig:intPNDelta2Bpll} reveal the relevance of $B_\mathrm{PLL}$ and its interplay with the subcarrier spacing $\Delta f$. Fig.~\ref{fig:intPN} shows the integrated \ac{PN} level as a function of the frequency offset, revealing that a \ac{PN} level of $\SI{-52.78}{dBc}$, i.e., around $\SI{4.88}{dB}$ below the total \ac{PN} level of $\SI{-47.90}{dBc}$, is accumulated by a frequency offset equal to $B_\mathrm{PLL}$ as expected from the discussion in Section~\ref{sec:sysModel}. This level is calculated by integrating the \ac{PN} \ac{PSD} occupied bandwidth assumed alongside the parameters listed in Table~\ref{tab:ofdmParameters}, i.e., \mbox{$B=\SI{1}{\giga\hertz}$}. In Fig.~\ref{fig:Delta_f}, the subcarrier spacing $\Delta f$ is shown as a function of the assumed numbers of subcarriers $N$ in Table~\ref{tab:ofdmParameters}. For better visualization, the $N$ axis is shown in $\log_2$ scale, which results in $\log_2\left(\{256, 512, 1024, 2048, 4096, 8192, 16384\}\right)=\{8,9,10,11,12,13,14\}$. It is seen that only \mbox{$N\leq 4096$} yield higher $\Delta f$ than $B_\mathrm{PLL}$. A more relevant ratio, which is the one between half of the subcarrier spacing, i.e., $\Delta f/2$, and $B_\mathrm{PLL}$ as discussed in Section~\ref{sec:sysModel}, is shown in Fig.~\ref{fig:Delta2Bpll}. It can be observed that \mbox{$N\in\{256,512,1024,2048\}$} yield \mbox{$(\Delta f/2)/B_\mathrm{PLL}>1$}, which implies that \ac{CPE} is the dominant \ac{PN}-induced effect for these numbers of subcarriers. However, for \mbox{$N\in\{4096,8192,16384\}$}, it holds that \mbox{$(\Delta f/2)/B_\mathrm{PLL}<1$}. Consequently, the \ac{PN}-induced \ac{ICI} becomes the dominant effect. This happens since the subcarrier spacing $\Delta f$ decreases with increasing $N$, leading to more significant leakage of each subcarrier, which follows the shape of the \ac{PSD} from Fig.~\ref{fig:PNpsd}, on the frequency bins of its neighbors.
	
	To evaluate the actual effect of \ac{PN} on the performance of \ac{OFDM}-based \ac{ISAC} systems, simulation results focusing on the distortion of the constellation and radar sensing performance are presented in Sections~\ref{subsec:constDist} and \ref{subsec:sensPerf}, respectively. Finally, remarks on the obtained simulation results in the two aforementioned sections are given in Section~\ref{subsec:discussion}.
	
	\begin{figure}[!t]
		\centering
		\subfloat[ ]{
			\psfrag{8}[c][c]{\scriptsize $8$}
			\psfrag{9}[c][c]{\scriptsize $9$}
			\psfrag{10}[c][c]{\scriptsize $10$}
			\psfrag{11}[c][c]{\scriptsize $11$}
			\psfrag{12}[c][c]{\scriptsize $12$}
			\psfrag{13}[c][c]{\scriptsize $13$}
			\psfrag{14}[c][c]{\scriptsize $14$}
			
			\psfrag{-60}[c][c]{\scriptsize -$60$}
			\psfrag{-50}[c][c]{\scriptsize -$50$}
			\psfrag{-40}[c][c]{\scriptsize -$40$}
			\psfrag{-30}[c][c]{\scriptsize -$30$}
			\psfrag{-20}[c][c]{\scriptsize -$20$}
			\psfrag{-10}[c][c]{\scriptsize -$10$}
			\psfrag{0}[c][c]{\scriptsize $0$}
			
			\psfrag{log2N}[c][c]{\footnotesize $\log_2(N)$}
			\psfrag{EVM (dB)}[c][c]{\footnotesize $\mathrm{EVM (dB)}$}
			
			\includegraphics[width=4.25cm]{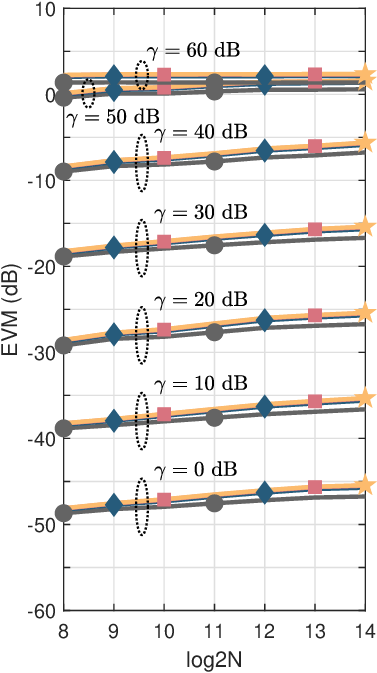}
		}
		\subfloat[ ]{
			\psfrag{8}[c][c]{\scriptsize $8$}
			\psfrag{9}[c][c]{\scriptsize $9$}
			\psfrag{AA}[c][c]{\scriptsize $10$}
			\psfrag{11}[c][c]{\scriptsize $11$}
			\psfrag{12}[c][c]{\scriptsize $12$}
			\psfrag{13}[c][c]{\scriptsize $13$}
			\psfrag{14}[c][c]{\scriptsize $14$}
			
			\psfrag{-60}[c][c]{}
			\psfrag{-50}[c][c]{}
			\psfrag{-40}[c][c]{}
			\psfrag{-30}[c][c]{}
			\psfrag{-20}[c][c]{}
			\psfrag{-10}[c][c]{}
			\psfrag{0}[c][c]{}
			\psfrag{10}[c][c]{}
			
			\psfrag{log2N}[c][c]{\footnotesize $\log_2(N)$}
			\psfrag{EVM (dB)}[c][c]{}
			
			\includegraphics[width=4.25cm]{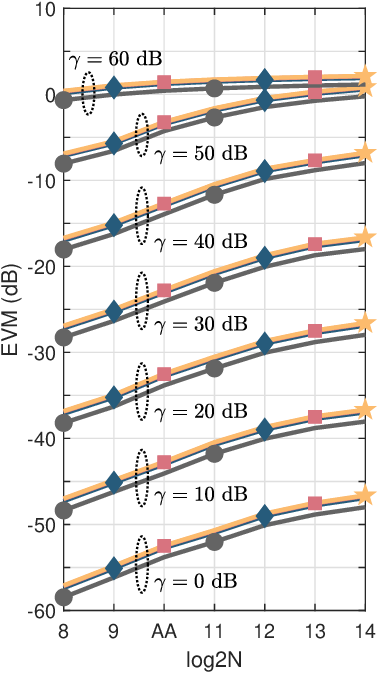}
		}
		\captionsetup{justification=raggedright,labelsep=period,singlelinecheck=false}
		\caption{\ EVM for $M=128$ OFDM symbols, different numbers of subcarriers $N$ and combined transmit and receive PN levels of $\SI{-44.90}{dBc}$, $\SI{-34.90}{dBc}$, $\SI{-24.90}{dBc}$, $\SI{-14.90}{dBc}$, $\SI{-4.90}{dBc}$, $\SI{4.90}{dBc}$, and $\SI{14.90}{dBc}$ ($\gamma= \SI{0}{dB}, \SI{10}{dB}, \SI{20}{dB}, \SI{30}{dB}, \SI{40}{dB}, \SI{50}{dB}, \text{and}~\SI{60}{dB}$, respectively). In all cases, and \ac{QPSK} ({\color[rgb]{0.3922,0.3922,0.3922}$\CIRCLE$}), \mbox{16-\ac{QAM}} ({\color[rgb]{0.1490,0.3569,0.4824}$\blacklozenge$}), \mbox{64-\ac{QAM}} ({\color[rgb]{0.8471,0.4510,0.4980}$\blacksquare$}), and \mbox{256-\ac{QAM}} ({\color[rgb]{0.9882,0.7333,0.4275}$\bigstar$}) modulations were considered. In (a), both ICI and CPE are entirely present, whre as in (b) CPE was estimated and corrected.}\label{fig:phaseNoise_EVM}
		
	\end{figure}
	
	\subsection{PN-induced Distortion of Digital Modulation Constellation}\label{subsec:constDist}
	
	For the analysis on the constellation distortion in an \ac{OFDM}-based \ac{ISAC} system under \ac{PN} influence, a communication scenario where transmit and receive \ac{PN} are uncorrelated is assumed. Assuming that the aforementioned \ac{PN} contributions have the same level, their incoherent combination results in a $\SI{3}{dB}$ higher \ac{PN} level as they are assumed to be uncorrelated Gaussian processes. In addition, an ideal channel model with a single path ($P=1$) and null delay and Doppler shift (i.e., $\tau_0=\SI{0}{\second}$ and $f_{\mathrm{D},0}=\SI{0}{\hertz}$) is adopted. Since no \ac{ISI} happens in this scenario, \mbox{$N_\mathrm{CP}=0$} is assumed.
	
	Disregarding \ac{AWGN}, Fig.~\ref{fig:phaseNoise_EVM}a shows the achieved mean \ac{EVM} over multiple \ac{PN} realizations as a function of the number of subcarriers $N$ for all digital modulations listed in Table~\ref{tab:ofdmParameters}. For the achieved results, the \ac{PN} model from Fig.~\ref{fig:PN_psd_dist} was considered, and its original level of $\SI{-47.90}{dBc}$ was increased by a factor $\gamma\in\mathbb{R}_{\geq0}$ to showcase the induced degradation by \ac{PN} at higher levels. Focusing on the $\gamma=\SI{0}{dB}$ case, i.e. the original \ac{PN} model from Fig.~\ref{fig:PN_psd_dist} with a combined transmit and receive \ac{PN} level of $\SI{-44.90}{dBc}$, it can be seen that the \ac{EVM} degrades, i.e., increases, linearly along with increasing $N$, which is due to the decreasing subcarrier spacing $\Delta f$ and resulting increased \ac{ICI} effect. This becomes more pronounced for higher-order modulations, i.e., \mbox{16-\ac{QAM}}, \mbox{64-\ac{QAM}}, and \mbox{256-\ac{QAM}}, when compared to \ac{QPSK}. This is due to the fact that the constellation points are relatively closer to one another in these modulations, and also because they do not have a constant envelope. Consequently, neighbor subcarriers may not have the same energy, leading to a more severe \ac{ICI} effect. For \ac{PN} increase factors of $\gamma=\SI[parse-numbers = false]{\{0, 10, 20, 30, 40, 50\}}{dB}$, which result in combined transmit and receive \ac{PN} levels of $\SI{-44.90}{dBc}$, $\SI{-34.90}{dBc}$, $\SI{-24.90}{dBc}$, $\SI{-14.90}{dBc}$, $\SI{-4.90}{dBc}$, $\SI{4.90}{dBc}$, respectively, the \ac{EVM} is linearly degraded, i.e., increased, by the same factor $\gamma$. The \ac{EVM} degradation only stops following the previously described pattern at a combined transmit and receive \ac{PN} level of $\SI{14.90}{dBc}$, which corresponds to $\gamma=\SI{60}{dB}$. At this \ac{PN} level, the \ac{EVM} is constant regardless of the adopted number of subcarriers. In addition, the \ac{EVM} increase trend observed for lower $\gamma$ is not present for $\gamma=\SI{60}{dB}$, as the \ac{EVM} difference w.r.t. $\gamma=\SI{50}{dB}$ is smaller than $\SI{10}{dB}$ for all considered $N$ values. This happens since even the highest $\Delta f$ cannot prevent severe \ac{ICI} at such \ac{PN} level.
	Still considering an ideal channel model as well as the ideal case where all subcarriers are known at the receiver side, the \ac{EVM} after \ac{CPE} estimation and correction based on \eqref{eq:CPE_est} and \eqref{eq:CPE_corr}, respectively, is shown as a function of the number of subcarriers $N$ is shown in Fig.~\ref{fig:phaseNoise_EVM}b. It is observed that a significant \ac{EVM} improvement of up to around $\SI{9.78}{dB}$ is observed for decreasing $N$ in comparison with the results without \ac{CPE} correction in Fig.~\ref{fig:phaseNoise_EVM}a. This happens since \ac{CPE} is more relevant for lower $N$, while higher $N$, and therefore lower subcarrier spacing $\Delta f$, leads to more pronounced \ac{PN}-induced \ac{ICI} that cannot be compensated with the adopted method.
	
	\begin{figure}[!t]
		\centering
		\subfloat[ ]{
			
			\psfrag{-45}[c][c]{\scriptsize -$45$}
			\psfrag{-30}[c][c]{\scriptsize -$30$}
			\psfrag{-15}[c][c]{\scriptsize -$15$}
			\psfrag{0}[c][c]{\scriptsize $0$}
			\psfrag{15}[c][c]{\scriptsize $15$}
			\psfrag{30}[c][c]{\scriptsize $30$}
			\psfrag{45}[c][c]{\scriptsize $45$}
			\psfrag{60}[c][c]{\scriptsize $60$}
			
			\psfrag{Phase noise level (dBc)}[c][c]{\footnotesize PN level (dBc)}
			\psfrag{SIR (dB)}[c][c]{\footnotesize $\mathrm{SIR~(dB)}$}	
			
			\includegraphics[width=3.75cm]{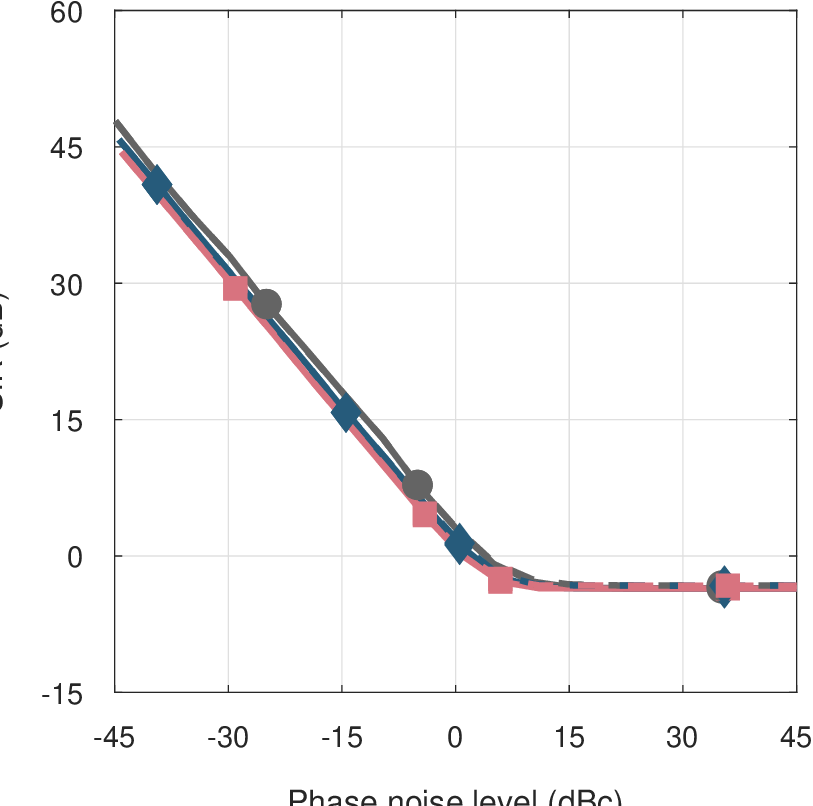}\label{fig:SIR_PN_woCPEcorr}			
		}\hspace{0.1cm}
		\subfloat[ ]{
			
			\psfrag{-45}[c][c]{\scriptsize -$45$}
			\psfrag{-30}[c][c]{\scriptsize -$30$}
			\psfrag{-15}[c][c]{\scriptsize -$15$}
			\psfrag{0}[c][c]{\scriptsize $0$}
			\psfrag{15}[c][c]{\scriptsize $15$}
			\psfrag{30}[c][c]{\scriptsize $30$}
			\psfrag{45}[c][c]{\scriptsize $45$}
			\psfrag{60}[c][c]{\scriptsize $60$}
			
			\psfrag{Phase noise level (dBc)}[c][c]{\footnotesize PN level (dBc)}
			\psfrag{SIR (dB)}[c][c]{\footnotesize $\mathrm{SIR~(dB)}$}
			
			\includegraphics[width=3.75cm]{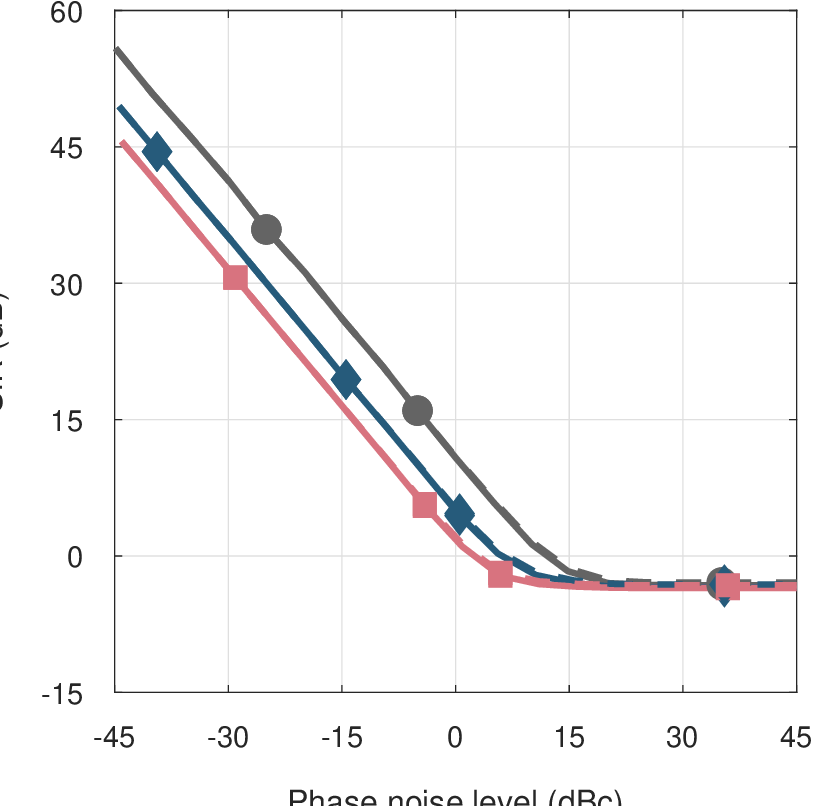}\label{fig:SIR_PN_wCPEcorr}			
		}
		
		\captionsetup{justification=raggedright,labelsep=period,singlelinecheck=false}
		\caption{\ SIR as a function of the combined transmit and receive PN level (a) without and (b) with CPE estimation and correction. Simulation results are shown assuming \mbox{$N=256$} and QPSK \mbox{({\color[rgb]{0.3922,0.3922,0.392}\textbf{\textemdash}} and {\color[rgb]{0.3922,0.3922,0.3922}$\CIRCLE$})}, \mbox{$N=2048$} and QPSK \mbox{({\color[rgb]{0.1490,0.3569,0.4824}\textbf{\textemdash}} and {\color[rgb]{0.1490,0.3569,0.4824}$\blacklozenge$})}, and \mbox{$N=16384$} and QPSK \mbox{({\color[rgb]{0.8471,0.4510,0.4980}\textbf{\textemdash}} and {\color[rgb]{0.8471,0.4510,0.4980}$\blacksquare$})}. Results are also shown for \mbox{256-QAM} and the aforementioned number of subcarriers, i.e., \mbox{$N=256$} \mbox{({\color[rgb]{0.3922,0.3922,0.392}\textbf{\textendash~\textendash}} and {\color[rgb]{0.3922,0.3922,0.3922}$\CIRCLE$})}, \mbox{$N=2048$} \mbox{({\color[rgb]{0.1490,0.3569,0.4824}\textbf{\textendash~\textendash}} and {\color[rgb]{0.1490,0.3569,0.4824}$\blacklozenge$})}, and \mbox{$N=16384$} \mbox{({\color[rgb]{0.8471,0.4510,0.4980}\textbf{\textendash~\textendash}} and {\color[rgb]{0.8471,0.4510,0.4980}$\blacksquare$})}. In both cases, $M=128$ OFDM symbols were considered.}\label{fig:SIR_PN}
		
	\end{figure}
	\begin{figure*}[!t]
		\centering
		\subfloat[ ]{
			
			\psfrag{-1.5}[c][c]{\scriptsize -$1.5$}
			\psfrag{-0.5}[c][c]{\scriptsize -$0.5$}
			\psfrag{0.5}[c][c]{\scriptsize $0.5$}
			\psfrag{1.5}[c][c]{\scriptsize $1.5$}
			
			\psfrag{AA}[c][c]{\scriptsize -$1.5$}
			\psfrag{BB}[c][c]{\scriptsize -$0.5$}
			\psfrag{CC}[c][c]{\scriptsize $0.5$}
			\psfrag{DD}[c][c]{\scriptsize $1.5$}
			
			\psfrag{I}[c][c]{\scriptsize $I$}
			\psfrag{Q}[c][c]{\scriptsize $Q$}
			
			\includegraphics[width=2.4cm]{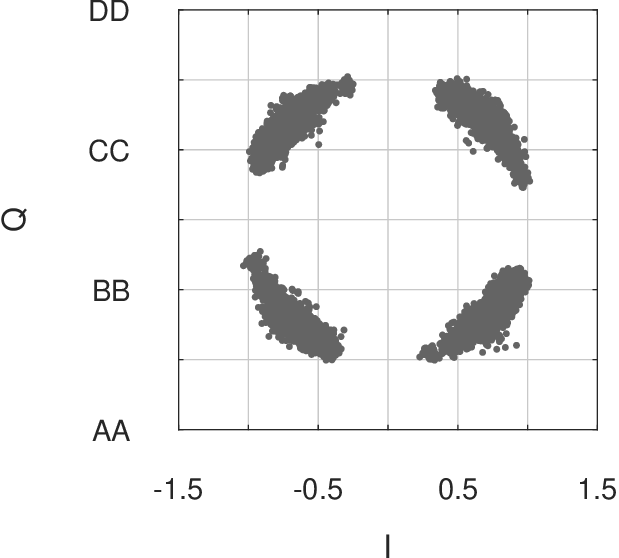}
		}
		\subfloat[ ]{
			
			\psfrag{-1.5}[c][c]{\scriptsize -$1.5$}
			\psfrag{-0.5}[c][c]{\scriptsize -$0.5$}
			\psfrag{0.5}[c][c]{\scriptsize $0.5$}
			\psfrag{1.5}[c][c]{\scriptsize $1.5$}
			
			\psfrag{AA}[c][c]{ }
			\psfrag{BB}[c][c]{ }
			\psfrag{CC}[c][c]{ }
			\psfrag{DD}[c][c]{ }
			
			\psfrag{I}[c][c]{\scriptsize $I$}
			\psfrag{Q}[c][c]{ }
			
			\includegraphics[width=2.4cm]{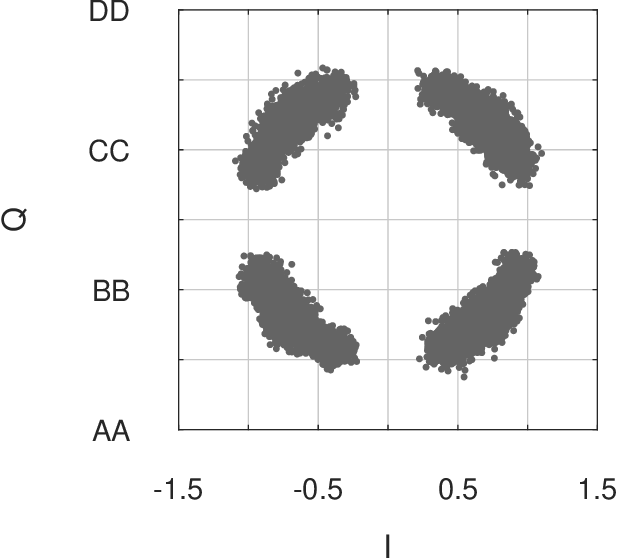}
		}
		\subfloat[ ]{
			
			\psfrag{-1.5}[c][c]{\scriptsize -$1.5$}
			\psfrag{-0.5}[c][c]{\scriptsize -$0.5$}
			\psfrag{0.5}[c][c]{\scriptsize $0.5$}
			\psfrag{1.5}[c][c]{\scriptsize $1.5$}
			
			\psfrag{AA}[c][c]{ }
			\psfrag{BB}[c][c]{ }
			\psfrag{CC}[c][c]{ }
			\psfrag{DD}[c][c]{ }
			
			\psfrag{I}[c][c]{\scriptsize $I$}
			\psfrag{Q}[c][c]{ }
			
			\includegraphics[width=2.4cm]{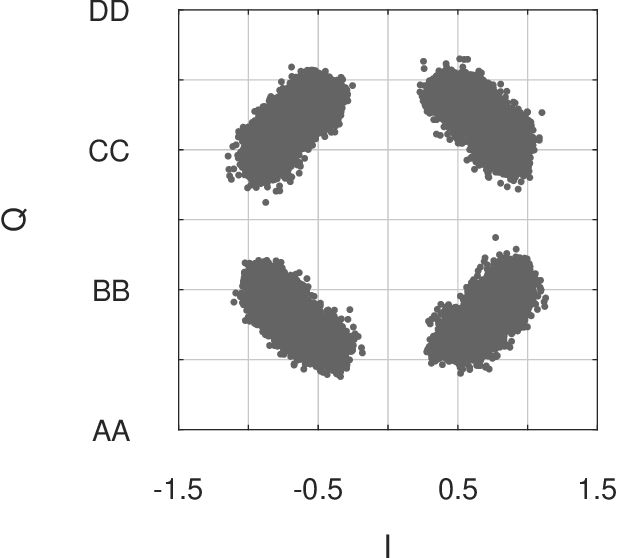}
		}
		\subfloat[ ]{
			
			\psfrag{-1.5}[c][c]{\scriptsize -$1.5$}
			\psfrag{-0.5}[c][c]{\scriptsize -$0.5$}
			\psfrag{0.5}[c][c]{\scriptsize $0.5$}
			\psfrag{1.5}[c][c]{\scriptsize $1.5$}
			
			\psfrag{AA}[c][c]{ }
			\psfrag{BB}[c][c]{ }
			\psfrag{CC}[c][c]{ }
			\psfrag{DD}[c][c]{ }
			
			\psfrag{I}[c][c]{\scriptsize $I$}
			\psfrag{Q}[c][c]{ }
			
			\includegraphics[width=2.4cm]{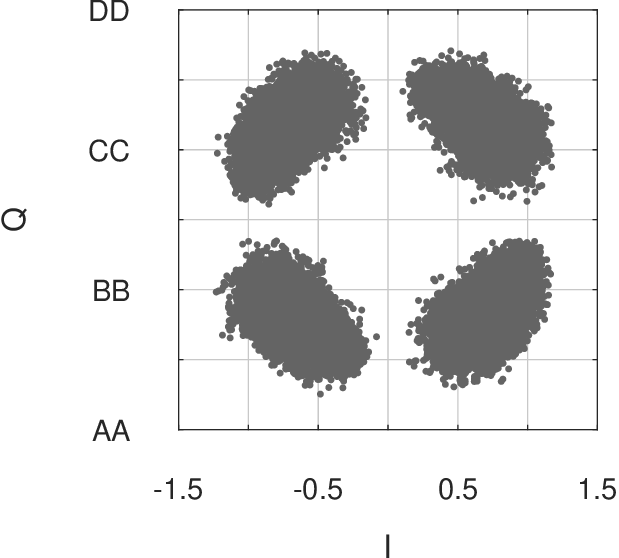}
		}
		\subfloat[ ]{
			
			\psfrag{-1.5}[c][c]{\scriptsize -$1.5$}
			\psfrag{-0.5}[c][c]{\scriptsize -$0.5$}
			\psfrag{0.5}[c][c]{\scriptsize $0.5$}
			\psfrag{1.5}[c][c]{\scriptsize $1.5$}
			
			\psfrag{AA}[c][c]{ }
			\psfrag{BB}[c][c]{ }
			\psfrag{CC}[c][c]{ }
			\psfrag{DD}[c][c]{ }
			
			\psfrag{I}[c][c]{\scriptsize $I$}
			\psfrag{Q}[c][c]{ }
			
			\includegraphics[width=2.4cm]{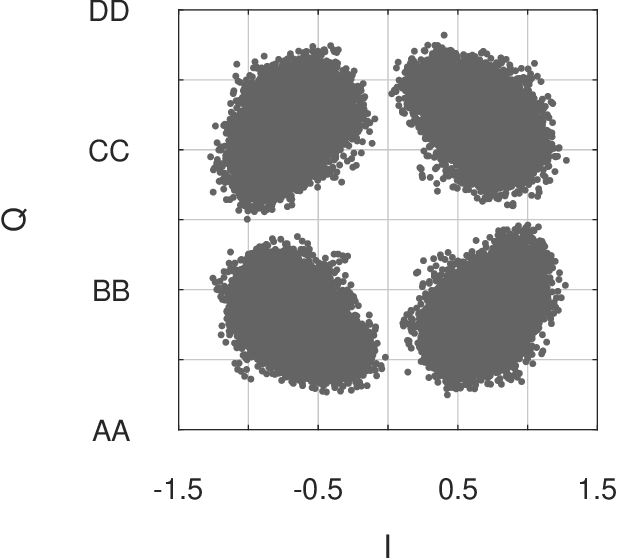}
		}
		\subfloat[ ]{
			
			\psfrag{-1.5}[c][c]{\scriptsize -$1.5$}
			\psfrag{-0.5}[c][c]{\scriptsize -$0.5$}
			\psfrag{0.5}[c][c]{\scriptsize $0.5$}
			\psfrag{1.5}[c][c]{\scriptsize $1.5$}
			
			\psfrag{AA}[c][c]{ }
			\psfrag{BB}[c][c]{ }
			\psfrag{CC}[c][c]{ }
			\psfrag{DD}[c][c]{ }
			
			\psfrag{I}[c][c]{\scriptsize $I$}
			\psfrag{Q}[c][c]{ }
			
			\includegraphics[width=2.4cm]{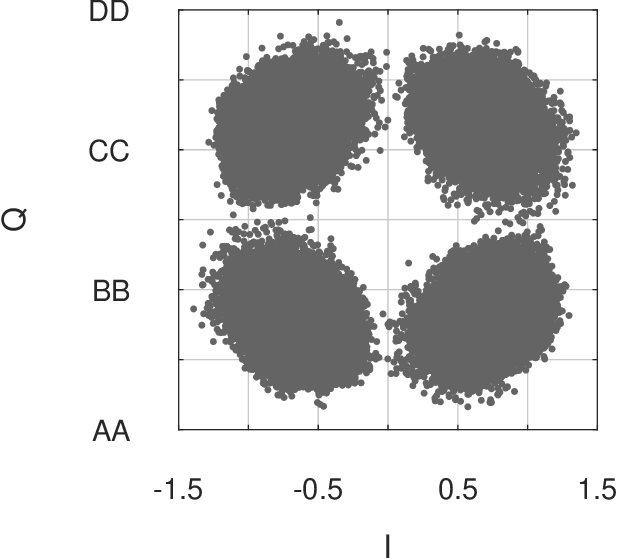}
		}
		\subfloat[ ]{
			
			\psfrag{-1.5}[c][c]{\scriptsize -$1.5$}
			\psfrag{-0.5}[c][c]{\scriptsize -$0.5$}
			\psfrag{0.5}[c][c]{\scriptsize $0.5$}
			\psfrag{1.5}[c][c]{\scriptsize $1.5$}
			
			\psfrag{AA}[c][c]{ }
			\psfrag{BB}[c][c]{ }
			\psfrag{CC}[c][c]{ }
			\psfrag{DD}[c][c]{ }
			
			\psfrag{I}[c][c]{\scriptsize $I$}
			\psfrag{Q}[c][c]{ }
			
			\includegraphics[width=2.4cm]{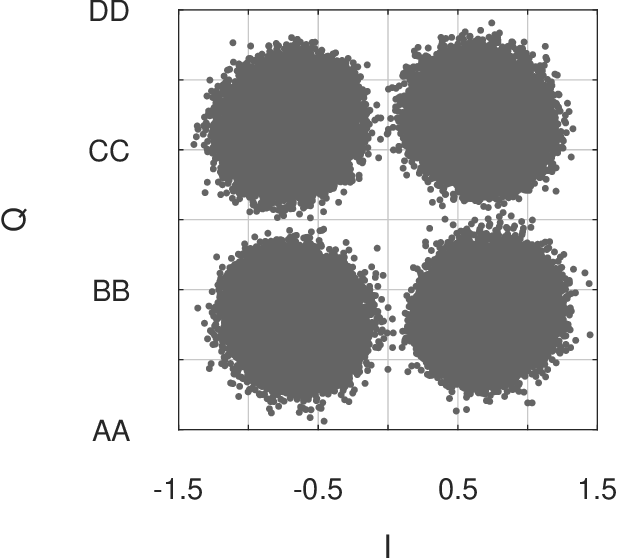}
		}
		
		\captionsetup{justification=raggedright,labelsep=period,singlelinecheck=false}
		\caption{\ Normalized receive QPSK constellations obtained for a combined transmit and receive PN level of $\SI{-14.90}{dBc}$ \mbox{($\gamma=\SI{30}{dB}$)} and different numbers of subcarriers: (a) $N=256$, (b) $N=512$, (c) $N=1024$, (d) $N=2048$, (e) $N=4096$, (f) $N=8192$, and (g) $N=16384$. In all cases, $M=128$ OFDM symbols were considered.}\label{fig:QPSKconst_psd30}
		
	\end{figure*}
	
	\begin{figure*}[!t]
		\centering
		\subfloat[ ]{
			
			\psfrag{-1.5}[c][c]{\scriptsize -$1.5$}
			\psfrag{-0.5}[c][c]{\scriptsize -$0.5$}
			\psfrag{0.5}[c][c]{\scriptsize $0.5$}
			\psfrag{1.5}[c][c]{\scriptsize $1.5$}
			
			\psfrag{AA}[c][c]{\scriptsize -$1.5$}
			\psfrag{BB}[c][c]{\scriptsize -$0.5$}
			\psfrag{CC}[c][c]{\scriptsize $0.5$}
			\psfrag{DD}[c][c]{\scriptsize $1.5$}
			
			\psfrag{I}[c][c]{\scriptsize $I$}
			\psfrag{Q}[c][c]{\scriptsize $Q$}
			
			\includegraphics[width=2.4cm]{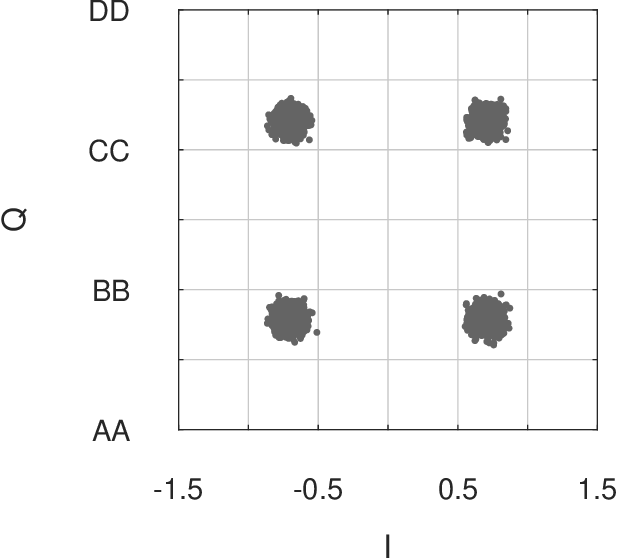}
		}
		\subfloat[ ]{
			
			\psfrag{-1.5}[c][c]{\scriptsize -$1.5$}
			\psfrag{-0.5}[c][c]{\scriptsize -$0.5$}
			\psfrag{0.5}[c][c]{\scriptsize $0.5$}
			\psfrag{1.5}[c][c]{\scriptsize $1.5$}
			
			\psfrag{AA}[c][c]{ }
			\psfrag{BB}[c][c]{ }
			\psfrag{CC}[c][c]{ }
			\psfrag{DD}[c][c]{ }
			
			\psfrag{I}[c][c]{\scriptsize $I$}
			\psfrag{Q}[c][c]{ }
			
			\includegraphics[width=2.4cm]{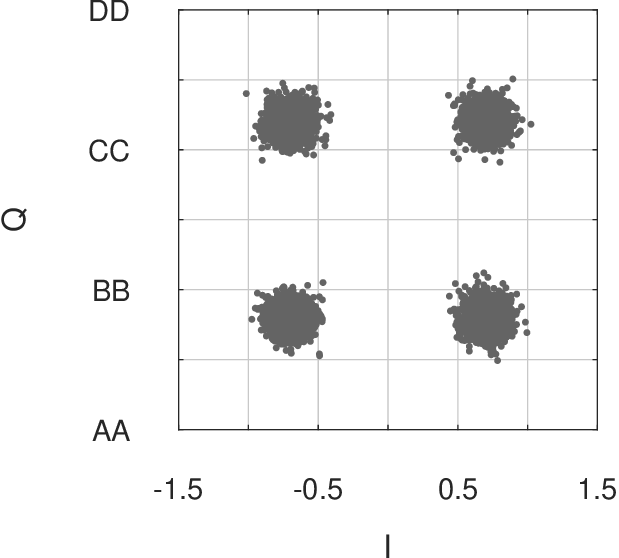}
		}
		\subfloat[ ]{
			
			\psfrag{-1.5}[c][c]{\scriptsize -$1.5$}
			\psfrag{-0.5}[c][c]{\scriptsize -$0.5$}
			\psfrag{0.5}[c][c]{\scriptsize $0.5$}
			\psfrag{1.5}[c][c]{\scriptsize $1.5$}
			
			\psfrag{AA}[c][c]{ }
			\psfrag{BB}[c][c]{ }
			\psfrag{CC}[c][c]{ }
			\psfrag{DD}[c][c]{ }
			
			\psfrag{I}[c][c]{\scriptsize $I$}
			\psfrag{Q}[c][c]{ }
			
			\includegraphics[width=2.4cm]{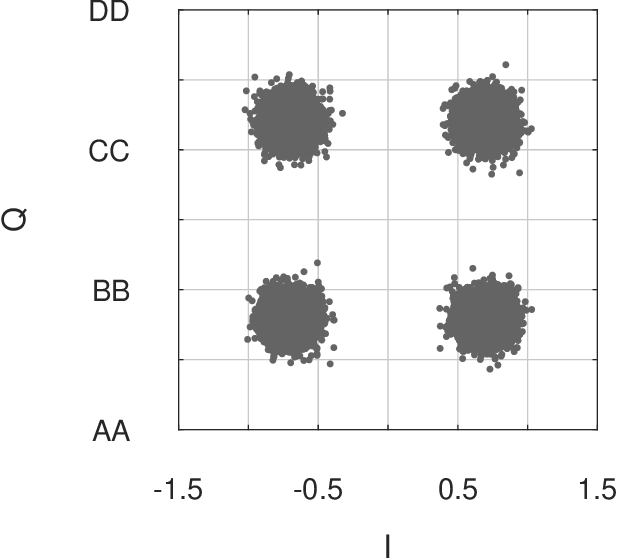}
		}
		\subfloat[ ]{
			
			\psfrag{-1.5}[c][c]{\scriptsize -$1.5$}
			\psfrag{-0.5}[c][c]{\scriptsize -$0.5$}
			\psfrag{0.5}[c][c]{\scriptsize $0.5$}
			\psfrag{1.5}[c][c]{\scriptsize $1.5$}
			
			\psfrag{AA}[c][c]{ }
			\psfrag{BB}[c][c]{ }
			\psfrag{CC}[c][c]{ }
			\psfrag{DD}[c][c]{ }
			
			\psfrag{I}[c][c]{\scriptsize $I$}
			\psfrag{Q}[c][c]{ }
			
			\includegraphics[width=2.4cm]{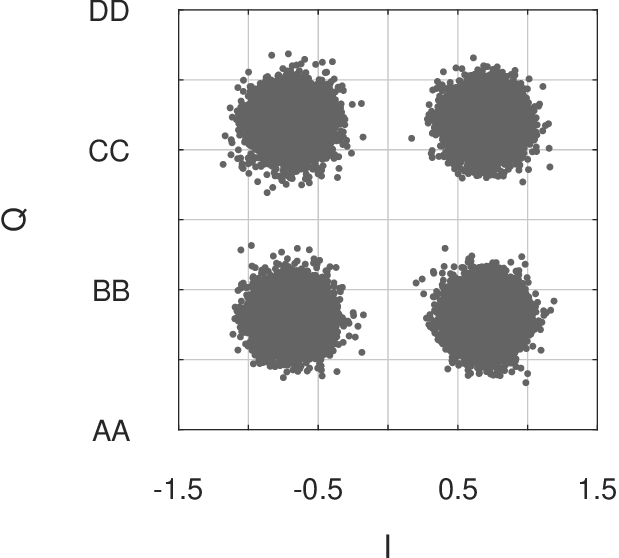}
		}
		\subfloat[ ]{
			
			\psfrag{-1.5}[c][c]{\scriptsize -$1.5$}
			\psfrag{-0.5}[c][c]{\scriptsize -$0.5$}
			\psfrag{0.5}[c][c]{\scriptsize $0.5$}
			\psfrag{1.5}[c][c]{\scriptsize $1.5$}
			
			\psfrag{AA}[c][c]{ }
			\psfrag{BB}[c][c]{ }
			\psfrag{CC}[c][c]{ }
			\psfrag{DD}[c][c]{ }
			
			\psfrag{I}[c][c]{\scriptsize $I$}
			\psfrag{Q}[c][c]{ }
			
			\includegraphics[width=2.4cm]{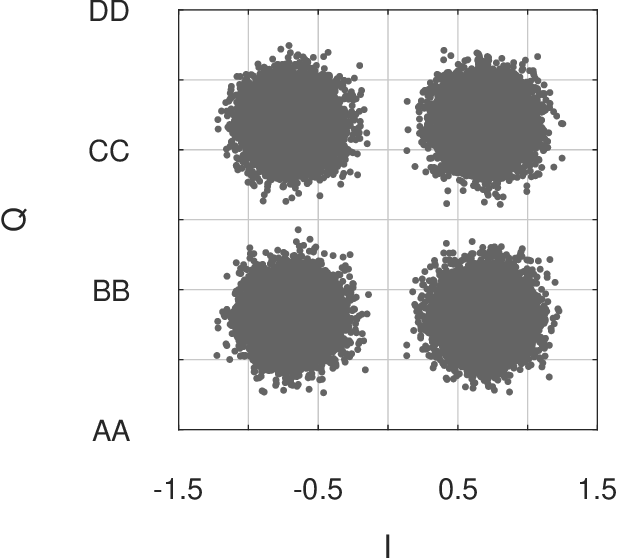}
		}
		\subfloat[ ]{
			
			\psfrag{-1.5}[c][c]{\scriptsize -$1.5$}
			\psfrag{-0.5}[c][c]{\scriptsize -$0.5$}
			\psfrag{0.5}[c][c]{\scriptsize $0.5$}
			\psfrag{1.5}[c][c]{\scriptsize $1.5$}
			
			\psfrag{AA}[c][c]{ }
			\psfrag{BB}[c][c]{ }
			\psfrag{CC}[c][c]{ }
			\psfrag{DD}[c][c]{ }
			
			\psfrag{I}[c][c]{\scriptsize $I$}
			\psfrag{Q}[c][c]{ }
			
			\includegraphics[width=2.4cm]{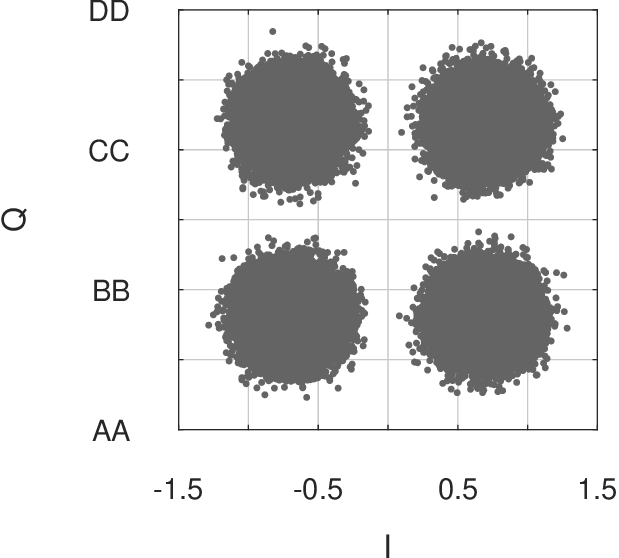}
		}
		\subfloat[ ]{
			
			\psfrag{-1.5}[c][c]{\scriptsize -$1.5$}
			\psfrag{-0.5}[c][c]{\scriptsize -$0.5$}
			\psfrag{0.5}[c][c]{\scriptsize $0.5$}
			\psfrag{1.5}[c][c]{\scriptsize $1.5$}
			
			\psfrag{AA}[c][c]{ }
			\psfrag{BB}[c][c]{ }
			\psfrag{CC}[c][c]{ }
			\psfrag{DD}[c][c]{ }
			
			\psfrag{I}[c][c]{\scriptsize $I$}
			\psfrag{Q}[c][c]{ }
			
			\includegraphics[width=2.4cm]{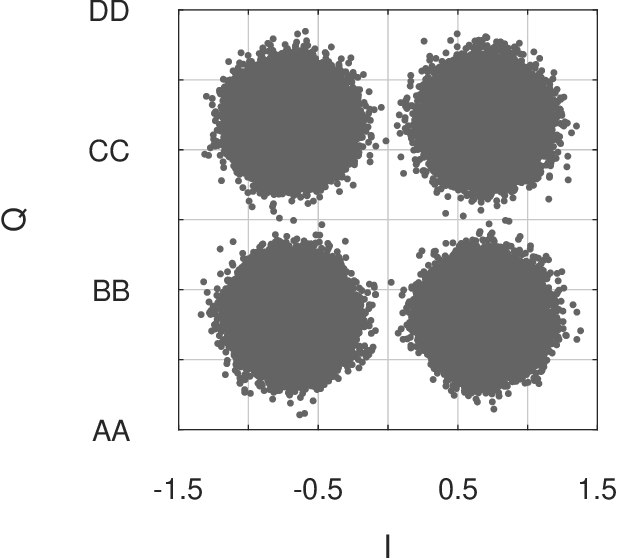}
		}
		
		\captionsetup{justification=raggedright,labelsep=period,singlelinecheck=false}
		\caption{\ Normalized receive QPSK constellations obtained after CPE estimation and correction for a combined transmit and receive PN level of $\SI{-14.90}{dBc}$ \mbox{($\gamma=\SI{30}{dB}$)}, $M=128$ OFDM symbols, and number of subcarriers equal to: (a) $N=256$, (b) $N=512$, (c) $N=1024$, (d) $N=2048$, (e) $N=4096$, (f) $N=8192$, and (g) $N=16384$.}\label{fig:QPSKconst_wCPEcorr_psd30}
		
	\end{figure*}
	
	Next, Fig.~\ref{fig:SIR_PN} shows the experienced \ac{SIR} as a function of the combined transmit and receive \ac{PN} level. For the performed simulations, \mbox{$N\in\{256,2048,16384\}$}, as well as \ac{QPSK} and \mbox{256-\ac{QAM}} modulations were adopted. If \ac{CPE} is not compensated, all considered $N$ values and modulations yield similar performance, with \ac{SIR} decreasing linearly from $\SI{45.72}{dB}$ on average at the \ac{PN} level of $\SI{-45}{dBc}$ until an average of $\SI{1.5}{dB}$ at the \ac{PN} level of $\SI{0}{dBc}$. In this zone, $N=256$ performs around $\SI{1.5}{dB}$ better than $N=2048$ and $\SI{3}{dB}$ better than $N=16384$. At \ac{PN} levels slightly higher than the latter, the linear behavior of the \ac{SIR} is no longer observed, and a constant level is reached. In the case where \ac{CPE} is estimated and corrected, higher \ac{SIR} is observed for lower $N$ in the linear \ac{SIR} zone, which was already expected based on the \ac{EVM} results from Fig.~\ref{fig:phaseNoise_EVM}b. A significant \ac{SIR} improvement, however, is only observed for $N=256$ among the considered values. The maximum \ac{SIR} improvement in the linear zone, which is of around $\SI{7.90}{dB}$, is observed for $N=256$. As for the other $N$ values, lower \ac{SIR} improvement is observed after \ac{CPE} estimation and correction, being the difference for $N=16384$ negligible. As in the case of Fig.~\ref{fig:phaseNoise_EVM}b, the better performance improvement for lower $N$ is due to the fact that the subcarrier spacing is higher and the \ac{ICI} less pronounced in these cases. For all considered settings, a \ac{SIR} of at least $\SI{15}{dB}$ is observed for combined transmit and receive \ac{PN} levels equal to or lower than $\SI{-15}{dBc}$. At this \ac{PN} level, reasonable communication performance in \ac{OFDM}-based systems can still be achieved, e.g., as shown in \cite{ismail2023}. For lower \ac{PN} levels, critical \ac{SIR} may be experience, leading to significant communication performance degradation. This is especially critical for bistatic \ac{ISAC}, where reliable communication is require to reconstruct the transmit data and enable high-performance sensing \cite{giroto2023_EuMW,brunner2024}. 
	
	Examples on how the \ac{EVM} and \ac{SIR} degradation is related to changes in the constellation shape are shown for \ac{QPSK} modulation considering $\SI{-14.90}{dBc}$ \mbox{($\gamma=\SI{30}{dB}$)} in Figs.~\ref{fig:QPSKconst_psd30} and \ref{fig:QPSKconst_wCPEcorr_psd30}, which respectively address the cases without and with \ac{CPE} estimation and correction. As expected from the results in Fig.~\ref{fig:intPNDelta2Bpll}, the obtained constellations show that the \ac{CPE} is the dominating \ac{PN}-induced degradation for \mbox{$N\in\{256,512,1024,2048\}$}, whereas \mbox{$N\in\{4096,8192,16384\}$} results in more significant \ac{ICI}, which is evidenced by a more spurious, noise-like spreading of the constellation clusters.

	\subsection{Effects of PN on Radar Sensing}\label{subsec:sensPerf}
	
\begin{figure}[!t]
		\centering
		\subfloat[ ]{
			
			\psfrag{0}[c][c]{\scriptsize $0$}
			\psfrag{0.5}[c][c]{\scriptsize $0.5$}
			\psfrag{1}[c][c]{\scriptsize $1$}
			\psfrag{1.5}[c][c]{\scriptsize $1.5$}
			
			\psfrag{-52}[c][c]{\scriptsize -$52$}
			\psfrag{-50}[c][c]{\scriptsize -$50$}
			\psfrag{-48}[c][c]{\scriptsize -$48$}
			\psfrag{-46}[c][c]{\footnotesize -$46$}
			\psfrag{-44}[c][c]{\footnotesize -$44$}
			
			\psfrag{Monostatic range (km)}[c][c]{\footnotesize Monostatic range (km)}
			\psfrag{Phase noise level (dBc)}[c][c]{\footnotesize PN level (dBc)}
			
			\includegraphics[width=3.75cm]{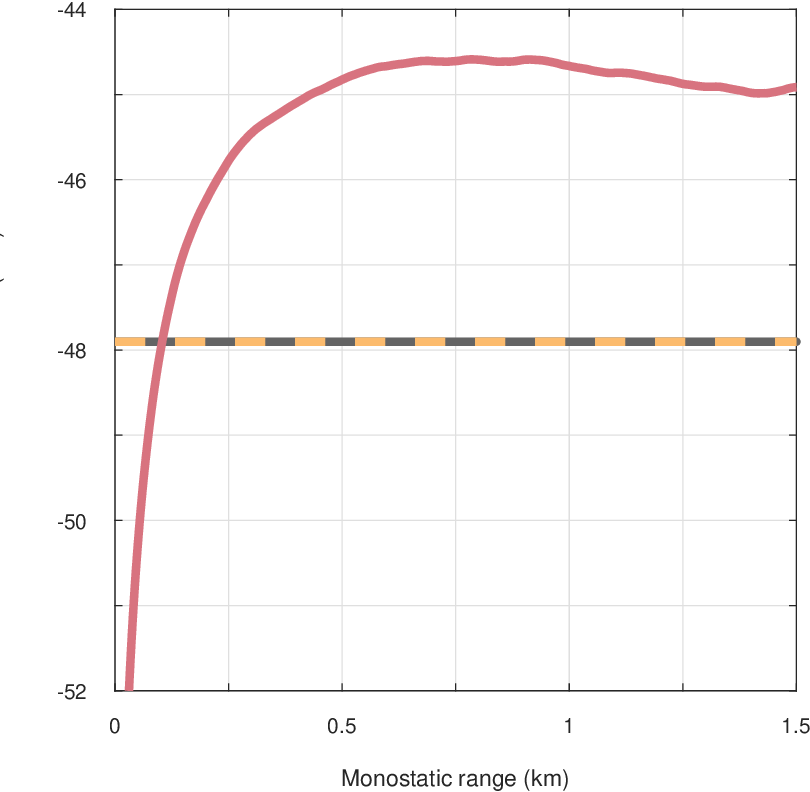}\label{fig:monostPNcorr}			
		}
		\subfloat[ ]{
			
			\psfrag{0}[c][c]{\scriptsize $0$}
			\psfrag{1}[c][c]{\scriptsize $1$}
			\psfrag{2}[c][c]{\scriptsize $2$}
			\psfrag{3}[c][c]{\scriptsize $3$}
			
			\psfrag{-52}[c][c]{ }
			\psfrag{-50}[c][c]{ }
			\psfrag{-48}[c][c]{ }
			\psfrag{-46}[c][c]{ }
			\psfrag{-44}[c][c]{ }
			
			\psfrag{Bistatic range (km)}[c][c]{\footnotesize Bistatic range (km)}
			\psfrag{Phase noise level (dBc)}[c][c]{ }
			
			\includegraphics[width=3.75cm]{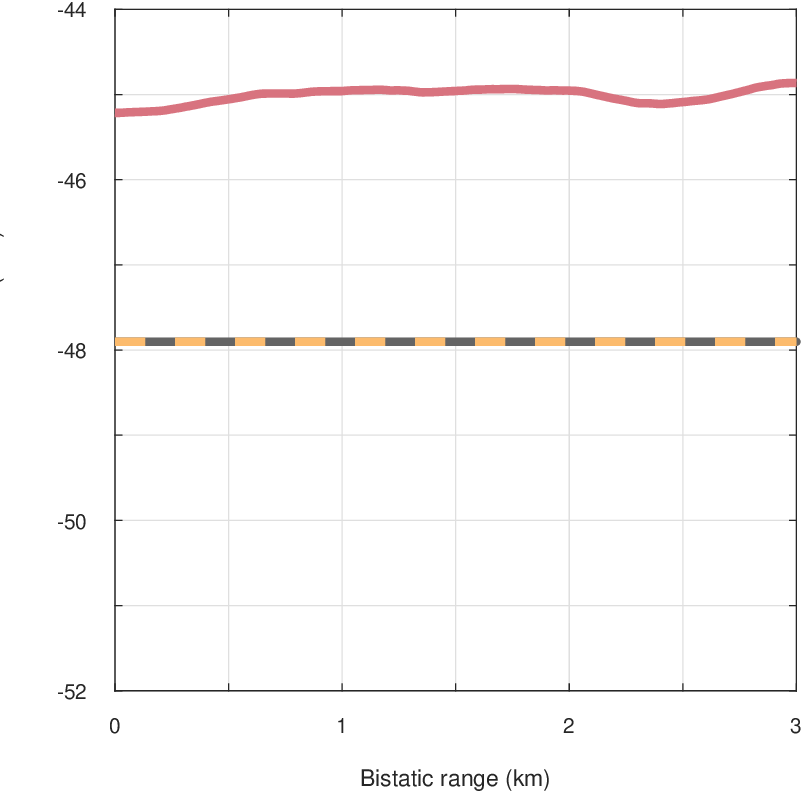}\label{fig:bistPNcorr}
		}
		
		\captionsetup{justification=raggedright,labelsep=period,singlelinecheck=false}
		\caption{\ Combined ({\color[rgb]{0.8471,0.4510,0.4980}\textbf{\textemdash}}) transmit ({\color[rgb]{0.3922,0.3922,0.3922}\textbf{\textemdash}}) and receive ({\color[rgb]{0.9882,0.7333,0.4275}\textbf{\textendash~\textendash}}) PN level as a function of range: (a) monostatic sensing with correlated PN at transmitter and receiver and (b) bistatic sensing with uncorrelated PN. For communication, the same PN level is obtained as for bistatic sensing.}\label{fig:PNcorr}
		
	\end{figure}

\begin{figure}[!t]
		\centering
		
		\psfrag{8}[c][c]{\scriptsize $8$}
		\psfrag{9}[c][c]{\scriptsize $9$}
		\psfrag{10}[c][c]{\scriptsize $10$}
		\psfrag{11}[c][c]{\scriptsize $11$}
		\psfrag{12}[c][c]{\scriptsize $12$}
		\psfrag{13}[c][c]{\scriptsize $13$}
		\psfrag{14}[c][c]{\scriptsize $14$}
		
		\psfrag{AA}[c][c]{\scriptsize $10^{\text{-}2}$}
		\psfrag{BB}[c][c]{\scriptsize $10^{\text{-}1}$}
		\psfrag{CC}[c][c]{\scriptsize $10^0$}
		\psfrag{DD}[c][c]{\scriptsize $10^1$}
		
		\psfrag{log2N}[c][c]{\footnotesize $\log_2(N)$}
		\psfrag{Max. ua. range (km)}[c][c]{\footnotesize Max. ua. range (km)}
		
		\includegraphics[width=3.75cm]{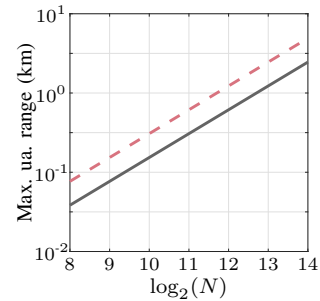}
		
		\captionsetup{justification=raggedright,labelsep=period,singlelinecheck=false}
		\caption{\ Maximum unambiguous range as a function of the number of subcarriers $N$: monostatic sensing ({\color[rgb]{0.3922,0.3922,0.3922}\textbf{\textemdash}}) and bistatic sensing ({\color[rgb]{0.8471,0.4510,0.4980}\textbf{\textendash~\textendash}}).}\label{fig:MUR_N}
		
	\end{figure}

	Considering the interaction of the \ac{PN} at both transmitter and receiver as discussed in Section~\ref{sec:sysModel}, the total level of combined transmit and receive \ac{PN} is shown in Fig~\ref{fig:PNcorr} for both monostatic and bistatic \ac{ISAC} architectures previously depicted in Fig.~\ref{fig:isacArchitectures}. Fig.~\ref{fig:monostPNcorr} shows the combined \ac{PN} level for monostatic radar sensing as a function of the monostatic range. For monostatic radar sensing, the transmit \ac{PN} $\theta^\mathrm{PN}_\mathrm{Tx}(t)$ and the receive \ac{PN} $\theta^\mathrm{PN}_\mathrm{Rx}(t)$ are correlated. As discussed in Section.~\ref{sec:sysModel}, the transmit \ac{PN} contribution to the overall \ac{PN} is \mbox{$\theta^\mathrm{PN}_\mathrm{Tx}(t-\tau)$}, where \mbox{$\tau=2R/c_0$} is the propagation delay associated with the monostatic range $R$. The receive \ac{PN} contribution is in turn \mbox{$\theta^\mathrm{PN}_\mathrm{Rx}(t)=\theta^\mathrm{PN}_\mathrm{Tx}(t)$}. This results in a combined transmit and receive \ac{PN} level that tends to $\SI{0}{rad^2}$ or $\SI[parse-numbers = false]{-\infty}{dBc}$ for decreasing ranges or propagation delays, as
	\begin{equation}\label{eq:eq_lim}
		\lim_{\tau\to 0} \left[\theta^\mathrm{PN}_\mathrm{Tx}(t-\tau)-\theta^\mathrm{PN}_\mathrm{Tx}(t)\right] = 0.
	\end{equation}
	For increasing ranges or delays, however, the $\theta^\mathrm{PN}_\mathrm{Tx}(t)$ and $\theta^\mathrm{PN}_\mathrm{Rx}(t)$ tend to become uncorrelated and their \ac{PN} levels, which are both equal to $\SI{-47.90}{dBc}$, are simply added, resulting in the same combined transmit and receive \ac{PN} level of $\SI{-44.90}{dBc}$ observed in Section~\ref{subsec:constDist}. The bistatic sensing case in Fig.~\ref{fig:bistPNcorr} shows a different behavior. Since $\theta^\mathrm{PN}_\mathrm{Tx}(t)$ and $\theta^\mathrm{PN}_\mathrm{Rx}(t)$ are uncorrelated as in the communication case discussed in Section~\ref{subsec:constDist}, since distinct osccilators are used by transmitter and receiver, the transmit and receive \ac{PN} level add to $\SI{-44.90}{dBc}$ regardless of the experienced range or propagation delay. For comparison, the achievable maximum unambiguous range for both monostatic and bistatic radar sensing, as a function of the number of subcarriers $N$ and assuming $B=\SI{1}{\giga\hertz}$ as in Table~\ref{tab:ofdmParameters} is shown in Fig.~\ref{fig:MUR_N}. The maximum unambiguous range is defined as \mbox{$R_\mathrm{max,ua}=N\Delta R$}, where $\Delta R$ is the range resolution. For monostatic radar sensing, the maximum unambiguous range is defined as \mbox{$R^\mathrm{monost}_\mathrm{max,ua}=N\Delta R^\mathrm{monost}=N[c_0/(2B)]$} \cite{giroto2021_tmtt}, while for bistatic radar sensing it is \mbox{$R^\mathrm{bist}_\mathrm{max,ua}=N\Delta R^\mathrm{bist}=N(c_0/B)$} \cite{giroto2023_EuMW}. The achieved results in Fig.~\ref{fig:MUR_N} show that the maximum unambiguous range spans from $\SI{0.04}{\kilo\meter}$ to $\SI{2.45}{\kilo\meter}$ for the monostatic case, and from $\SI{0.08}{\kilo\meter}$ to $\SI{4.91}{\kilo\meter}$ for the bistatic case. This reveals that, for $N\leq4096$, the whole achieved monostatic maximum unambiugous ranges profit from the transmit and receive \ac{PN} correlation to a certain extent. Consequently, lower combined transmit and receive \ac{PN} levels are experienced than in both the monostatic case for $N>4096$ and the bistatic case regardless of $N$.
	
	\begin{figure*}[!t]
		\centering
		\subfloat[ ]{
			
			\psfrag{-45}[c][c]{\scriptsize -$45$}
			\psfrag{-30}[c][c]{\scriptsize -$30$}
			\psfrag{-15}[c][c]{\scriptsize -$15$}
			\psfrag{0}[c][c]{\scriptsize $0$}
			\psfrag{15}[c][c]{\scriptsize $15$}
			\psfrag{30}[c][c]{\scriptsize $30$}
			\psfrag{45}[c][c]{\scriptsize $45$}
			
			\psfrag{-60}[c][c]{\scriptsize -$60$}
			\psfrag{-50}[c][c]{\scriptsize -$50$}
			\psfrag{-40}[c][c]{\scriptsize -$40$}
			\psfrag{-30}[c][c]{\scriptsize -$30$}
			\psfrag{-20}[c][c]{\scriptsize -$20$}
			\psfrag{-10}[c][c]{\scriptsize -$10$}
			\psfrag{0}[c][c]{\scriptsize $0$}
			
			\psfrag{Phase noise level (dBc)}[c][c]{\footnotesize PN level (dBc)}
			\psfrag{PPLR (dB)}[c][c]{\footnotesize $\mathrm{PPLR~(dB)}$}			
			
			\includegraphics[width=3.25cm]{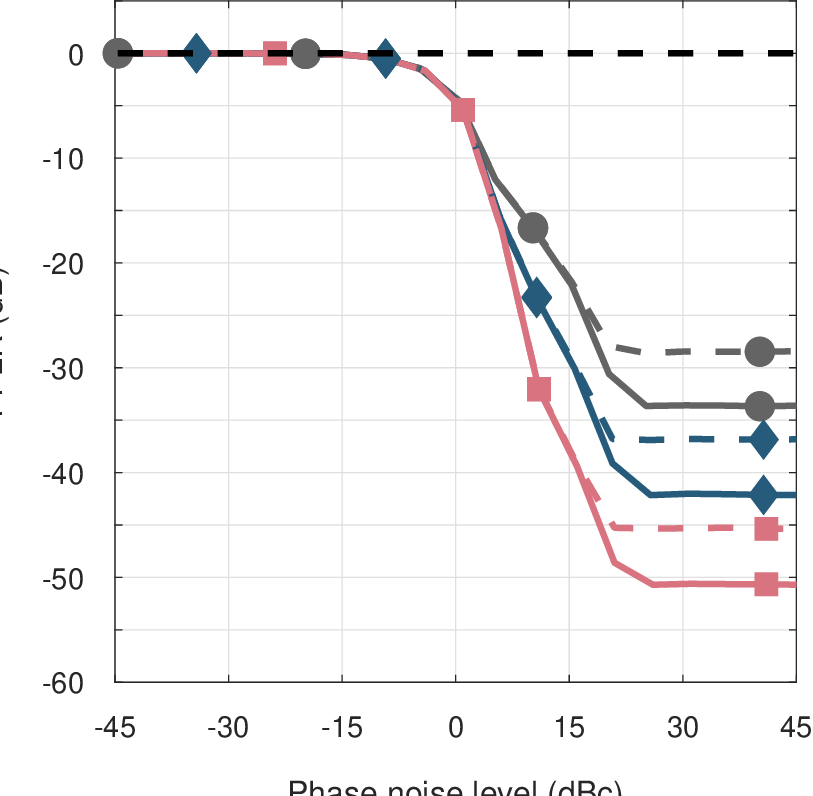}			
		}\hspace{0.05cm}
		\subfloat[ ]{
			
			\psfrag{-45}[c][c]{\scriptsize -$45$}
			\psfrag{-30}[c][c]{\scriptsize -$30$}
			\psfrag{-15}[c][c]{\scriptsize -$15$}
			\psfrag{0}[c][c]{\scriptsize $0$}
			\psfrag{15}[c][c]{\scriptsize $15$}
			\psfrag{30}[c][c]{\scriptsize $30$}
			\psfrag{45}[c][c]{\scriptsize $45$}
			
			\psfrag{-15}[c][c]{\scriptsize -$15$}
			\psfrag{-12}[c][c]{\scriptsize -$12$}
			\psfrag{-9}[c][c]{\scriptsize -$9$}
			\psfrag{-6}[c][c]{\scriptsize -$6$}
			\psfrag{-3}[c][c]{\scriptsize -$3$}
			\psfrag{0}[c][c]{\scriptsize $0$}
			
			\psfrag{Phase noise level (dBc)}[c][c]{\footnotesize PN level (dBc)}
			\psfrag{PSLR (dB)}[c][c]{\footnotesize Range $\mathrm{PSLR~(dB)}$}	
			
			\includegraphics[width=3.25cm]{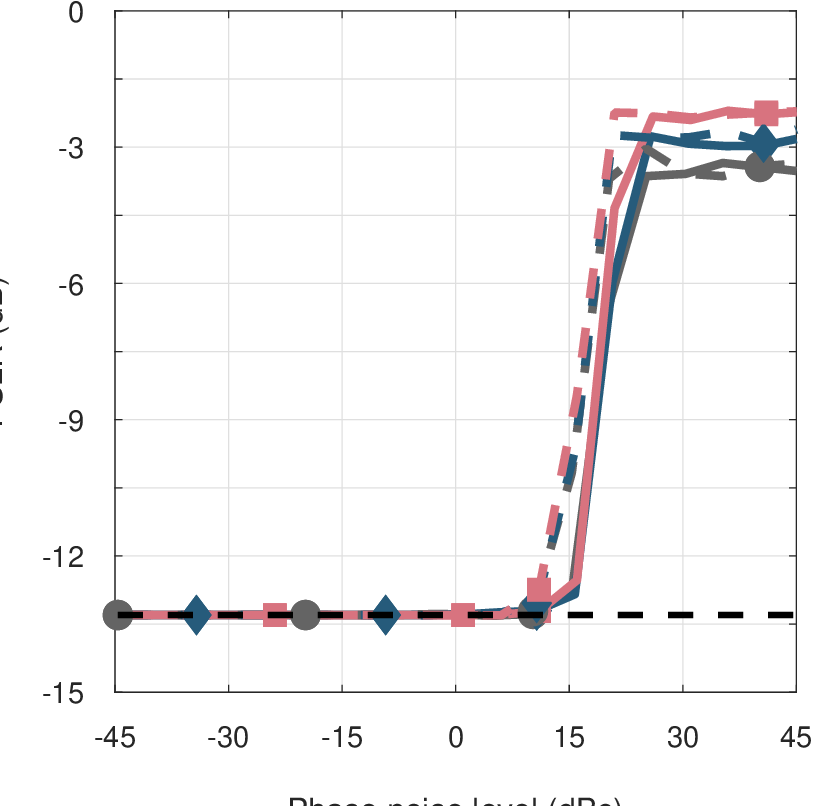}	
			
		}\hspace{0.05cm}
		\subfloat[ ]{
			
			\psfrag{-45}[c][c]{\scriptsize -$45$}
			\psfrag{-30}[c][c]{\scriptsize -$30$}
			\psfrag{-15}[c][c]{\scriptsize -$15$}
			\psfrag{0}[c][c]{\scriptsize $0$}
			\psfrag{15}[c][c]{\scriptsize $15$}
			\psfrag{30}[c][c]{\scriptsize $30$}
			\psfrag{45}[c][c]{\scriptsize $45$}
			
			\psfrag{-12}[c][c]{\scriptsize -$12$}
			\psfrag{-4}[c][c]{\scriptsize -$4$}
			\psfrag{4}[c][c]{\scriptsize $4$}
			\psfrag{12}[c][c]{\scriptsize $12$}
			\psfrag{20}[c][c]{\scriptsize $20$}
			\psfrag{28}[c][c]{\scriptsize $28$}
			\psfrag{36}[c][c]{\scriptsize $36$}
			
			\psfrag{Phase noise level (dBc)}[c][c]{\footnotesize PN level (dBc)}
			\psfrag{ISLR (dB)}[c][c]{\footnotesize Range $\mathrm{ISLR~(dB)}$}	
			
			\includegraphics[width=3.25cm]{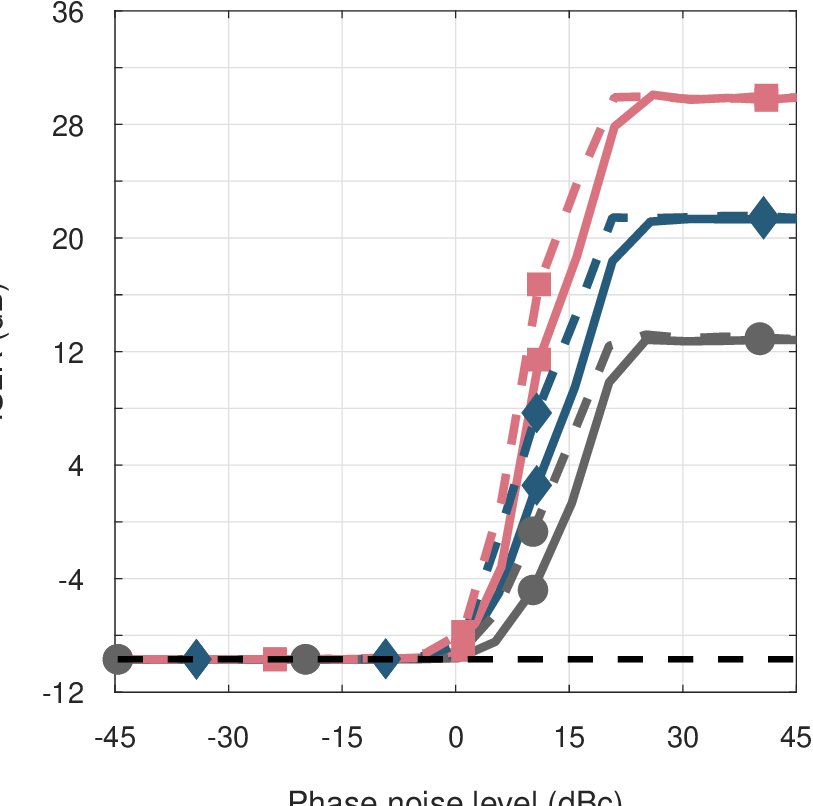}
			
		}\hspace{0.05cm}
		\subfloat[ ]{
			
			\psfrag{-45}[c][c]{\scriptsize -$45$}
			\psfrag{-30}[c][c]{\scriptsize -$30$}
			\psfrag{-15}[c][c]{\scriptsize -$15$}
			\psfrag{0}[c][c]{\scriptsize $0$}
			\psfrag{15}[c][c]{\scriptsize $15$}
			\psfrag{30}[c][c]{\scriptsize $30$}
			\psfrag{45}[c][c]{\scriptsize $45$}
			
			\psfrag{-15}[c][c]{\scriptsize -$15$}
			\psfrag{-12}[c][c]{\scriptsize -$12$}
			\psfrag{-9}[c][c]{\scriptsize -$9$}
			\psfrag{-6}[c][c]{\scriptsize -$6$}
			\psfrag{-3}[c][c]{\scriptsize -$3$}
			\psfrag{0}[c][c]{\scriptsize $0$}
			
			\psfrag{Phase noise level (dBc)}[c][c]{\footnotesize PN level (dBc)}
			\psfrag{PSLR (dB)}[c][c]{\footnotesize Doppler $\mathrm{PSLR~(dB)}$}	
			
			\includegraphics[width=3.25cm]{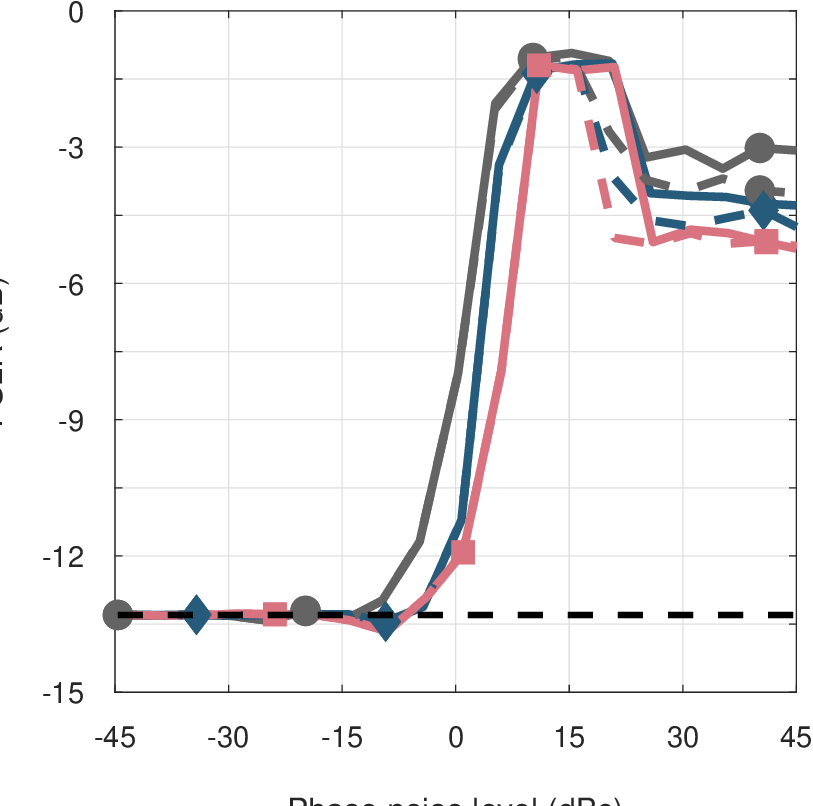}	
			
		}\hspace{0.05cm}
		\subfloat[ ]{
			
			\psfrag{-45}[c][c]{\scriptsize -$45$}
			\psfrag{-30}[c][c]{\scriptsize -$30$}
			\psfrag{-15}[c][c]{\scriptsize -$15$}
			\psfrag{0}[c][c]{\scriptsize $0$}
			\psfrag{15}[c][c]{\scriptsize $15$}
			\psfrag{30}[c][c]{\scriptsize $30$}
			\psfrag{45}[c][c]{\scriptsize $45$}
			
			\psfrag{-12}[c][c]{\scriptsize -$12$}
			\psfrag{-4}[c][c]{\scriptsize -$4$}
			\psfrag{4}[c][c]{\scriptsize $4$}
			\psfrag{12}[c][c]{\scriptsize $12$}
			\psfrag{20}[c][c]{\scriptsize $20$}
			\psfrag{28}[c][c]{\scriptsize $28$}
			\psfrag{36}[c][c]{\scriptsize $36$}
			
			\psfrag{Phase noise level (dBc)}[c][c]{\footnotesize PN level (dBc)}
			\psfrag{ISLR (dB)}[c][c]{\footnotesize Doppler $\mathrm{ISLR~(dB)}$}
			
			\includegraphics[width=3.25cm]{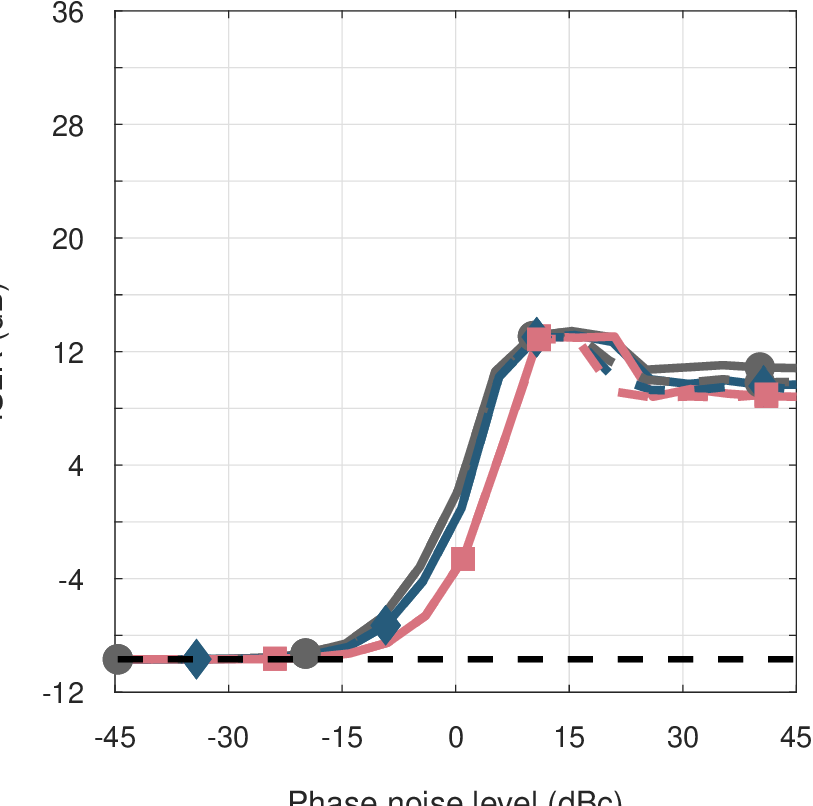}		
			
		}
		
		\captionsetup{justification=raggedright,labelsep=period,singlelinecheck=false}
		\caption{\ PPLR (a), range PSLR (b) and ISLR (c), and Doppler shift PSLR (e) and ISLR (d) as functions of the combined transmit and receive PN level. Simulation results are shown for a static target assuming \mbox{$N=256$} and QPSK \mbox{({\color[rgb]{0.3922,0.3922,0.392}\textbf{\textemdash}} and {\color[rgb]{0.3922,0.3922,0.3922}$\CIRCLE$})}, \mbox{$N=2048$} and QPSK \mbox{({\color[rgb]{0.1490,0.3569,0.4824}\textbf{\textemdash}} and {\color[rgb]{0.1490,0.3569,0.4824}$\blacklozenge$})}, and \mbox{$N=16384$} and QPSK \mbox{({\color[rgb]{0.8471,0.4510,0.4980}\textbf{\textemdash}} and {\color[rgb]{0.8471,0.4510,0.4980}$\blacksquare$})}. In addition, results are also shown for \mbox{256-QAM} combined with the same aforementioned number of subcarriers, i.e., \mbox{$N=256$} \mbox{({\color[rgb]{0.3922,0.3922,0.392}\textbf{\textendash~\textendash}} and {\color[rgb]{0.3922,0.3922,0.3922}$\CIRCLE$})}, \mbox{$N=2048$} \mbox{({\color[rgb]{0.1490,0.3569,0.4824}\textbf{\textendash~\textendash}} and {\color[rgb]{0.1490,0.3569,0.4824}$\blacklozenge$})}, and \mbox{$N=16384$} \mbox{({\color[rgb]{0.8471,0.4510,0.4980}\textbf{\textendash~\textendash}} and {\color[rgb]{0.8471,0.4510,0.4980}$\blacksquare$})}. In all subfigures, a CP length of $N_\mathrm{CP}=N$ and $M=128$ OFDM symbols were considered. In addition, the PPLR, PSLR and ISLR values for simulations without PN are also shown for comparison ({\color[rgb]{0,0,0}\textbf{\textendash~\textendash}}) and are the same regardless of the adopted modulation.}\label{fig:sidelobe_PSD}
		
	\end{figure*}
	
	Next, to quantify the \ac{PN}-induced degradation on monostatic and bistatic sensing in \ac{OFDM}-based \ac{ISAC} systems, the \ac{PPLR}, \ac{PSLR}, and \ac{ISLR} parameters are used \cite{lellouch2016,giroto2021_tmtt,liao2024}. The first of the aforementioned parameters expresses the ratio between the main lobe power under \ac{PN} influence and the ideally expected main lobe power without \ac{PN} \cite{overdevest2020} after experiencing the radar processing gain of ideally $NM$ under absence of \ac{PN} \cite{giroto2021_tmtt,giroto2023_EuMW}. The latter parameters, i.e., \ac{PSLR} and \ac{ISLR},  express the ratio between the peak power of the highest sidelobe and the main lobe and the ratio between the integrated sidelobe and main lobe powers \cite{lellouch2016}, respectively. All three parameters are used to evaluate the \ac{PN} effect on target reflections in a radar image under different settings. In addition, analysis using image \ac{SIR}, which provides a measure between a target peak and the \ac{PN}-induced interference in the radar image, as a performance parameter are also presented. The obtained results are discussed in Sections~\ref{subsubsec:sensN} through \ref{subsubsec:3gpp}, for which parameter sets Table~\ref{tab:ofdmParameters} are considered. The specifically adopted parameter values are explicitly mentioned in each of the aforementioned sections. Since the constellation distortion and \ac{SIR} were analyzed in detail in Sec.~\ref{subsec:constDist} for all $N$ and modulation alphabets from Table~\ref{tab:ofdmParameters}, only $N\in\{256,2048,16384\}$ as well as \ac{QPSK} and \mbox{256-\ac{QAM}} are henceforth considered for conciseness. This covers the cases \mbox{$\Delta f/2\gg B_\mathrm{PLL}$}, \mbox{$\Delta f/2\approx B_\mathrm{PLL}$}, and \mbox{$\Delta f/2\ll B_\mathrm{PLL}$}, respectively, as well as the two extremes of the considered modulation alphabets.
	
	To focus only on the \ac{PN}-induced impairments, self-interference is disregarded in the monostatic sensing case, and full knowledge of the transmit \ac{OFDM} frame as well as perfect time, frequency and sampling frequency synchronization is assumed for bistatic sensing. For a more insightful analysis, \ac{PPLR} and \ac{ISLR} are calculated in both range and Doppler shift directions and, as in Section~\ref{subsec:constDist}, \ac{AWGN} is not considered to highlight the \ac{PN} effects. As in Section~\ref{subsec:constDist}, multiple \ac{PN} level increase factors $\gamma$ were considered in the simulations.
	
	\subsubsection{Main and sidelobe distortion for different numbers of subcarriers and modulation alphabets}\label{subsubsec:sensN}
	
	To enable analyzing the radar sensing performance as a function of the number of subcarriers $N$ and the modulation alphabet for a wide range of combined transmit and receive \ac{PN} level, Fig.~\ref{fig:sidelobe_PSD} shows the \ac{PPLR}, range \ac{PSLR} and \ac{ISLR}, as well as Doppler shift \ac{PSLR} and \ac{ISLR} obtained for a static target. Since the monostatic and bistatic cases only differ in the transmit and receive \ac{PN} correlation, which ultimately yields different combined \ac{PN} levels as shown in Fig.~\ref{fig:PNcorr}, a generic case was simulated and the \ac{PN} level was varied. This is later also done in Sections~\ref{subsubsec:sensM} and \ref{subsubsec:sensDoppler}. In addition, \mbox{$N\in\{256,2048,16384\}$}, \mbox{$N_\mathrm{CP}=N$} to ensure that the maximum unambiguous range is measurable, as well as \ac{QPSK} and 256-\ac{QAM} were considered, besides \mbox{$M=128$} as in Section~\ref{subsec:constDist}. The obtained results in Fig.~\ref{fig:sidelobe_PSD}a reveal that the combined \ac{PN} level result in negligible \ac{PPLR} until around $\SI{-9.25}{dBc}$. Afterwards, the \ac{PPLR} degrades linearly until around $\SI{20.26}{dBc}$, reaching a constant level that depends on the adopted $N$ and modulation. These are $\SI{-33.67}{dB}$, $\SI{-42.13}{dB}$, and $\SI{-50.66}{dB}$ for $N=256$, $N=2048$, and $N=16384$, respectively, and \ac{QPSK} modulation. For \mbox{256-\ac{QAM}}, the obtained \ac{PPLR} values for high combined \ac{PN} level were $\SI{5.28}{dB}$ higher, i.e., $\SI{-28.47}{dB}$, $\SI{-36.85}{dB}$, and $\SI{-45.38}{dB}$ for $N=256$, $N=2048$, and $N=16384$, respectively. From $N=256$ to $N=2048$ as well as from $N=2048$ to $N=16384$, the \ac{PPLR} degrades $\SI{8.47}{dB}$ on average, i.e., it is worsened by a factor of approximately $7$. This is close to the ratio $8$ between $N=16384$ and $N=2048$, and also $N=2048$ and $N=256$, which also indicates the factor by which the subcarrier spacing $\Delta f$ is decreased when increasing $N$, leading to more significant \ac{PN}-induced \ac{ICI} effect. An exact \ac{PPLR} degradation by a factor of $8$ is only not observed due to the form in which the \ac{PN} level is spread over frequency as shown in Fig.~\ref{fig:PNpsd}.
		
	Regarding range \ac{PSLR}, Fig.~\ref{fig:sidelobe_PSD}b shows that apart from a transition zone of a combined \ac{PN} level between approximately $\SI{10.21}{dBc}$ and $\SI{25.73}{dBc}$, where \mbox{256-\ac{QAM}} starts degrading at \ac{PN} levels around $\SI{3}{dB}$ lower than \ac{QPSK}, the choice of the modulation order does not influence the achieved \ac{PSLR}. For  combined transmit and receive \ac{PN} levels from $\SI{25.73}{dBc}$, $N=256$, $N=2048$, and $N=16384$ result in constant range \ac{PSLR} values of $\SI{-3.42}{dB}$, $\SI{-2.97}{dB}$, and $\SI{-2.28}{dB}$, respectively. In terms of range \ac{ISLR}, a similar transition zone to the aforementioned one is observed in Fig.~\ref{fig:sidelobe_PSD}c. Afterwards, the same result is achieved regardless of the adopted modulation, and $N=256$, $N=2048$, and $N=16384$ yield \ac{ISLR} values of around $\SI{12.99}{dB}$, $\SI{21.53}{dB}$, and $\SI{30.02}{dB}$, respectively. The difference of about $\SI{8.5}{dB}$, which corresponds to worse \ac{ISLR} by a factor of $7$ between $N=16384$ and $N=2048$, as well as between $N=2048$ and $N=256$, is due to the increase in the number of samples used to represent the range axis of the radar image. Since the main lobe width is the same for all three aforementioned cases but the amount of samples containing sidelobes increases, the \ac{ISLR} naturally degrades. This is, however, only observed since the high \ac{PN} level results in significant \ac{ICI} that severely increases the sidelobe level. Otherwise approximately the same \ac{ISLR} is observed for all considered $N$ values as is the case for combined \ac{PN} levels equal to or smaller than about $\SI{9.65}{dBc}$ in Fig.~\ref{fig:sidelobe_PSD}c. The achieved values show that the range \ac{PSLR} and \ac{ISLR} are only affected by rather high \ac{PN} levels. This indicates that the \ac{PN}-induced \ac{ICI} can be suppressed by the radar processing gain as it is a Gaussian noise-like effect, and that \ac{CPE} is the dominating source of \ac{PN}-induced degradation of the sensing performance in \ac{OFDM}-based \ac{ISAC} for reasonable \ac{PN} levels.
	
	\begin{figure}[!t]
		\centering
		\psfrag{0}[c][c]{\scriptsize $0$}
		\psfrag{-10}[c][c]{\scriptsize -$10$}
		\psfrag{-20}[c][c]{\scriptsize -$20$}
		\psfrag{-30}[c][c]{\scriptsize -$30$}
		
		\psfrag{-0.01}[c][c]{\scriptsize -$0.01$}
		\psfrag{-0.005}[c][c]{\scriptsize -$0.005$}
		\psfrag{0}[c][c]{\scriptsize $0$}
		\psfrag{0.005}[c][c]{\scriptsize $0.005$}
		\psfrag{0.01}[c][c]{\scriptsize $0.01$}
		
		\psfrag{fD Df}[c][c]{\footnotesize $f_\mathrm{D}/\Delta f$}
		\psfrag{Norm. mag. (dB)}[c][c]{\footnotesize Norm. mag. (dB)}
		
		\includegraphics[width=3.75cm]{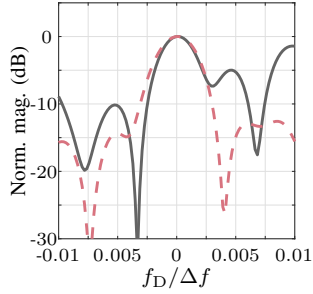}
		
		\captionsetup{justification=raggedright,labelsep=period,singlelinecheck=false}
		\caption{\ Cut along the Doppler shift direction of bistatic range-Doppler radar images generated for a static target assuming $N=2048$, $N_\mathrm{CP}=512$, $M=128$, QPSK modulation, and rectangular windowing in both range and Doppler shift directions. Results are shown for $\SI{10}{dBc}$ \mbox{($\gamma=\SI{54.90}{dB}$)} ({\color[rgb]{0.3922,0.3922,0.3922}\textbf{\textemdash}}) and $\SI{35}{dBc}$ \mbox{($\gamma=\SI{79.90}{dB}$)} ({\color[rgb]{0.8471,0.4510,0.4980}\textbf{\textendash~\textendash}}). Since the considered PN levels lead to different PPLR and therefore peak heights in the radar image, the magnitudes are normalized to the respective static target peaks located at \mbox{$f_D/\Delta f=0$}. }\label{fig:dopplerCuts_zoom}
		
	\end{figure}

\begin{figure}[!t]
		\centering
		\subfloat[ ]{
			
			\psfrag{-15}[c][c]{\scriptsize -$15$}
			\psfrag{-7.5}[c][c]{\scriptsize -$7.5$}
			\psfrag{0}[c][c]{\scriptsize $0$}
			\psfrag{7.5}[c][c]{\scriptsize $7.5$}
			\psfrag{15}[c][c]{\scriptsize $15$}
			
			\psfrag{0}[c][c]{\scriptsize $0$}
			\psfrag{0.04}[c][c]{\scriptsize $0.04$}
			\psfrag{0.08}[c][c]{\scriptsize $0.08$}
			\psfrag{0.12}[c][c]{\scriptsize $0.12$}
			\psfrag{0.16}[c][c]{\scriptsize $0.16$}
			
			\psfrag{Phase noise (rad)}[c][c]{\footnotesize Phase offset (rad)}
			\psfrag{Probability density}[c][c]{\vspace{-2cm}\footnotesize Probability density}
			
			\includegraphics[width=3.75cm]{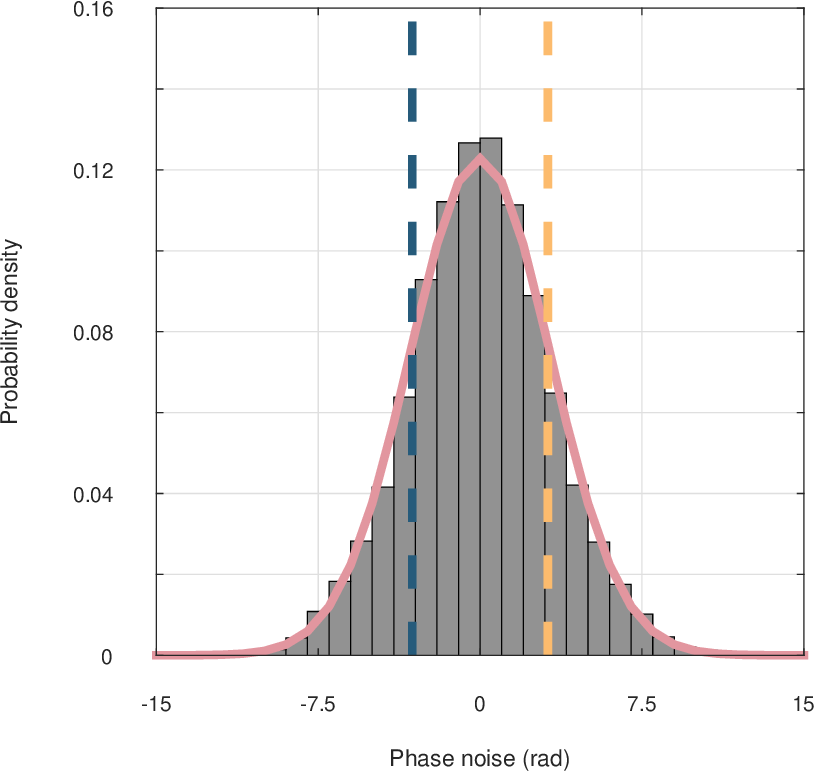}				
		}\hspace{0.1cm}
		\subfloat[ ]{
			
			\psfrag{-300}[c][c]{\scriptsize -$300$}
			\psfrag{-150}[c][c]{\scriptsize -$150$}
			\psfrag{0}[c][c]{\scriptsize $0$}
			\psfrag{150}[c][c]{\scriptsize $150$}
			\psfrag{300}[c][c]{\scriptsize $300$}
			
			\psfrag{AAA}[c][c]{\scriptsize $0$}
			\psfrag{BBB}[c][c]{\hspace{-0.2cm}\scriptsize $2\cdot10^{\text{-}3}$}
			\psfrag{CCC}[c][c]{\hspace{-0.2cm}\scriptsize $4\cdot10^{\text{-}3}$}
			\psfrag{DDD}[c][c]{\hspace{-0.2cm}\scriptsize $6\cdot10^{\text{-}3}$}
			\psfrag{EEE}[c][c]{\hspace{-0.2cm}\scriptsize $8\cdot10^{\text{-}3}$}
			
			
			\psfrag{Phase noise (rad)}[c][c]{\footnotesize Phase offset (rad)}
			\psfrag{Probability density}[c][c]{ }
			
			\includegraphics[width=3.75cm]{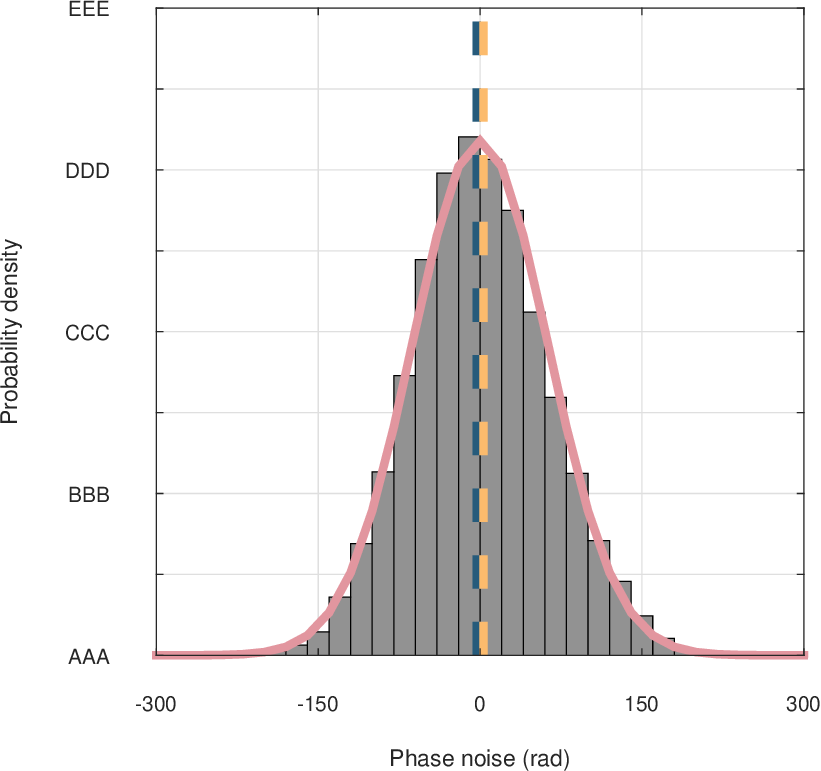}				
		}\\
		\subfloat[ ]{
			
			\psfrag{-2pi}[c][c]{\scriptsize -$2\pi$}
			\psfrag{-3pi2}[c][c]{ }
			\psfrag{-pi}[c][c]{\scriptsize -$\pi$}
			\psfrag{-pi2}[c][c]{ }
			\psfrag{0}[c][c]{\scriptsize $0$}
			\psfrag{pi2}[c][c]{ }
			\psfrag{pi}[c][c]{\scriptsize $\pi$}
			\psfrag{3pi2}[c][c]{ }
			\psfrag{2pi}[c][c]{\scriptsize $2\pi$}
			
			\psfrag{0}[c][c]{\scriptsize $0$}
			\psfrag{0.05}[c][c]{\scriptsize $0.05$}
			\psfrag{0.1}[c][c]{\scriptsize $0.10$}
			\psfrag{0.15}[c][c]{\scriptsize $0.15$}
			\psfrag{0.2}[c][c]{\scriptsize $0.20$}
			
			\psfrag{Phase noise (rad)}[c][c]{\footnotesize PN (rad)}
			\psfrag{Probability density}[c][c]{\footnotesize Probability density}
			
			\includegraphics[width=3.75cm]{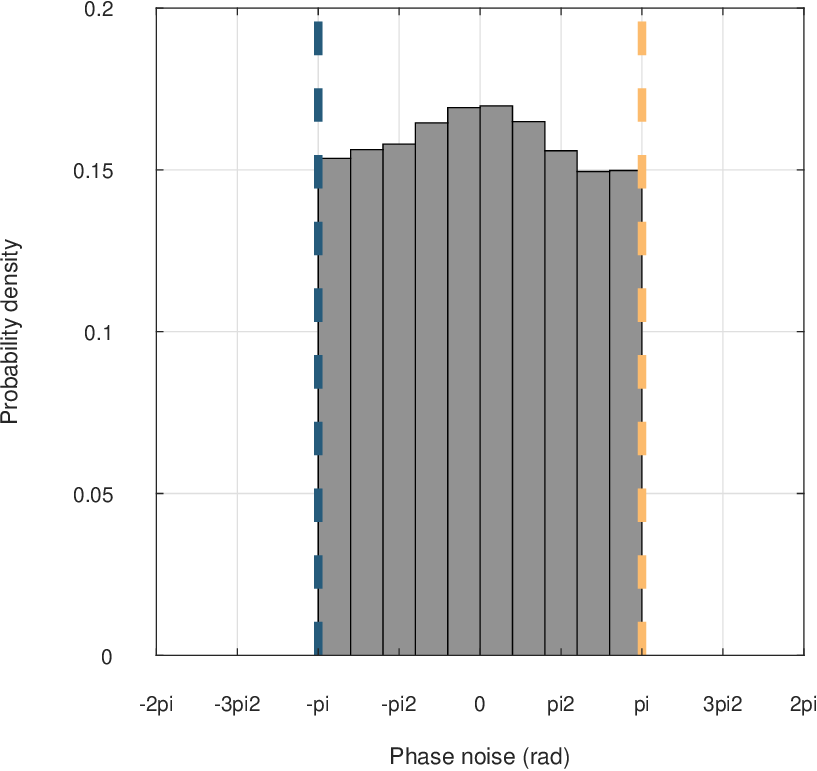}				
		}\hspace{0.1cm}
		\subfloat[ ]{
			
			\psfrag{-2pi}[c][c]{\scriptsize -$2\pi$}
			\psfrag{-3pi2}[c][c]{ }
			\psfrag{-pi}[c][c]{\scriptsize -$\pi$}
			\psfrag{-pi2}[c][c]{ }
			\psfrag{0}[c][c]{\scriptsize $0$}
			\psfrag{pi2}[c][c]{ }
			\psfrag{pi}[c][c]{\scriptsize $\pi$}
			\psfrag{3pi2}[c][c]{ }
			\psfrag{2pi}[c][c]{\scriptsize $2\pi$}
			
			\psfrag{0}[c][c]{\scriptsize $0$}
			\psfrag{0.05}[c][c]{\scriptsize $0.05$}
			\psfrag{0.1}[c][c]{\scriptsize $0.10$}
			\psfrag{0.15}[c][c]{\scriptsize $0.15$}
			\psfrag{0.2}[c][c]{\scriptsize $0.20$}
			
			\psfrag{Phase noise (rad)}[c][c]{\footnotesize PN (rad)}
			\psfrag{Probability density}[c][c]{ }
			
			\includegraphics[width=3.75cm]{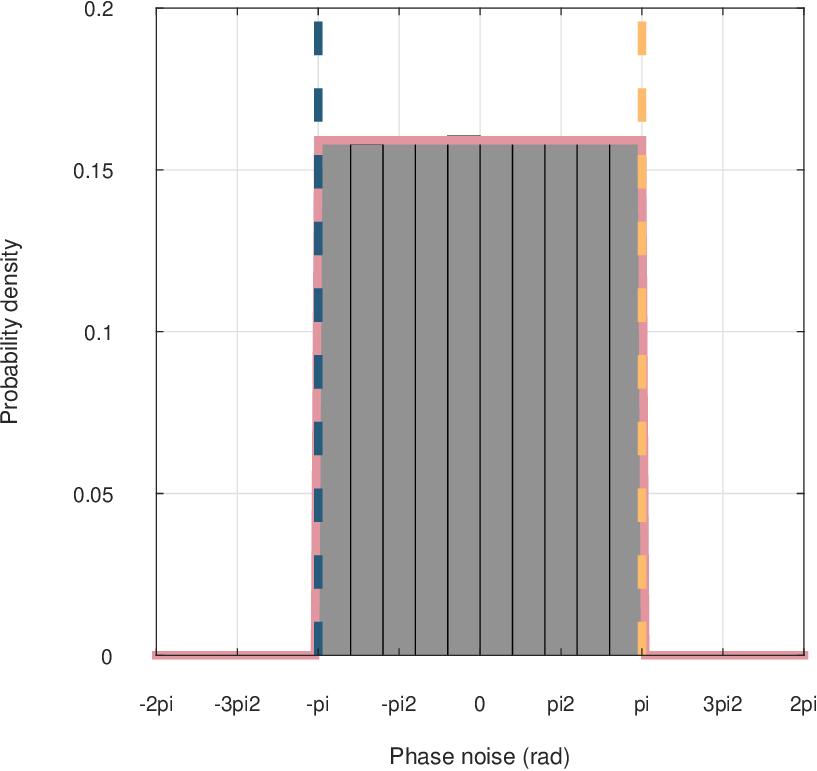}				
		}
		
		\captionsetup{justification=raggedright,labelsep=period,singlelinecheck=false}
		\caption{\ PDF of the PN in the time domain: (a) $\SI{9}{dBc}$ \mbox{($\gamma=\SI{53.90}{dB}$)} and (b) $\SI{34}{dBc}$ \mbox{($\gamma=\SI{78.90}{dB}$)}. In (c), and (d), the same cases as in (a) and (b) are respectively shown, after wrapping the phase between $-\pi$ ({\color[rgb]{0.1490,0.3569,0.4824}\textbf{\textendash~\textendash}}) and $\pi$ ({\color[rgb]{0.9882,0.7333,0.4275}\textbf{\textendash~\textendash}}). While the PDFs in both (a) and (b) can be properly fitted by Gaussian distributions, and (d) by an uniform distribution, the PDF in (c) rather presents the form of a wrapped Gaussian distribution. For convenience, all obtained distribution fits are marked ({\color[rgb]{0.8471,0.4510,0.4980}\textbf{\textemdash}}).}\label{fig:PN_dist12}
		
	\end{figure}

	Finally, in terms of Doppler shift \ac{PSLR} and \ac{ISLR}, better performance is achieved by \ac{QPSK} and smaller $N$ in the region between $\SI{-14.55}{dBc}$ and $\SI{11}{dBc}$, where both \ac{PSLR} and \ac{ISLR} linearly degrade. The influence of the adopted modulation order on this tendency is small and becomes negligible with increasing $N$. Afterwards, both parameters improve with increasing combined transmit and receive \ac{PN} level of up to about $\SI{26.06}{dBc}$, where a constant level is achieved. This is illustrated in Fig.~\ref{fig:dopplerCuts_zoom}, where cuts along the Doppler shift direction of the radar images used to generate the results from Fig.~\ref{fig:sidelobe_PSD} are shown. It can be observed that a combined \ac{PN} level of $\SI{10}{dBc}$ \mbox{($\gamma=\SI{54.90}{dB}$)} results in higher sidelobes and a sligthly narrower main lobe due to interaction with the sidelobes than in the case for $\SI{35}{dBc}$ \mbox{($\gamma=\SI{79.90}{dB}$)}. As a result, this ultimately leads to worse \ac{PSLR} and \ac{ISLR} performances. The unexpected improved radar sensing performance with increasing \ac{PN} level is due to the fact that the resulting \ac{PN} \ac{PDF} presents phase values outside the $(-\pi,\pi)$ range as shown in Fig.~\ref{fig:PN_dist12}. Consequently, the \ac{PDF} folds around and is becomes a wrapped Gaussian distribution. This results in a changing \ac{PSLR} and \ac{ISLR} performance until the folding of the \ac{PDF} becomes strong enough to give the \ac{PN} \ac{PDF} an uniform shape, which is when the \ac{PSLR} and \ac{ISLR} parameters present a constant level. It is, however, worth mentioning that such high combined transmit and receive \ac{PN} levels should not be expected in practice, especially for \ac{mmWave} systems. The aforementioned achieved \ac{PSLR} levels are of approximately $\SI{-3.02}{dB}$, $\SI{-4.07}{dB}$, and $\SI{-5.07}{dB}$ for $N=256$, $N=2048$, and $N=16384$, respectively. As for \ac{ISLR}, the aforementioned $N$ values yield $\SI{10.88}{dB}$, $\SI{9.57}{dB}$, and $\SI{8.97}{dB}$, respectively. The slightly better \ac{PSLR} and \ac{ISLR} along with $N$ is due to the increasing \ac{ICI} relevance in this case, which tends to make the \ac{PN}-induced \ac{CPE} less severe, although still significant, therefore resulting in less degradation of the \ac{DFT}-based Doppler shift estimation that relies on the phases of consecutive \ac{OFDM} symbols resulting from the target's Doppler shift.
	
	\begin{figure*}[!t]
		\centering
		\subfloat[ ]{
			
			\psfrag{-45}[c][c]{\scriptsize -$45$}
			\psfrag{-30}[c][c]{\scriptsize -$30$}
			\psfrag{-15}[c][c]{\scriptsize -$15$}
			\psfrag{0}[c][c]{\scriptsize $0$}
			\psfrag{15}[c][c]{\scriptsize $15$}
			\psfrag{30}[c][c]{\scriptsize $30$}
			\psfrag{45}[c][c]{\scriptsize $45$}
			
			\psfrag{-60}[c][c]{\scriptsize -$60$}
			\psfrag{-50}[c][c]{\scriptsize -$50$}
			\psfrag{-40}[c][c]{\scriptsize -$40$}
			\psfrag{-30}[c][c]{\scriptsize -$30$}
			\psfrag{-20}[c][c]{\scriptsize -$20$}
			\psfrag{-10}[c][c]{\scriptsize -$10$}
			\psfrag{0}[c][c]{\scriptsize $0$}
			
			\psfrag{Phase noise level (dBc)}[c][c]{\footnotesize PN level (dBc)}
			\psfrag{PPLR (dB)}[c][c]{\footnotesize $\mathrm{PPLR~(dB)}$}			
			
			\includegraphics[width=3.25cm]{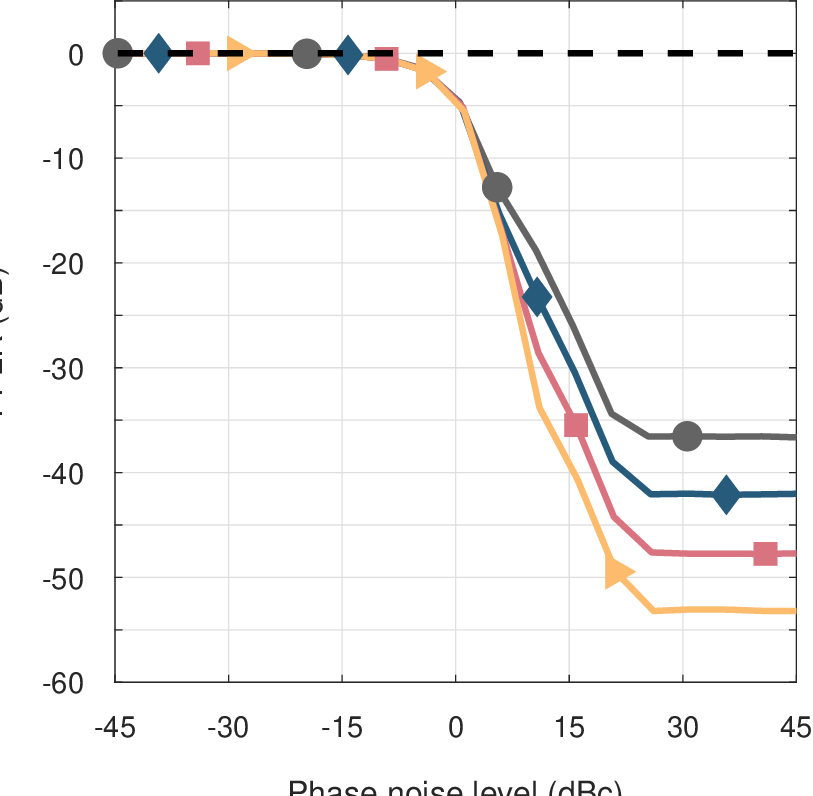}			
		}\hspace{0.05cm}
		\subfloat[ ]{
			
			\psfrag{-45}[c][c]{\scriptsize -$45$}
			\psfrag{-30}[c][c]{\scriptsize -$30$}
			\psfrag{-15}[c][c]{\scriptsize -$15$}
			\psfrag{0}[c][c]{\scriptsize $0$}
			\psfrag{15}[c][c]{\scriptsize $15$}
			\psfrag{30}[c][c]{\scriptsize $30$}
			\psfrag{45}[c][c]{\scriptsize $45$}
			
			\psfrag{-15}[c][c]{\scriptsize -$15$}
			\psfrag{-12}[c][c]{\scriptsize -$12$}
			\psfrag{-9}[c][c]{\scriptsize -$9$}
			\psfrag{-6}[c][c]{\scriptsize -$6$}
			\psfrag{-3}[c][c]{\scriptsize -$3$}
			\psfrag{0}[c][c]{\scriptsize $0$}
			
			\psfrag{Phase noise level (dBc)}[c][c]{\footnotesize PN level (dBc)}
			\psfrag{PSLR (dB)}[c][c]{\footnotesize Range $\mathrm{PSLR~(dB)}$}	
			
			\includegraphics[width=3.25cm]{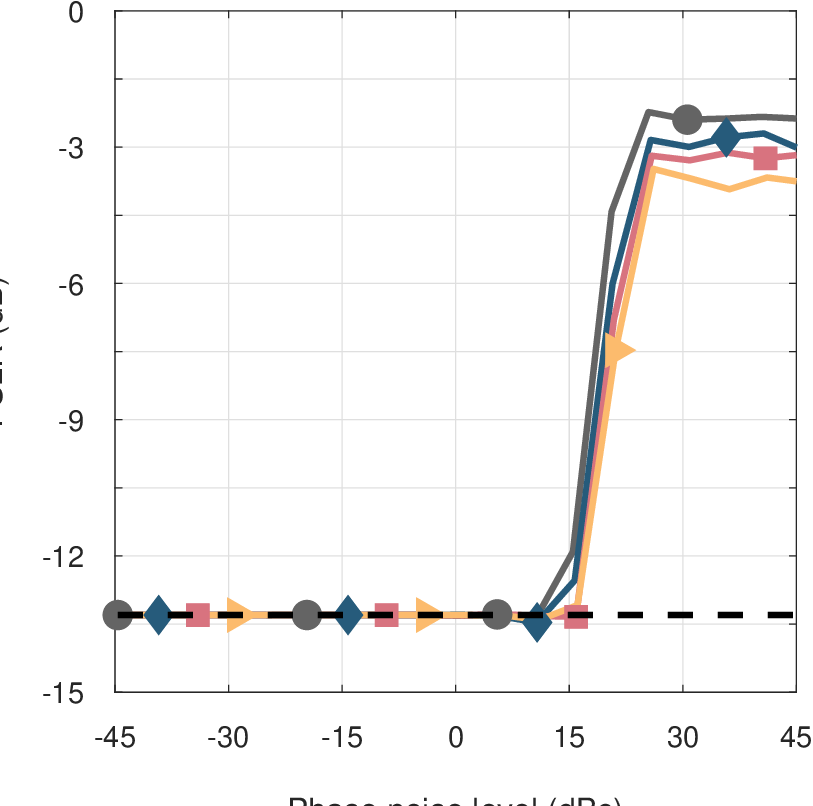}	
			
		}\hspace{0.05cm}
		\subfloat[ ]{
			
			\psfrag{-45}[c][c]{\scriptsize -$45$}
			\psfrag{-30}[c][c]{\scriptsize -$30$}
			\psfrag{-15}[c][c]{\scriptsize -$15$}
			\psfrag{0}[c][c]{\scriptsize $0$}
			\psfrag{15}[c][c]{\scriptsize $15$}
			\psfrag{30}[c][c]{\scriptsize $30$}
			\psfrag{45}[c][c]{\scriptsize $45$}
			
			\psfrag{-12}[c][c]{\scriptsize -$12$}
			\psfrag{-4}[c][c]{\scriptsize -$4$}
			\psfrag{4}[c][c]{\scriptsize $4$}
			\psfrag{12}[c][c]{\scriptsize $12$}
			\psfrag{20}[c][c]{\scriptsize $20$}
			\psfrag{28}[c][c]{\scriptsize $28$}
			\psfrag{36}[c][c]{\scriptsize $36$}
			
			\psfrag{Phase noise level (dBc)}[c][c]{\footnotesize PN level (dBc)}
			\psfrag{ISLR (dB)}[c][c]{\footnotesize Range $\mathrm{ISLR~(dB)}$}	
			
			\includegraphics[width=3.25cm]{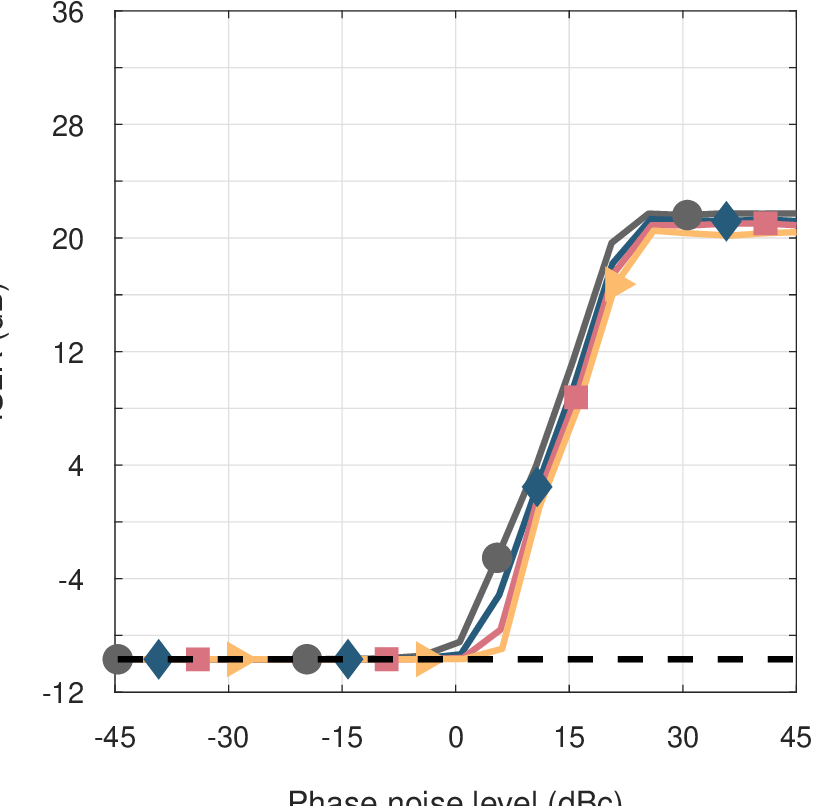}
			
		}\hspace{0.05cm}
		\subfloat[ ]{
			
			\psfrag{-45}[c][c]{\scriptsize -$45$}
			\psfrag{-30}[c][c]{\scriptsize -$30$}
			\psfrag{-15}[c][c]{\scriptsize -$15$}
			\psfrag{0}[c][c]{\scriptsize $0$}
			\psfrag{15}[c][c]{\scriptsize $15$}
			\psfrag{30}[c][c]{\scriptsize $30$}
			\psfrag{45}[c][c]{\scriptsize $45$}
			
			\psfrag{-15}[c][c]{\scriptsize -$15$}
			\psfrag{-12}[c][c]{\scriptsize -$12$}
			\psfrag{-9}[c][c]{\scriptsize -$9$}
			\psfrag{-6}[c][c]{\scriptsize -$6$}
			\psfrag{-3}[c][c]{\scriptsize -$3$}
			\psfrag{0}[c][c]{\scriptsize $0$}
			
			\psfrag{Phase noise level (dBc)}[c][c]{\footnotesize PN level (dBc)}
			\psfrag{PSLR (dB)}[c][c]{\footnotesize Doppler $\mathrm{PSLR~(dB)}$}	
			
			\includegraphics[width=3.25cm]{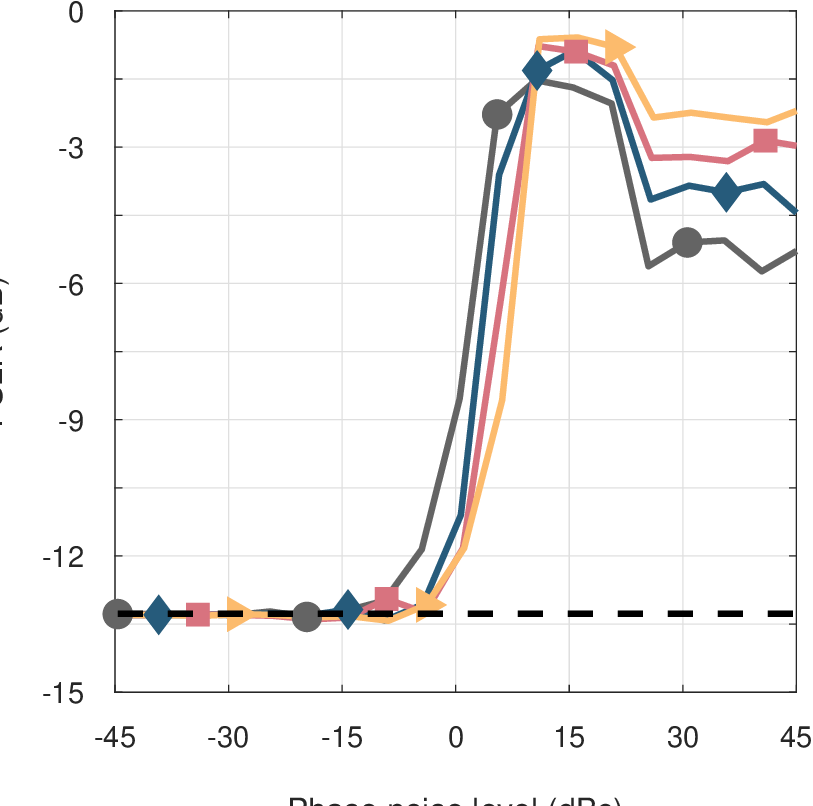}	
			
		}\hspace{0.05cm}
		\subfloat[ ]{
			
			\psfrag{-45}[c][c]{\scriptsize -$45$}
			\psfrag{-30}[c][c]{\scriptsize -$30$}
			\psfrag{-15}[c][c]{\scriptsize -$15$}
			\psfrag{0}[c][c]{\scriptsize $0$}
			\psfrag{15}[c][c]{\scriptsize $15$}
			\psfrag{30}[c][c]{\scriptsize $30$}
			\psfrag{45}[c][c]{\scriptsize $45$}
			
			\psfrag{-12}[c][c]{\scriptsize -$12$}
			\psfrag{-4}[c][c]{\scriptsize -$4$}
			\psfrag{4}[c][c]{\scriptsize $4$}
			\psfrag{12}[c][c]{\scriptsize $12$}
			\psfrag{20}[c][c]{\scriptsize $20$}
			\psfrag{28}[c][c]{\scriptsize $28$}
			\psfrag{36}[c][c]{\scriptsize $36$}
			
			\psfrag{Phase noise level (dBc)}[c][c]{\footnotesize PN level (dBc)}
			\psfrag{ISLR (dB)}[c][c]{\footnotesize Doppler $\mathrm{ISLR~(dB)}$}
			
			\includegraphics[width=3.25cm]{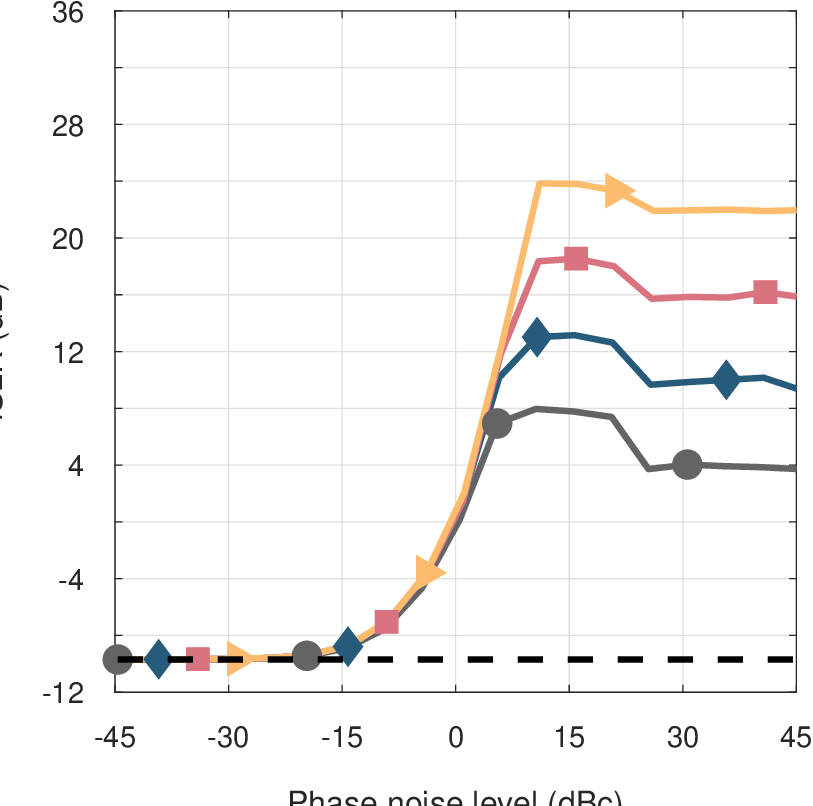}		
			
		}
		
		\captionsetup{justification=raggedright,labelsep=period,singlelinecheck=false}
		\caption{\ PPLR (a), range PSLR (b) and ISLR (c), and Doppler shift PSLR (e) and ISLR (d) as functions of the combined transmit and receive PN level. Simulation results are shown for a static target assuming \mbox{$N=2048$}, QPSK modulation, and $M=32$ \mbox{({\color[rgb]{0.3922,0.3922,0.3922}$\CIRCLE$})}, $M=128$ \mbox{({\color[rgb]{0.1490,0.3569,0.4824}$\blacklozenge$})}, $M=512$ \mbox{({\color[rgb]{0.8471,0.4510,0.4980}$\blacksquare$})}, and $M=2048$ \mbox{({\color[rgb]{0.9882,0.7333,0.4275}$\blacktriangleright$})}. In all subfigures, a CP length of $N_\mathrm{CP}=N$ was considered, and the PPLR, PSLR and ISLR values for simulations without PN are also shown ({\color[rgb]{0,0,0}\textbf{\textendash~\textendash}}).}\label{fig:sidelobe_M}
		
	\end{figure*}	
	\begin{figure}[!t]
		\centering
		
		\psfrag{-45}[c][c]{\scriptsize -$45$}
		\psfrag{-30}[c][c]{\scriptsize -$30$}
		\psfrag{-15}[c][c]{\scriptsize -$15$}
		\psfrag{0}[c][c]{\scriptsize $0$}
		\psfrag{15}[c][c]{\scriptsize $15$}
		\psfrag{30}[c][c]{\scriptsize $30$}
		\psfrag{45}[c][c]{\scriptsize $45$}
		
		\psfrag{-60}[c][c]{\scriptsize -$60$}
		\psfrag{-50}[c][c]{\scriptsize -$50$}
		\psfrag{-40}[c][c]{\scriptsize -$40$}
		\psfrag{-30}[c][c]{\scriptsize -$30$}
		\psfrag{-20}[c][c]{\scriptsize -$20$}
		\psfrag{-10}[c][c]{\scriptsize -$10$}
		\psfrag{0}[c][c]{\scriptsize $0$}
		
		\psfrag{Phase noise level (dBc)}[c][c]{\footnotesize PN level (dBc)}
		\psfrag{PPLR (dB)}[c][c]{\footnotesize $\mathrm{PPLR~(dB)}$}			
		
		\includegraphics[width=3.75cm]{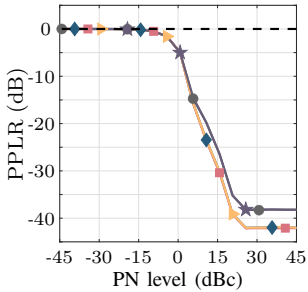}
		
		\captionsetup{justification=raggedright,labelsep=period,singlelinecheck=false}
		\caption{\ PPLR as a function of the combined transmit and receive PN level. Simulation results are shown for a moving target assuming \mbox{$N=2048$}, QPSK modulation, and \mbox{$f_\mathrm{D}=-0.5\Delta f$} \mbox{({\color[rgb]{0.3922,0.3922,0.3922}$\CIRCLE$})}, \mbox{$f_\mathrm{D}=-0.1\Delta f$} \mbox{({\color[rgb]{0.1490,0.3569,0.4824}$\blacklozenge$})}, \mbox{$f_\mathrm{D}=0$} \mbox{({\color[rgb]{0.8471,0.4510,0.4980}$\blacksquare$})}, \mbox{$f_\mathrm{D}=0.1\Delta f$} \mbox{({\color[rgb]{0.9882,0.7333,0.4275}$\blacktriangleright$})}, and \mbox{$f_\mathrm{D}=0.5\Delta f$} \mbox{({\color[rgb]{0.4078,0.3647,0.4745}$\bigstar$})}. A CP length of $N_\mathrm{CP}=N$ and $M=128$ OFDM symbols were considered. For comparison, the PPLR values for simulations without PN or Doppler shift are also shown ({\color[rgb]{0,0,0}\textbf{\textendash~\textendash}}).}\label{fig:sidelobe_Doppler}
		
	\end{figure}
	 
	\begin{figure*}[!t]
		\centering
		
		\psfrag{55}{(a)}
		\psfrag{22}{(b)}
		\psfrag{33}{(c)}
		
		\psfrag{7.5}[c][c]{\scriptsize $7.5$}
		\psfrag{10}[c][c]{\scriptsize $10$}
		\psfrag{12.5}[c][c]{\scriptsize $12.5$}
		\psfrag{15}[c][c]{\scriptsize $15$}	
		\psfrag{17.5}[c][c]{\scriptsize $17.5$}
		
		\psfrag{-0.1}[c][c]{\scriptsize -$0.1$}
		\psfrag{0}[c][c]{\scriptsize $0$}
		\psfrag{0.1}[c][c]{\scriptsize $0.1$}
		\psfrag{0.2}[c][c]{\scriptsize $0.2$}
		
		\psfrag{0}[c][c]{\scriptsize $0$}
		\psfrag{-15}[c][c]{\scriptsize -$15$}
		\psfrag{-30}[c][c]{\scriptsize -$30$}
		\psfrag{-45}[c][c]{\scriptsize -$45$}
		\psfrag{-60}[c][c]{\scriptsize -$60$}
		
		\psfrag{fD Df}[c][c]{\footnotesize $f_\mathrm{D}/\Delta f$}
		\psfrag{Bistatic range (m)}[c][c]{\footnotesize Bistatic range (m)}
		\psfrag{Norm. mag. (dB)}[c][c]{\footnotesize Norm. mag. (dB)}

		\includegraphics[width=12cm]{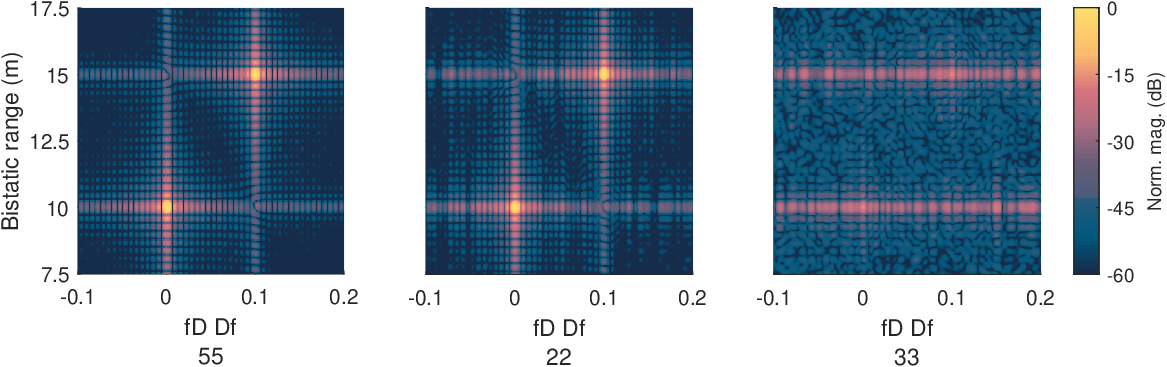}
		\captionsetup{justification=raggedright,labelsep=period,singlelinecheck=false}
		\caption{\ Bistatic range-Doppler radar images obtained with $N=2048$, $N_\mathrm{CP}=512$, $M=128$, QPSK modulation, and rectangular windowing in both range and Doppler shift directions. Two targets were simulated, the first at $\SI{10}{\meter}$ and with a Doppler shift of $\SI{0}{\kilo\hertz}$ and the second at $\SI{15}{\meter}$ and with a Doppler shift of $0.1\Delta f$ (which results in negligible \ac{ICI} \cite{giroto2021_tmtt}). The considered combined transmit and receive PN levels in the simulations were (a) $\SI{-44.90}{dBc}$ \mbox{($\gamma=\SI{0}{dB}$)}, (b) $\SI{-14.90}{dBc}$ \mbox{($\gamma=\SI{30}{dB}$)}, and (c) $\SI{4.90}{dBc}$ \mbox{($\gamma=\SI{50}{dB}$)}.}\label{fig:I_rD}
		
	\end{figure*}
	\begin{figure*}[!t]
		\centering
		
		\psfrag{55}{(a)}
		\psfrag{22}{(b)}
		\psfrag{33}{(c)}
		
		\psfrag{7.5}[c][c]{\scriptsize $7.5$}
		\psfrag{10}[c][c]{\scriptsize $10$}
		\psfrag{12.5}[c][c]{\scriptsize $12.5$}
		\psfrag{15}[c][c]{\scriptsize $15$}	
		\psfrag{17.5}[c][c]{\scriptsize $17.5$}
		
		\psfrag{-0.1}[c][c]{\scriptsize -$0.1$}
		\psfrag{0}[c][c]{\scriptsize $0$}
		\psfrag{0.1}[c][c]{\scriptsize $0.1$}
		\psfrag{0.2}[c][c]{\scriptsize $0.2$}
		
		\psfrag{0}[c][c]{\scriptsize $0$}
		\psfrag{-15}[c][c]{\scriptsize -$15$}
		\psfrag{-30}[c][c]{\scriptsize -$30$}
		\psfrag{-45}[c][c]{\scriptsize -$45$}
		\psfrag{-60}[c][c]{\scriptsize -$60$}
		
		\psfrag{fD Df}[c][c]{\footnotesize $f_\mathrm{D}/\Delta f$}
		\psfrag{Bistatic range (m)}[c][c]{\footnotesize Bistatic range (m)}
		\psfrag{Norm. mag. (dB)}[c][c]{\footnotesize Norm. mag. (dB)}

		\includegraphics[width=12cm]{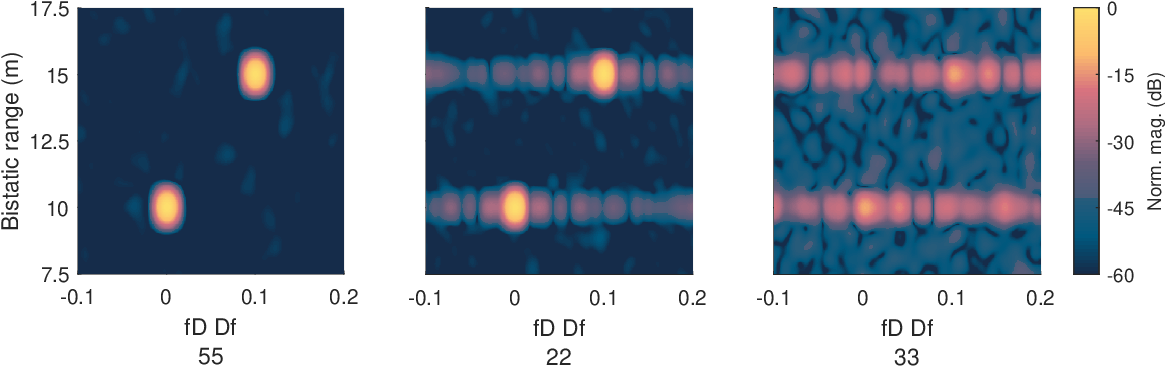}
		\captionsetup{justification=raggedright,labelsep=period,singlelinecheck=false}
		\caption{\ Bistatic range-Doppler radar images obtained with $N=2048$, $N_\mathrm{CP}=512$, $M=128$, QPSK modulation, and Chebyshev windowing with $\SI{100}{dB}$ sidelobe supression in both range and Doppler shift directions. Two targets were simulated, the first at $\SI{10}{\meter}$ and with a Doppler shift of $\SI{0}{\kilo\hertz}$ and the second at $\SI{15}{\meter}$ and with a Doppler shift of $0.1\Delta f$ (which results in negligible \ac{ICI} \cite{giroto2021_tmtt}). The considered combined transmit and receive PN levels in the simulations were (a) $\SI{-44.90}{dBc}$ \mbox{($\gamma=\SI{0}{dB}$)}, (b) $\SI{-14.90}{dBc}$ \mbox{($\gamma=\SI{30}{dB}$)}, and (c) $\SI{4.90}{dBc}$ \mbox{($\gamma=\SI{50}{dB}$)}.}\label{fig:I_rD_chebwin}
		
	\end{figure*}
	
	\subsubsection{Main and sidelobe distortion for different numbers of OFDM symbols}\label{subsubsec:sensM}
	
	Setting \mbox{$N=N_\mathrm{CP}=2048$} and focusing only on \ac{QPSK} modulation, Fig.~\ref{fig:sidelobe_M} shows results on the radar sensing performance degradation due to \ac{PN} for \mbox{$M\in\{32,128,512,2048\}$} \ac{OFDM} symbols. The obtained results show that increasing $M$ results in a worse overall performance. A slight improvement in range \ac{PSLR} and \ac{ISLR} can indeed be seen in Figs.~\ref{fig:sidelobe_M}b and \ref{fig:sidelobe_M}c, which is due to a higher radar processing gain against \ac{PN}-induced \ac{ICI}, which is the same for every \ac{OFDM} symbol. However, with increasing $M$ a higher number of \ac{OFDM} symbols with \ac{PN}-induced \ac{CPE} is processed, which leads to a degradation of the \ac{PPLR} proportional to $M$ as observed in Figs.~\ref{fig:sidelobe_M}a. After a linear degradation zone from around $\SI{-19.87}{dBc}$, the \ac{PPLR} reaches a constant level at $\SI{26.14}{dBc}$. This level is equal to $\SI{-36.58}{dB}$, $\SI{-42.08}{dB}$, $\SI{-47.76}{dB}$, and $\SI{-53.21}{dB}$ for $M=32$, $M=128$, $M=512$, and $M=2048$, respectively. This corresponds to an average \ac{PPLR} degradation of $\SI{5.58}{dB}$ (factor of $3.61$) from an $M$ value to the next higher one, and it is close to the ratio of $4$ between them. In addition, the Doppler shift \ac{PSLR} and \ac{ISLR} also degrade with increasing $M$ due to increased \ac{CPE} effect as seen in Figs.~\ref{fig:sidelobe_M}d and \ref{fig:sidelobe_M}e. It can be seen that both curves present a similar behavior as their counterparts in Fig.~\ref{fig:sidelobe_PSD}. In terms of \ac{PSLR}, constant levels of $\SI{-5.10}{dB}$, $\SI{-3.85}{dB}$, $\SI{-3.3}{dB}$, and $\SI{-2.24}{dB}$ is observed for the considered $M$ values in crescent order. The \ac{PSLR} difference between consecutive values is, in all cases, smaller than $\SI{2}{dB}$, and  the \ac{PSLR} values do not increase linearly with the ratio in which $M$ is increased. Regarding Doppler shift \ac{ISLR}, however, an average increasing step of $\SI{6.24}{dB}$ (factor $4.21$) is observed for increasing $M$, which is close to the ratio of $4$ between the considered consecutive $M$ values. The observed \ac{ISLR} values at the high \ac{PN} level regime for $M=32$, $M=128$, $M=512$, and $M=2048$ are $\SI{3.85}{dB}$, $\SI{10.01}{dB}$, $\SI{15.80}{dB}$, and $\SI{21.89}{dB}$, respectively.
	
	\subsubsection{Main and sidelobe distortion for moving targets}\label{subsubsec:sensDoppler}
	
	In Fig.~\ref{fig:sidelobe_Doppler}, radar sensing performance results are shown for \mbox{$N=N_\mathrm{CP}=2048$}, \ac{QPSK} modulation, and Doppler shifts of \mbox{$f_\mathrm{D}\in\{\SI{0}{\hertz},\pm0.1\Delta f,\pm0.5\Delta f\}$} to also cover scenarios with moving targets. Besides the static target case for \mbox{$f_\mathrm{D}=\SI{0}{\hertz}$}, which was already investigated in Fig.~\ref{fig:sidelobe_PSD} and is only shown here as a reference, Doppler shifts of \mbox{$f_\mathrm{D}=\pm0.1\Delta f$} are considered as they are assumed to result in tolerable Doppler-shift-induced \ac{ICI} \cite{giroto2021_tmtt}. The Doppler shifts of \mbox{$f_\mathrm{D}=\pm0.5\Delta f$}, in turn, correspond to the maximum achievable unambiguous Doppler shift interval, which is obtained in the case where \mbox{$N_\mathrm{CP}=0$} \cite{giroto2021_tmtt,giroto2023_EuMW}. Since the difference w.r.t. the range and Doppler shift \ac{PSLR} and \ac{ISLR} results from Fig.~\ref{fig:sidelobe_PSD} was negligible, only \ac{PPLR} is shown in Fig.~\ref{fig:sidelobe_Doppler}. The achieved results shown that the \ac{PN}-induced impairments combined have significantly more influence than the Doppler-shift-induced \ac{ICI} on the radar sensing performance of the \ac{OFDM}-based \ac{ISAC} system. Only a slightly better \ac{PPLR} is shown for \mbox{$f_\mathrm{D}=\pm0.5\Delta f$}, which is due to the fact that the Doppler-shift induced \ac{ICI} may not always add constructively to its \ac{PN}-induced counterpart. This difference, however, only becomes visible for relatively high combined \ac{PN} levels, i.e., above $\SI{5.70}{dBc}$, and it is of around $\SI{4}{dB}$ when a constant \ac{PPLR} level is achieved for a \ac{PN} level of $\SI{25.74}{dBc}$ or higher.
	
	\subsubsection{Main and sidelobe distortion for multiple moving targets}\label{subsubsec:sensMult}
	
	The previous subsections show how \ac{PN} degrades the \ac{PPLR}, as well as range and Doppler shift \ac{PSLR} and \ac{ISLR}, of a single target under different settings. These parameters allow quantifying both the power loss at the main lobe and the sidelobe level increase in range and Doppler shift directions. To illustrate how this impacts the radar sensing performance of an \ac{OFDM}-based \ac{ISAC} system in multi-target scenarios, Fig.~\ref{fig:I_rD} shows range-Doppler shift radar images for one static and one moving target. In addition $N=2048$, $N_\mathrm{CP}=512$, $M=128$, and \ac{QPSK} modulation were assumed. No \ac{AWGN} was considered, and rectangular windowing was used in both range and Doppler shift processing. It can be seen that the sinc-shaped sidelobe pattern expected even without \ac{PN} influence is clearly observable for a combined transmit and receive \ac{PN} level of $\SI{-44.90}{dBc}$ \mbox{($\gamma=\SI{0}{dB}$)}. For the considered higher combined \ac{PN} levels of $\SI{-14.90}{dBc}$ \mbox{($\gamma=\SI{30}{dB}$)} and $\SI{4.90}{dBc}$ \mbox{($\gamma=\SI{50}{dB}$)}, the main lobe of both targets suffer from power loss as expected from the \ac{PPLR}, and the Doppler shift sidelobes become much more relevant than their counterparts in the range direction, as expected from the analyses in the previous subsections.
	
	Next, Fig.~\ref{fig:I_rD_chebwin} shows range-Doppler shift radar images obtained with the same parameters adopted for Fig.~\ref{fig:I_rD}, but applying Chebyshev windows with $\SI{100}{dB}$ sidelobe supression in both range and Doppler shift directions. It can be seen in Fig.~\ref{fig:I_rD_chebwin} that the windowing is only effective in range direction, as the \ac{PN}-induced effect that plays a major role in range processing is \ac{ICI}. As discussed in the previous subsections, the \ac{ICI} produces a Gaussian noise-like effect that can be suppressed by the radar processing gain to some extent. Since this effect is additive, it does not reduce the effectiveness of windowing during range processing. As for windowing in the Doppler shift direction, it becomes little effective with increasing combined transmit and receive \ac{PN} level as shown for $\SI{-14.90}{dBc}$ \mbox{($\gamma=\SI{30}{dB}$)} and $\SI{4.90}{dBc}$ \mbox{($\gamma=\SI{50}{dB}$)}. The reason for this is the severe \ac{CPE}, which prevents the \ac{DFT}-based estimation to generate a sinc-shaped Doppler shift profile even with rectangular windowing, e.g., as previously shown in Fig.~\ref{fig:dopplerCuts_zoom}. Consequently, the distorted, non-decreasing Doppler shift sidelobe pattern cannot be effectively suppressed by Chebyshev windowing.
	
	\begin{figure}[!t]
		\centering
		
		\subfloat[ ]{
			
			\psfrag{-45}[c][c]{\scriptsize -$45$}
			\psfrag{-30}[c][c]{\scriptsize -$30$}
			\psfrag{-15}[c][c]{\scriptsize -$15$}
			\psfrag{0}[c][c]{\scriptsize $0$}
			\psfrag{15}[c][c]{\scriptsize $15$}
			\psfrag{30}[c][c]{\scriptsize $30$}
			\psfrag{45}[c][c]{\scriptsize $45$}
			
			\psfrag{-30}[c][c]{\scriptsize -$30$}
			\psfrag{30}[c][c]{\scriptsize $30$}
			\psfrag{60}[c][c]{\scriptsize $60$}
			\psfrag{90}[c][c]{\scriptsize $90$}
			\psfrag{120}[c][c]{\scriptsize $120$}
			
			\psfrag{Phase noise level (dBc)}[c][c]{\footnotesize PN level (dBc)}
			\psfrag{AAAA Image SIR (dB)}[c][c]{\footnotesize Mean image SIR (dB)}	
			
			\includegraphics[width=3.75cm]{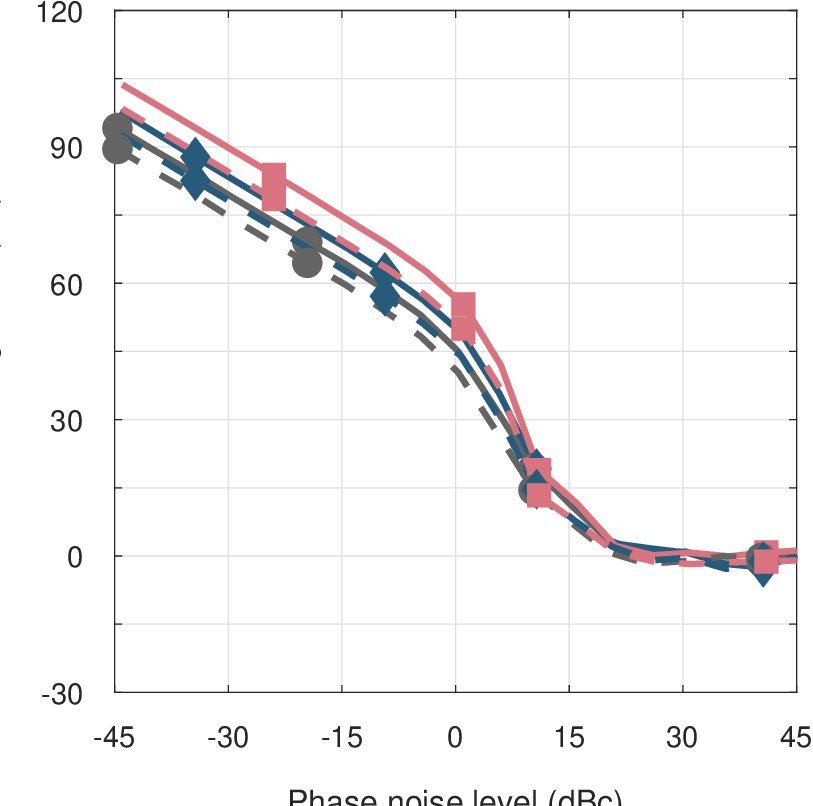}\label{fig:meanImageSIR_chebWin}
			
		}\hspace{0.1cm}
		\subfloat[ ]{
			
			\psfrag{-45}[c][c]{\scriptsize -$45$}
			\psfrag{-30}[c][c]{\scriptsize -$30$}
			\psfrag{-15}[c][c]{\scriptsize -$15$}
			\psfrag{0}[c][c]{\scriptsize $0$}
			\psfrag{15}[c][c]{\scriptsize $15$}
			\psfrag{30}[c][c]{\scriptsize $30$}
			\psfrag{45}[c][c]{\scriptsize $45$}
			
			\psfrag{-30}[c][c]{\scriptsize -$30$}
			\psfrag{30}[c][c]{\scriptsize $30$}
			\psfrag{60}[c][c]{\scriptsize $60$}
			\psfrag{90}[c][c]{\scriptsize $90$}
			\psfrag{120}[c][c]{\scriptsize $120$}
			
			\psfrag{Phase noise level (dBc)}[c][c]{\footnotesize PN level (dBc)}
			\psfrag{AAAA Image SIR (dB)}[c][c]{\footnotesize Min. image SIR (dB)}	
			
			\includegraphics[width=3.75cm]{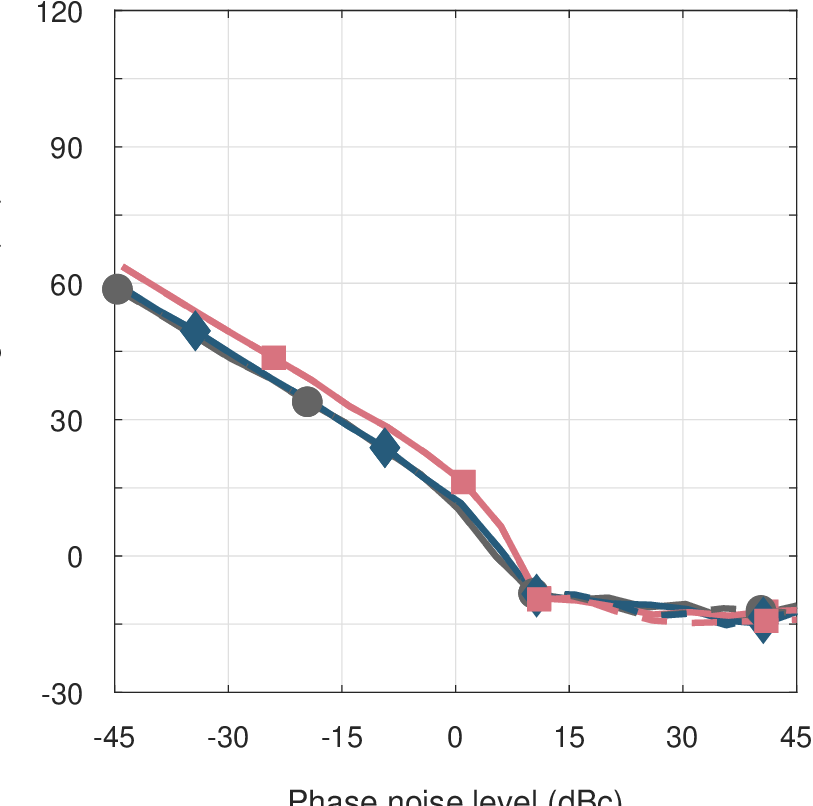}\label{fig:minImageSIR_chebWin}	
			
		}		
		
		\captionsetup{justification=raggedright,labelsep=period,singlelinecheck=false}
		\caption{\ Image SIR: (a) mean and (b) minimum values. In the performed simulations, a single static target was considered and \mbox{$N=256$} and QPSK \mbox{({\color[rgb]{0.3922,0.3922,0.392}\textbf{\textemdash}} and {\color[rgb]{0.3922,0.3922,0.3922}$\CIRCLE$})}, \mbox{$N=2048$} and QPSK \mbox{({\color[rgb]{0.1490,0.3569,0.4824}\textbf{\textemdash}} and {\color[rgb]{0.1490,0.3569,0.4824}$\blacklozenge$})}, and \mbox{$N=16384$} and QPSK \mbox{({\color[rgb]{0.8471,0.4510,0.4980}\textbf{\textemdash}} and {\color[rgb]{0.8471,0.4510,0.4980}$\blacksquare$})} were assumed. Results were also obtained for \mbox{256-QAM} with \mbox{$N=256$} \mbox{({\color[rgb]{0.3922,0.3922,0.392}\textbf{\textendash~\textendash}} and {\color[rgb]{0.3922,0.3922,0.3922}$\CIRCLE$})}, \mbox{$N=2048$} \mbox{({\color[rgb]{0.1490,0.3569,0.4824}\textbf{\textendash~\textendash}} and {\color[rgb]{0.1490,0.3569,0.4824}$\blacklozenge$})}, and \mbox{$N=16384$} \mbox{({\color[rgb]{0.8471,0.4510,0.4980}\textbf{\textendash~\textendash}} and {\color[rgb]{0.8471,0.4510,0.4980}$\blacksquare$})}. In all subfigures, a CP length of $N_\mathrm{CP}=N$ and $M=128$ OFDM symbols were considered.}\label{fig:imageSIR_chebWin}
		
	\end{figure}
	
	\subsubsection{Image signal-to-interference ratio}\label{subsubsec:imageSIR}
	
	To quantify the \ac{PN}-induced radar image distortion as observed in Section~\ref{subsubsec:sensMult}, the image \ac{SIR} as a function of the combined transmit and receive \ac{PN} level is analyzed in Fig.~\ref{fig:imageSIR_chebWin}. For simplicity, a single-target scenario is considered, and the same parameters from Section~\ref{subsubsec:sensN} are assumed. For the obtained results, Chebyshev window with $\SI{100}{dB}$ sidelobe supression was used and the \ac{SIR} was calculated as the magnitude ratio between the target reflection peak and the region outside the main lobe. In addition, both mean and minimum \ac{SIR} values were presented. These correspond to the average and minimum \ac{SIR} calculated between the peak and each point outside the mainlobe, respectively.
	
	The obtained mean \ac{SIR} is shown in Fig.~\ref{fig:imageSIR_chebWin}a. In this figure, a similar trend to what is experienced when performing communication over an ideal channel as shown in Fig.~\ref{fig:SIR_PN} is observed. However, the relationship between the \ac{SIR} at the subcarrier level and the image \ac{SIR} is not fully linear, as the \ac{PN}-induced \ac{CPE} and \ac{ICI} are differently supressed by the radar processing gain. Back to Fig.~\ref{fig:imageSIR_chebWin}a, an approximately linear degradation is seen until a combined transmit and receive \ac{PN} level of around $\SI{0}{dBc}$ is observed in Fig.~\ref{fig:imageSIR_chebWin}a. Afterwards, the \ac{SIR} decreases more steeply until a $\SI{0}{dB}$ floor is reached at around $\SI{20}{dBc}$ \ac{PN} level. As expected, the minimum \ac{SIR} shown in Fig.~\ref{fig:imageSIR_chebWin}b tends to be significantly lower than the mean \ac{SIR} in Fig.~\ref{fig:imageSIR_chebWin}a, including the minimum \ac{SIR} floor of around $\SI{-15}{dB}$ at high \ac{PN} levels. This happens since the minimum \ac{SIR} is dominated by the Doppler shift sidelobes that are more severely degraded by \ac{CPE} than the range sidelobes are by \ac{ICI}, e.g., as discussed in Section~\ref{subsubsec:sensMult}. Compared to the results on mainlobe and sidelobe distortion in the previous subsections, it is seen that the image \ac{SIR} is reduced even for rather low \ac{PN} levels, while \ac{PPLR} and range and Doppler \ac{PSLR} and \ac{ISLR} do not degrade before $\SI{-15}{dBc}$. At this \ac{PN} level, however, mean and minimum image \ac{SIR} values of approximately $\SI{60}{dB}$ and $\SI{30}{dB}$ or higher are respectively observed, which already represents non-negligible sensing performance degradation.
	
	\subsubsection{Sensing performance under 3GPP specifications}\label{subsubsec:3gpp}

\begin{figure}[!t]
		\centering
		
		\subfloat[ ]{
			
			\psfrag{22}[c][c]{\scriptsize $10^2$}
			\psfrag{44}[c][c]{\scriptsize $10^4$}
			\psfrag{66}[c][c]{\scriptsize $10^6$}
			\psfrag{88}[c][c]{\scriptsize $10^8$}
			
			\psfrag{-140}[c][c]{\scriptsize -$140$}
			\psfrag{-120}[c][c]{\scriptsize -$120$}
			\psfrag{-100}[c][c]{\scriptsize -$100$}
			\psfrag{-80}[c][c]{\scriptsize -$80$}
			\psfrag{-60}[c][c]{\scriptsize -$60$}
			\psfrag{-40}[c][c]{\scriptsize -$40$}
			
			\psfrag{Frequency offset (Hz)}[c][c]{\footnotesize Frequency offset (Hz)}
			\psfrag{Phase noise PSD (dBc/Hz)}[c][c]{\vspace{-0.1cm}\footnotesize PN PSD (dBc/Hz)}
			
			\includegraphics[width=3.75cm]{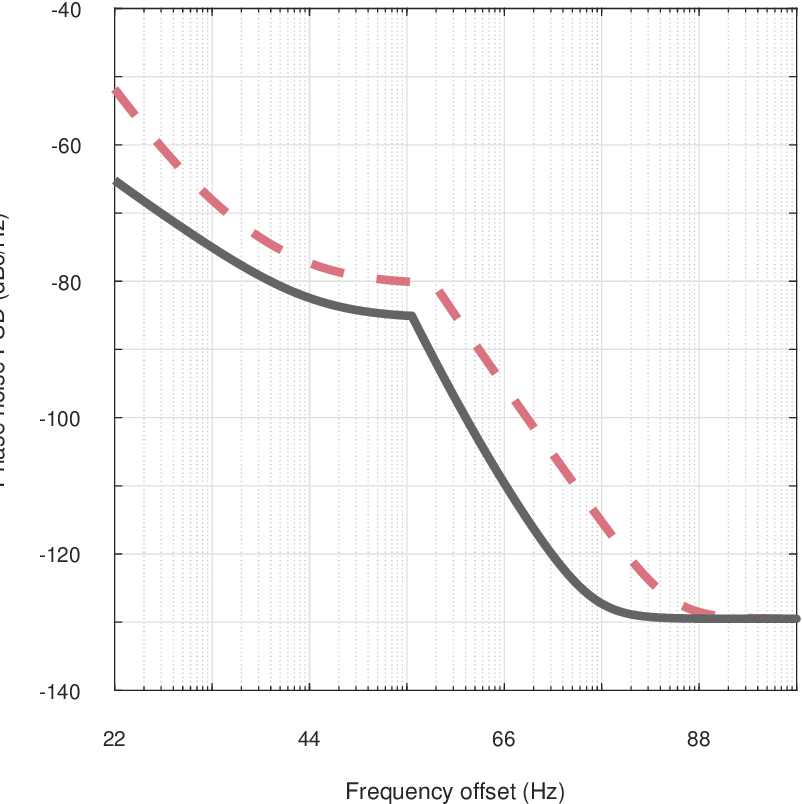}\label{fig:PNpsd_UE_BS}
			
		}\hspace{0.1cm}
		\subfloat[ ]{
			
			\psfrag{44}[c][c]{\scriptsize $10^4$}
			\psfrag{55}[c][c]{\scriptsize $10^5$}
			\psfrag{66}[c][c]{\scriptsize $10^6$}
			\psfrag{77}[c][c]{\scriptsize $10^7$}
			\psfrag{88}[c][c]{\scriptsize $10^8$}
			\psfrag{99}[c][c]{\scriptsize $10^9$}
			
			\psfrag{-40}[c][c]{\scriptsize -$40$}
			\psfrag{-35}[c][c]{\scriptsize -$35$}
			\psfrag{-30}[c][c]{\scriptsize -$30$}
			\psfrag{-25}[c][c]{\scriptsize -$25$}
			\psfrag{-20}[c][c]{\scriptsize -$20$}
			
			\psfrag{Frequency offset (Hz)}[c][c]{\footnotesize Frequency offset (Hz)}
			\psfrag{Int. phase noise level (dBc)}[c][c]{\footnotesize Int. PN level (dBc)}
			
			\includegraphics[width=3.75cm]{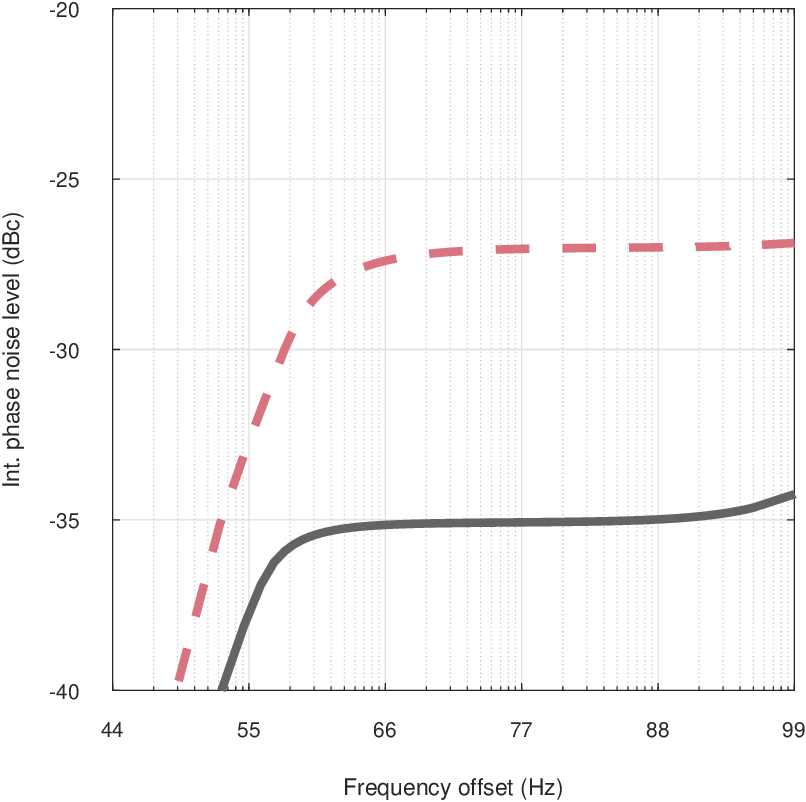}\label{fig:intPN_UE_BS}	
			
		}
		
		\captionsetup{justification=raggedright,labelsep=period,singlelinecheck=false}
		\caption{\ 3GPP TR 38.803 PN model for a gNB ({\color[rgb]{0.3922,0.3922,0.3922}\textbf{\textemdash}}) and a UE ({\color[rgb]{0.8471,0.4510,0.4980}\textbf{\textendash~\textendash}}) at $\SI{30}{\giga\hertz}$: (a) double-sided PN PSD and (b) integrated PN level as a function of frequency offset (also double-sided) for the frequency range of interest.}\label{fig:PNpsd_intPN_UE_BS}
		
	\end{figure}

	In the \ac{3GPP} technical report TR 38.803 \cite{3GPPTR38803}, \ac{PN} models for \acp{gNB} and \acp{UE} at $\SI{30}{\giga\hertz}$ were proposed. Based on these models, the double-sided \ac{PN} \acp{PSD} shown in Fig.~\ref{fig:PNpsd_UE_BS} and the integrated \ac{PN} level were obtained. To analyze the sensing performance degradation imposed by these \ac{PN} \ac{PSD} profiles, four scenarios are considered. These are namely monostatic sensing at a single \ac{gNB}, bistatic sensing between two \acp{gNB}, bistatic sensing with a \ac{gNB} acting as a transmitter and a \ac{UE} as a receiver, and bistatic sensing between two \acp{UE}, which are henceforth named Scenarios \#1 to \#4, respectively. For all scenarios, a single static target was considered, and the adopted \ac{OFDM} signal parameters are listed in Table~\ref{tab:3GPPParameters}. The parameters for Scenario \#1 were suggested in \cite{mandelli2023survey}, while those for Scenarios \#2 to \#4 were used for measurement-based \ac{ISAC} demonstrations in \cite{henninger2023,henninger2024,tosi2024}. Moreover, all of the adopted parameter sets comply with \ac{3GPP} specifications. Since there is only a slight difference between the adopted carrier frequencies and $\SI{30}{\giga\hertz}$, the \ac{PN} models from Fig.~\ref{fig:PNpsd_UE_BS} were still used for Scenarios \#1 to \#4.
	
	\begin{table}[!t]
		\renewcommand{\arraystretch}{1.5}
		\arrayrulecolor[HTML]{708090}
		\setlength{\arrayrulewidth}{.1mm}
		\setlength{\tabcolsep}{4pt}
		
		\centering
		\captionsetup{width=43pc,justification=centering,labelsep=newline}
		\caption{\textsc{Adopted OFDM Signal Parameters for the Scenarios of\\ Monostatic Sensing at a Single gNB (\#1), Bistatic Sensing\\ Between Two gNBs (\#2), Bistatic Sensing With a\\ gNB as a Transmitter and a UE as a Receiver (\#3),\\ and Bistatic Sensing Between Two UEs (\#4)}}
		\label{tab:3GPPParameters}
		\resizebox{\columnwidth}{!}{
			\begin{tabular}{|c|c|c|c|c|}
				\hhline{|=====|}
				\textbf{}                                   & \textbf{Scenario \#1}   & \textbf{Scenario \#2}    & \textbf{Scenario \#3} & \textbf{Scenario \#4}    \\ \hhline{|=====|}
				\textbf{Carrier frequency ($f_\mathrm{c}$)} & $\SI{28}{\giga\hertz}$  & $\SI{27.4}{\giga\hertz}$ & $\SI{27.4}{\giga\hertz}$ & $\SI{27.4}{\giga\hertz}$ \\ \hline
				\textbf{Frequency bandwidth ($B$)}          & $\SI{1.6}{\giga\hertz}$ & $\SI{190}{\mega\hertz}$  & $\SI{190}{\mega\hertz}$ & $\SI{190}{\mega\hertz}$  \\ \hline
				\textbf{No. of subcarriers ($N$)}           & $12672$                 & $1584$                   & $1584$ & $1584$                  \\ \hline
				\textbf{CP length ($N_\mathrm{CP}$)}        & $1600$                  & $112$                    & $112$  & $112$                   \\ \hline
				\textbf{No. of OFDM symbols ($M$)}          & $384$                   & $1120$                   & $1120$ & $1120$                   \\ \hline
				\textbf{Digital modulation}                 & QPSK                    & QPSK                     & QPSK & QPSK \\ \hhline{|=====|}
			\end{tabular}
		}
	\end{table}	
	
	The resulting transmit, receive, and combined transmit and receive \ac{PN} levels for each scenario, as well as the sensing performance degradation measured by \ac{PPLR}, range \ac{PSLR} and \ac{ISLR}, Doppler shift \ac{PSLR} and \ac{ISLR}, as well as mean an minimum image \ac{SIR} obtained for all four scenarios are listed in Table~\ref{tab:3GPPresults}. The achieved results show that the lowest combined transmit and receive \ac{PN} level was $\SI{-308.50}{dBc}$ in Scenario \#1, which is due to the fact that monostatic sensing is performed and the transmit and receive \ac{PN} correlation reduces the effectively experienced \ac{PN} level. The achieved \ac{PN} level was $\SI{-26.99}{dBc}$ in Scenario \#4, where bistatic sensing between two \acp{UE} is performed.
	
	Overall, all considered scenarios yielded combined transmit and receive \ac{PN} levels are all below $\SI{-15}{dBc}$, below which only negligible main and sidelobe level degradation is expected according to the presented results in Sections~\ref{subsubsec:sensN} to \ref{subsubsec:sensMult}. This is confirmed by the remaining results in Table~\ref{tab:3GPPresults}, which show that negligible \ac{PPLR}, range \ac{PSLR}, and \ac{ISLR} degradation in Scenarios \#2 to \#4 w.r.t. to the virtually \ac{PN}-free case in Scenario \#1 is observed. Regarding Doppler shift sidelobes, only a small variation in terms of \ac{PSLR} is observed, besides a slight increase in \ac{ISLR} from Scenarios \#1 to \#4 due to the increasing \ac{PN} level. In terms of mean image \ac{SIR}, high values are observed in all scenarios, which means that only negligible noise floor increase in radar images is expected. Regarding minimum \ac{SIR}, only Scenario \#1 presents no relevant \ac{SIR} degradation due to non-suppressed Doppler shift sidelobes, which is simply explained by the rather low \ac{PN} level in this case. As for Scenarios \#2 to \#4, the obtained minimum \ac{SIR} values of $\SI{56.92}{dB}$, $\SI{53.72}{dB}$, and $\SI{51.51}{dB}$ show that non-negligible degradation in the Doppler shift direction is expected, although not severe.
	
	Since only tolerable sensing performance degradation is observed in all four considered scenarios, it can be concluded that \ac{OFDM}-based \ac{ISAC} systems in the lower \ac{FR2} bands with \ac{3GPP}-compliant \ac{OFDM} signal parameterization and \ac{PN} levels equal to or lower than the models from TR 38.803 \cite{3GPPTR38803} present \ac{PN}-robust sensing performance in both monostatic and bistatic architectures.
	
	\begin{table}[!t]
		\renewcommand{\arraystretch}{1.5}
		\arrayrulecolor[HTML]{708090}
		\setlength{\arrayrulewidth}{.1mm}
		\setlength{\tabcolsep}{4pt}
		
		\centering
		\captionsetup{width=43pc,justification=centering,labelsep=newline}
		\caption{\textsc{Resulting PN-induced Sensing Performance\\ Degradation for Scenarios \#1,  \#2, \#3, and \#4}}
		\label{tab:3GPPresults}
		\resizebox{\columnwidth}{!}{
			\begin{tabular}{|c|c|c|c|c|}
				\hhline{|=====|}
				\textbf{}                                   & \textbf{Scenario \#1}   & \textbf{Scenario \#2}    & \textbf{Scenario \#3} & \textbf{Scenario \#4}    \\ \hhline{|=====|}
				\textbf{Tx PN level} & $\SI{-34.45}{dBc}$  & $\SI{-35.07}{dBc}$ & $\SI{-35.07}{dBc}$ & $\SI{-27.01}{dBc}$ \\ \hline
				\textbf{Rx PN level} & $\SI{-34.45}{dBc}$  & $\SI{-35.07}{dBc}$ & $\SI{-27.01}{dBc}$ & $\SI{-27.01}{dBc}$ \\ \hline
				\textbf{Comb. Tx and Rx PN level} & $\SI{-308.60}{dBc}$  & $\SI{-32.09}{dBc}$ & $\SI{-26.43}{dBc}$ & $\SI{-23.98}{dBc}$ \\ \hline
				\textbf{PPLR} & $\SI{0}{dB}$  & $\SI{0}{dB}$ & $\SI{-0.01}{dB}$ & $\SI{-0.02}{dB}$ \\ \hline
				\textbf{Range PSLR} & $\SI{-13.30}{dB}$  & $\SI{-13.30}{dB}$ & $\SI{-13.30}{dB}$ & $\SI{-13.30}{dB}$ \\ \hline
				\textbf{Range ISLR} & $\SI{-9.68}{dB}$  & $\SI{-9.68}{dB}$ & $\SI{-9.68}{dB}$ & $\SI{-9.68}{dB}$ \\ \hline
				\textbf{Doppler PSLR} & $\SI{-13.30}{dB}$  & $\SI{-13.27}{dB}$ & $\SI{-13.27}{dB}$ & $\SI{-13.26}{dB}$ \\ \hline
				\textbf{Doppler ISLR} & $\SI{-9.68}{dB}$  & $\SI{-9.67}{dB}$ & $\SI{-9.65}{dB}$ & $\SI{-9.62}{dB}$ \\ \hline
				\textbf{Mean image SIR} & $\SI{148.28}{dB}$  & $\SI{92.23}{dB}$ & $\SI{86.15}{dB}$ & $\SI{83.53}{dB}$ \\ \hline
				\textbf{Min. image SIR} & $\SI{100}{dB}$  & $\SI{56.92}{dB}$ & $\SI{53.72}{dB}$ & $\SI{51.51}{dB}$ \\ \hhline{|=====|}
			\end{tabular}
		}
	\end{table}
	
	\begin{figure*}[!t]
		\centering
		\subfloat[ ]{
			
			\psfrag{-45}[c][c]{\scriptsize -$45$}
			\psfrag{-30}[c][c]{\scriptsize -$30$}
			\psfrag{-15}[c][c]{\scriptsize -$15$}
			\psfrag{0}[c][c]{\scriptsize $0$}
			\psfrag{15}[c][c]{\scriptsize $15$}
			\psfrag{30}[c][c]{\scriptsize $30$}
			\psfrag{45}[c][c]{\scriptsize $45$}
			
			\psfrag{-60}[c][c]{\scriptsize -$60$}
			\psfrag{-50}[c][c]{\scriptsize -$50$}
			\psfrag{-40}[c][c]{\scriptsize -$40$}
			\psfrag{-30}[c][c]{\scriptsize -$30$}
			\psfrag{-20}[c][c]{\scriptsize -$20$}
			\psfrag{-10}[c][c]{\scriptsize -$10$}
			\psfrag{0}[c][c]{\scriptsize $0$}
			
			\psfrag{Phase noise level (dBc)}[c][c]{\footnotesize PN level (dBc)}
			\psfrag{PPLR (dB)}[c][c]{\footnotesize $\mathrm{PPLR~(dB)}$}			
			
			\includegraphics[width=3.25cm]{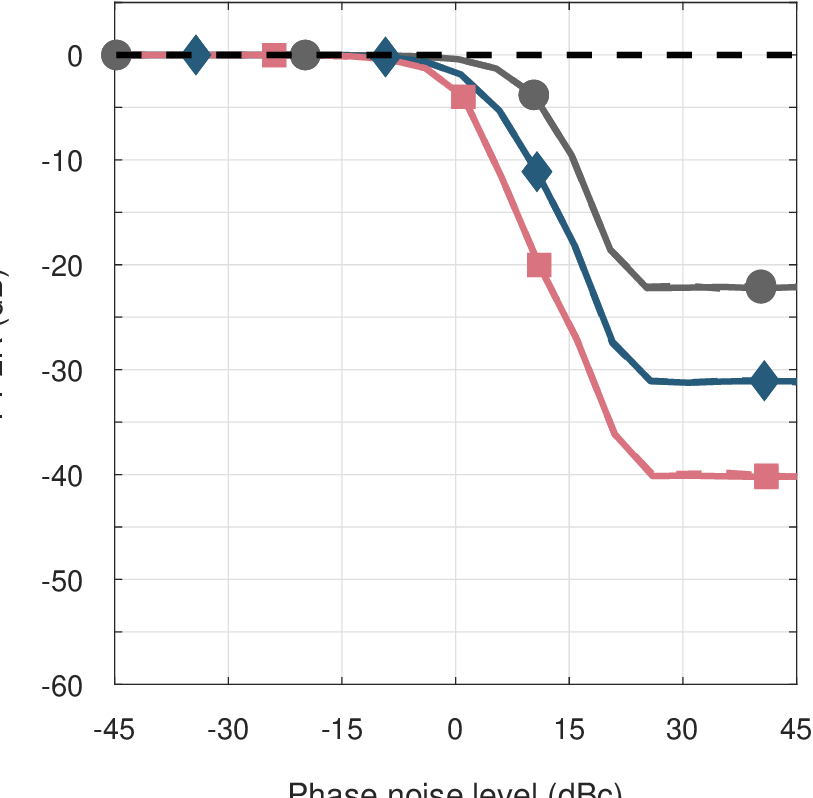}			
		}\hspace{0.05cm}
		\subfloat[ ]{
			
			\psfrag{-45}[c][c]{\scriptsize -$45$}
			\psfrag{-30}[c][c]{\scriptsize -$30$}
			\psfrag{-15}[c][c]{\scriptsize -$15$}
			\psfrag{0}[c][c]{\scriptsize $0$}
			\psfrag{15}[c][c]{\scriptsize $15$}
			\psfrag{30}[c][c]{\scriptsize $30$}
			\psfrag{45}[c][c]{\scriptsize $45$}
			
			\psfrag{-15}[c][c]{\scriptsize -$15$}
			\psfrag{-12}[c][c]{\scriptsize -$12$}
			\psfrag{-9}[c][c]{\scriptsize -$9$}
			\psfrag{-6}[c][c]{\scriptsize -$6$}
			\psfrag{-3}[c][c]{\scriptsize -$3$}
			\psfrag{0}[c][c]{\scriptsize $0$}
			
			\psfrag{Phase noise level (dBc)}[c][c]{\footnotesize PN level (dBc)}
			\psfrag{PSLR (dB)}[c][c]{\footnotesize Range $\mathrm{PSLR~(dB)}$}	
			
			\includegraphics[width=3.25cm]{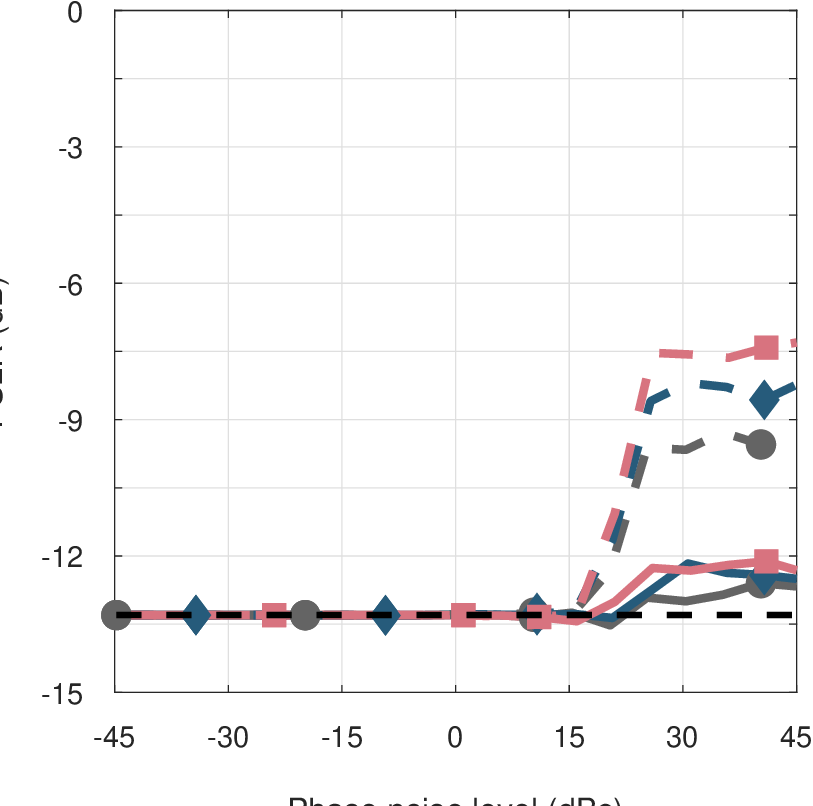}	
			
		}\hspace{0.05cm}
		\subfloat[ ]{
			
			\psfrag{-45}[c][c]{\scriptsize -$45$}
			\psfrag{-30}[c][c]{\scriptsize -$30$}
			\psfrag{-15}[c][c]{\scriptsize -$15$}
			\psfrag{0}[c][c]{\scriptsize $0$}
			\psfrag{15}[c][c]{\scriptsize $15$}
			\psfrag{30}[c][c]{\scriptsize $30$}
			\psfrag{45}[c][c]{\scriptsize $45$}
			
			\psfrag{-12}[c][c]{\scriptsize -$12$}
			\psfrag{-4}[c][c]{\scriptsize -$4$}
			\psfrag{4}[c][c]{\scriptsize $4$}
			\psfrag{12}[c][c]{\scriptsize $12$}
			\psfrag{20}[c][c]{\scriptsize $20$}
			\psfrag{28}[c][c]{\scriptsize $28$}
			\psfrag{36}[c][c]{\scriptsize $36$}
			
			\psfrag{Phase noise level (dBc)}[c][c]{\footnotesize PN level (dBc)}
			\psfrag{ISLR (dB)}[c][c]{\footnotesize Range $\mathrm{ISLR~(dB)}$}	
			
			\includegraphics[width=3.25cm]{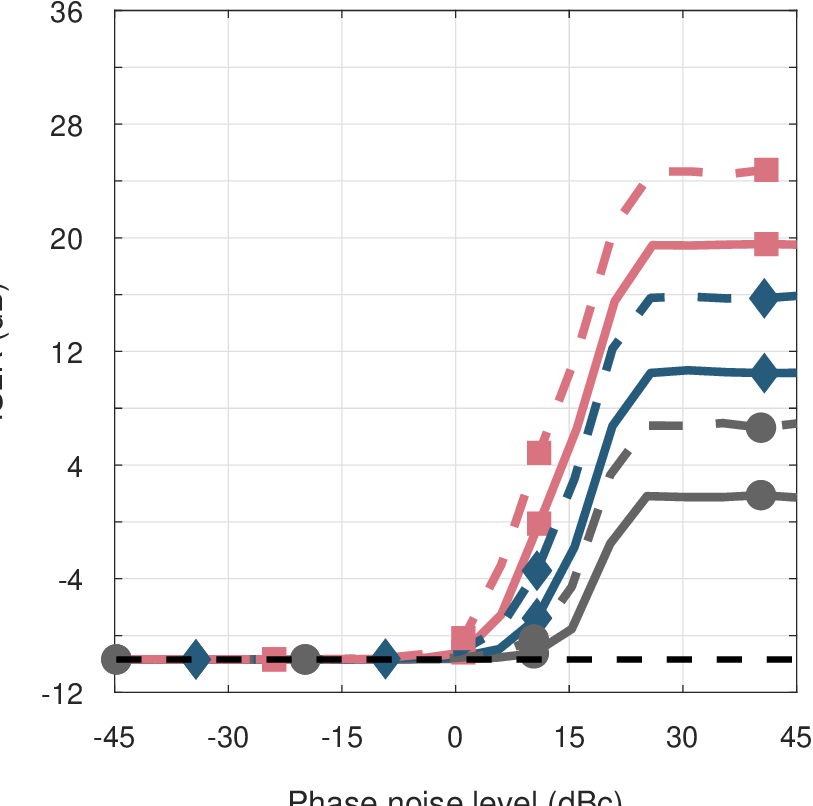}
			
		}\hspace{0.05cm}
		\subfloat[ ]{
			
			\psfrag{-45}[c][c]{\scriptsize -$45$}
			\psfrag{-30}[c][c]{\scriptsize -$30$}
			\psfrag{-15}[c][c]{\scriptsize -$15$}
			\psfrag{0}[c][c]{\scriptsize $0$}
			\psfrag{15}[c][c]{\scriptsize $15$}
			\psfrag{30}[c][c]{\scriptsize $30$}
			\psfrag{45}[c][c]{\scriptsize $45$}
			
			\psfrag{-15}[c][c]{\scriptsize -$15$}
			\psfrag{-12}[c][c]{\scriptsize -$12$}
			\psfrag{-9}[c][c]{\scriptsize -$9$}
			\psfrag{-6}[c][c]{\scriptsize -$6$}
			\psfrag{-3}[c][c]{\scriptsize -$3$}
			\psfrag{0}[c][c]{\scriptsize $0$}
			
			\psfrag{Phase noise level (dBc)}[c][c]{\footnotesize PN level (dBc)}
			\psfrag{PSLR (dB)}[c][c]{\footnotesize Doppler $\mathrm{PSLR~(dB)}$}	
			
			\includegraphics[width=3.25cm]{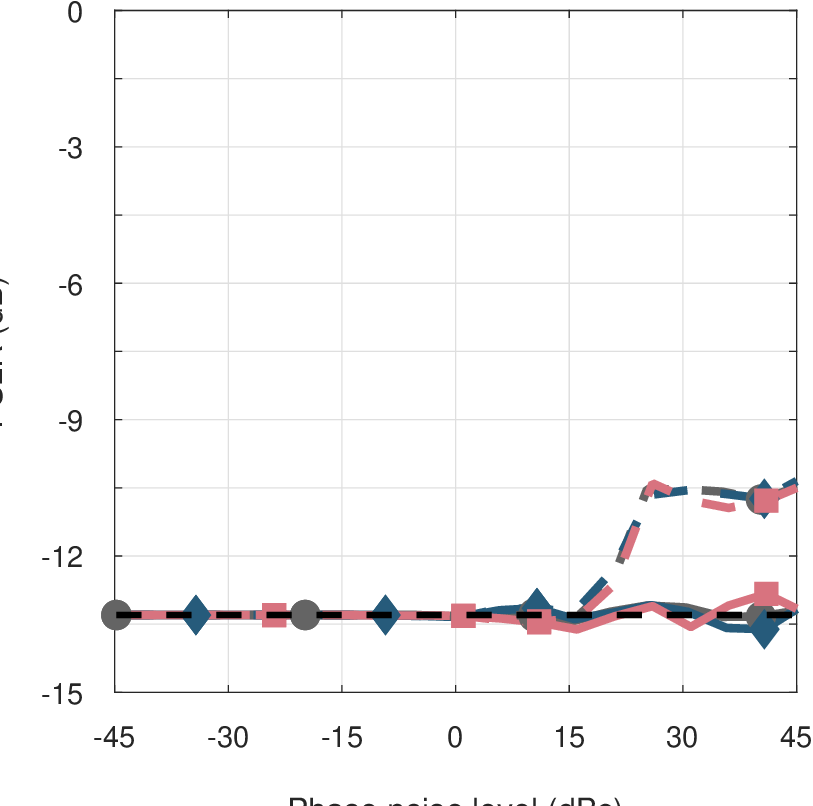}	
			
		}\hspace{0.05cm}
		\subfloat[ ]{
			
			\psfrag{-45}[c][c]{\scriptsize -$45$}
			\psfrag{-30}[c][c]{\scriptsize -$30$}
			\psfrag{-15}[c][c]{\scriptsize -$15$}
			\psfrag{0}[c][c]{\scriptsize $0$}
			\psfrag{15}[c][c]{\scriptsize $15$}
			\psfrag{30}[c][c]{\scriptsize $30$}
			\psfrag{45}[c][c]{\scriptsize $45$}
			
			\psfrag{-12}[c][c]{\scriptsize -$12$}
			\psfrag{-4}[c][c]{\scriptsize -$4$}
			\psfrag{4}[c][c]{\scriptsize $4$}
			\psfrag{12}[c][c]{\scriptsize $12$}
			\psfrag{20}[c][c]{\scriptsize $20$}
			\psfrag{28}[c][c]{\scriptsize $28$}
			\psfrag{36}[c][c]{\scriptsize $36$}
			
			\psfrag{Phase noise level (dBc)}[c][c]{\footnotesize PN level (dBc)}
			\psfrag{ISLR (dB)}[c][c]{\footnotesize Doppler $\mathrm{ISLR~(dB)}$}
			
			\includegraphics[width=3.25cm]{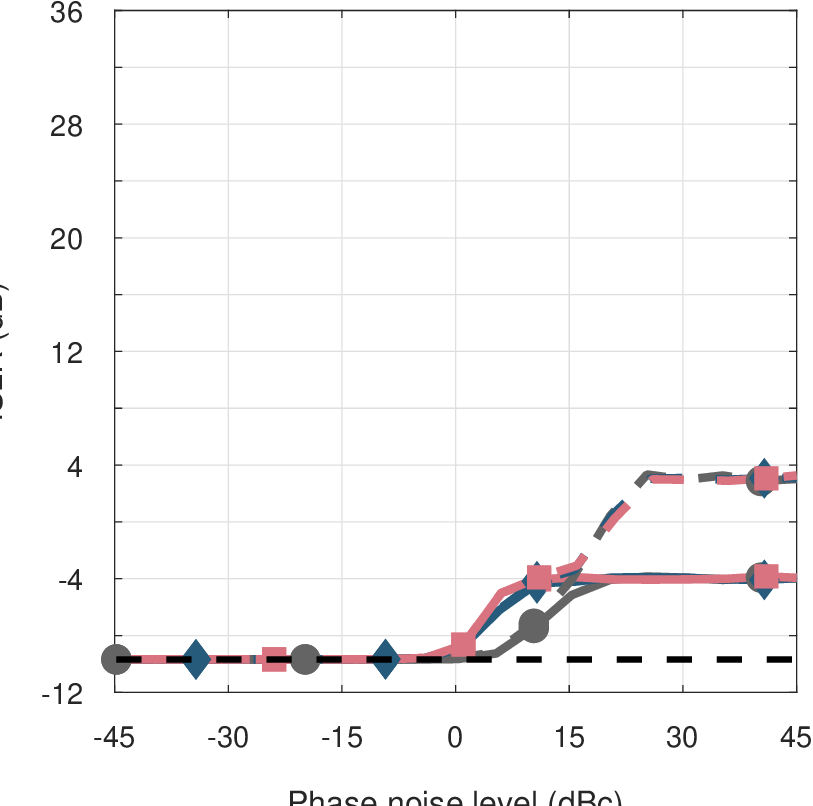}		
			
		}
		
		\captionsetup{justification=raggedright,labelsep=period,singlelinecheck=false}
		\caption{\ PPLR (a), range PSLR (b) and ISLR (c), and Doppler shift PSLR (e) and ISLR (d) as functions of the combined transmit and receive PN level considering CPE estimation and correction from \eqref{eq:CPE_est} and \eqref{eq:CPE_corr}, respectively. The same scenarios as in Fig.~\ref{fig:sidelobe_PSD} are considered, i.e., a single static target, CP length of $N_\mathrm{CP}=N$, and $M=128$ OFDM symbols are assumed, as well as \mbox{$N=256$} and QPSK \mbox{({\color[rgb]{0.3922,0.3922,0.392}\textbf{\textemdash}} and {\color[rgb]{0.3922,0.3922,0.3922}$\CIRCLE$})}, \mbox{$N=2048$} and QPSK \mbox{({\color[rgb]{0.1490,0.3569,0.4824}\textbf{\textemdash}} and {\color[rgb]{0.1490,0.3569,0.4824}$\blacklozenge$})}, and \mbox{$N=16384$} and QPSK \mbox{({\color[rgb]{0.8471,0.4510,0.4980}\textbf{\textemdash}} and {\color[rgb]{0.8471,0.4510,0.4980}$\blacksquare$})}. Results are also shown for \mbox{256-QAM} and the same aforementioned $N$ values, i.e., \mbox{$N=256$} \mbox{({\color[rgb]{0.3922,0.3922,0.392}\textbf{\textendash~\textendash}} and {\color[rgb]{0.3922,0.3922,0.3922}$\CIRCLE$})}, \mbox{$N=2048$} \mbox{({\color[rgb]{0.1490,0.3569,0.4824}\textbf{\textendash~\textendash}} and {\color[rgb]{0.1490,0.3569,0.4824}$\blacklozenge$})}, and \mbox{$N=16384$} \mbox{({\color[rgb]{0.8471,0.4510,0.4980}\textbf{\textendash~\textendash}} and {\color[rgb]{0.8471,0.4510,0.4980}$\blacksquare$})}. As in Fig.~\ref{fig:sidelobe_PSD}, the ideal PPLR, PSLR and ISLR values without PN are also shown ({\color[rgb]{0,0,0}\textbf{\textendash~\textendash}}).}\label{fig:sidelobe_PSD_CPEcorr}
		
	\end{figure*}
	
	\begin{figure}[!t]
		\centering
		
		\subfloat[ ]{
			
			\psfrag{-45}[c][c]{\scriptsize -$45$}
			\psfrag{-30}[c][c]{\scriptsize -$30$}
			\psfrag{-15}[c][c]{\scriptsize -$15$}
			\psfrag{0}[c][c]{\scriptsize $0$}
			\psfrag{15}[c][c]{\scriptsize $15$}
			\psfrag{30}[c][c]{\scriptsize $30$}
			\psfrag{45}[c][c]{\scriptsize $45$}
			
			\psfrag{-30}[c][c]{\scriptsize -$30$}
			\psfrag{30}[c][c]{\scriptsize $30$}
			\psfrag{60}[c][c]{\scriptsize $60$}
			\psfrag{90}[c][c]{\scriptsize $90$}
			\psfrag{120}[c][c]{\scriptsize $120$}
			
			\psfrag{Phase noise level (dBc)}[c][c]{\footnotesize PN level (dBc)}
			\psfrag{AAAA Image SIR (dB)}[c][c]{\footnotesize Mean image SIR (dB)}	
			
			\includegraphics[width=3.75cm]{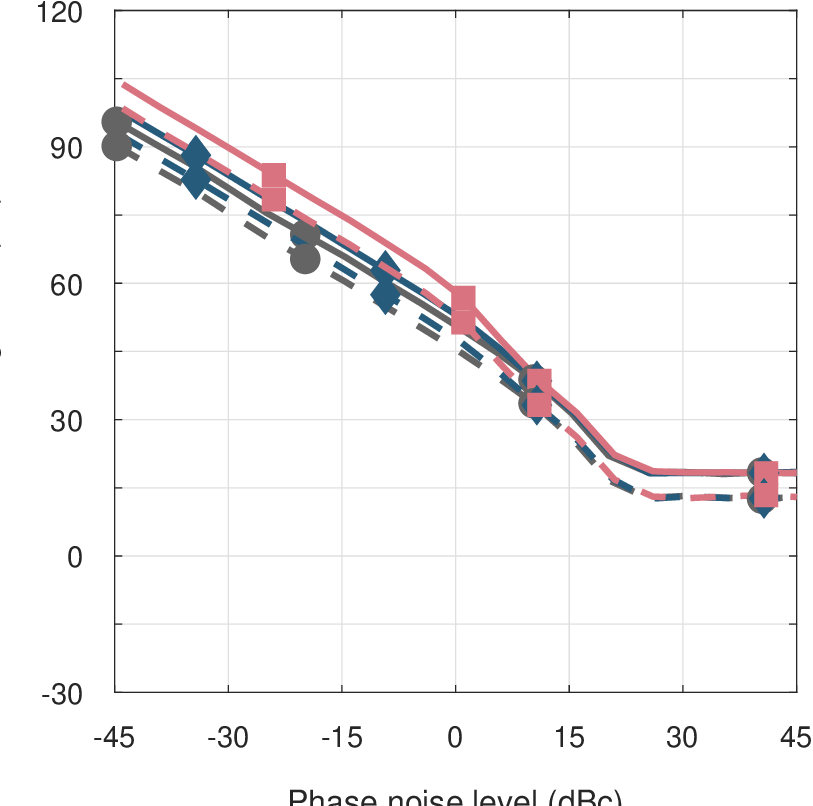}\label{fig:meanImageSIR_chebWin_CPEcorr}
			
		}\hspace{0.1cm}
		\subfloat[ ]{
			
			\psfrag{-45}[c][c]{\scriptsize -$45$}
			\psfrag{-30}[c][c]{\scriptsize -$30$}
			\psfrag{-15}[c][c]{\scriptsize -$15$}
			\psfrag{0}[c][c]{\scriptsize $0$}
			\psfrag{15}[c][c]{\scriptsize $15$}
			\psfrag{30}[c][c]{\scriptsize $30$}
			\psfrag{45}[c][c]{\scriptsize $45$}
			
			\psfrag{-30}[c][c]{\scriptsize -$30$}
			\psfrag{30}[c][c]{\scriptsize $30$}
			\psfrag{60}[c][c]{\scriptsize $60$}
			\psfrag{90}[c][c]{\scriptsize $90$}
			\psfrag{120}[c][c]{\scriptsize $120$}
			
			\psfrag{Phase noise level (dBc)}[c][c]{\footnotesize PN level (dBc)}
			\psfrag{AAAA Image SIR (dB)}[c][c]{\footnotesize Min. image SIR (dB)}	
			
			\includegraphics[width=3.75cm]{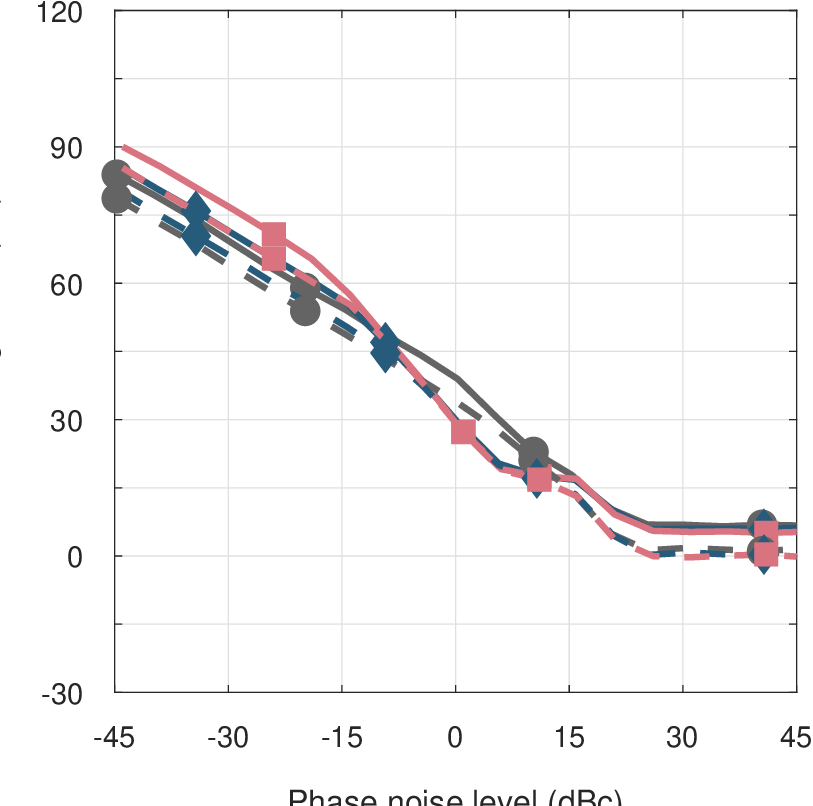}\label{fig:minImageSIR_chebWin_CPEcorr}	
			
		}

		\captionsetup{justification=raggedright,labelsep=period,singlelinecheck=false}
		\caption{\ Image SIR: (a) mean and (b) minimum values after CPE estimation and correction CPE estimation and correction from \eqref{eq:CPE_est} and \eqref{eq:CPE_corr}, respectively. As for Fig.~\ref{fig:imageSIR_chebWin_CPEcorr}, a single static target was considered and \mbox{$N=256$} and QPSK \mbox{({\color[rgb]{0.3922,0.3922,0.392}\textbf{\textemdash}} and {\color[rgb]{0.3922,0.3922,0.3922}$\CIRCLE$})}, \mbox{$N=2048$} and QPSK \mbox{({\color[rgb]{0.1490,0.3569,0.4824}\textbf{\textemdash}} and {\color[rgb]{0.1490,0.3569,0.4824}$\blacklozenge$})}, and \mbox{$N=16384$} and QPSK \mbox{({\color[rgb]{0.8471,0.4510,0.4980}\textbf{\textemdash}} and {\color[rgb]{0.8471,0.4510,0.4980}$\blacksquare$})} were assumed. Results were also obtained for \mbox{256-QAM} with \mbox{$N=256$} \mbox{({\color[rgb]{0.3922,0.3922,0.392}\textbf{\textendash~\textendash}} and {\color[rgb]{0.3922,0.3922,0.3922}$\CIRCLE$})}, \mbox{$N=2048$} \mbox{({\color[rgb]{0.1490,0.3569,0.4824}\textbf{\textendash~\textendash}} and {\color[rgb]{0.1490,0.3569,0.4824}$\blacklozenge$})}, and \mbox{$N=16384$} \mbox{({\color[rgb]{0.8471,0.4510,0.4980}\textbf{\textendash~\textendash}} and {\color[rgb]{0.8471,0.4510,0.4980}$\blacksquare$})}. In addition, a CP length of $N_\mathrm{CP}=N$ and $M=128$ OFDM symbols were considered.}\label{fig:imageSIR_chebWin_CPEcorr}
		
	\end{figure}
	
	\subsubsection{Sensing performance with CPE compensation}\label{subsubsec:CPEcomp}
	
	To analyze the effectiveness of the  \ac{CPE} estimation and correction according to \eqref{eq:CPE_est} and \eqref{eq:CPE_corr} sensing performance, a bistatic scenario with a dominant, reference path only is assumed due to the reasons discussed in Section~\ref{sec:sysModel}. In addition, the same parameters adopted for the results in Fig.~\ref{fig:sidelobe_PSD} discussed in Section~\ref{subsubsec:sensN} are considered. The obtained results in Fig.~\ref{fig:sidelobe_PSD_CPEcorr}. Comparing with the results without \ac{CPE} compensation from Fig.~\ref{fig:sidelobe_PSD}, it is seen that the \ac{PPLR} becomes negligible for combined transmit and receive \ac{PN} levels below $\SI{0}{dBc}$, whereas a \ac{PPLR} of $\SI{-5}{dB}$ is achieved without \ac{CPE} compensation. For higher \ac{PN} levels, a \ac{PPLR} floor is still reached, but less severe peak power losses are observed than in Fig.~\ref{fig:sidelobe_PSD}. On the one hand, it can also be seen that range \ac{PSLR} is significantly improved, with better nearly ideal results for \ac{QPSK} and smaller degradation than without \ac{CPE} compensation for \mbox{256-\ac{QAM}}. On the other hand, range \ac{ISLR} is not significantly improved as it is mainly caused by \ac{PN}-induced \ac{ICI}, which is not compensated. Finally, both Doppler shift \ac{PSLR} and \ac{ISLR} experience significant improvement, also with better performance for \ac{QPSK} compared to \mbox{256-\ac{QAM}}. As in the range case, some degradation is still experienced since the \ac{ICI} also plays a role, although not dominant, in the Doppler shift estimation.
	
	As for \ac{SIR} performance with the adopted \ac{CPE} estimation and correction approaches in the same scenario, the mean and minimum \ac{SIR} results from Figs.~\ref{fig:imageSIR_chebWin_CPEcorr}a and \ref{fig:imageSIR_chebWin_CPEcorr}b are respectively obtained. The results in Fig.~\ref{fig:imageSIR_chebWin_CPEcorr}a show that a slower \ac{SIR} degradation is observed after $\SI{0}{dBc}$ than without \ac{CPE} compensation previously shown in Fig.~\ref{fig:imageSIR_chebWin}a. In addition, a mean image \ac{SIR} floor of around $\SI{15}{dB}$ instead of $\SI{0}{dB}$ in the case without \ac{CPE} compensation is reached at around $\SI{20}{dBc}$. In Fig.~\ref{fig:imageSIR_chebWin_CPEcorr}b, significant \ac{SIR} improvement is also observed, with at least $\SI{15}{dB}$ higher values than in the case without \ac{CPE} correction in Fig.~\ref{fig:imageSIR_chebWin}b.
	
	\begin{figure}[!t]
		\centering
		
		\psfrag{0}[c][c]{\scriptsize $0$}
		\psfrag{0.05}[c][c]{\scriptsize $0.05$}
		\psfrag{0.1}[c][c]{\scriptsize $0.1$}
		\psfrag{0.15}[c][c]{\scriptsize $0.15$}
		\psfrag{0.2}[c][c]{\scriptsize $0.2$}
		
		\psfrag{000}[c][c]{\scriptsize $10^0$}
		\psfrag{111}[c][c]{\scriptsize $10^1$}
		\psfrag{222}[c][c]{\scriptsize $10^2$}
		\psfrag{333}[c][c]{\scriptsize $10^3$}
		\psfrag{444}[c][c]{\scriptsize $10^4$}
		
		\psfrag{Bistatic range (m)}[c][c]{\footnotesize Bistatic range (m)}
		\psfrag{CPE std. dev. (rad)}[c][c]{\footnotesize RMSE | Std. dev. (rad)}			
		
		\includegraphics[width=4cm]{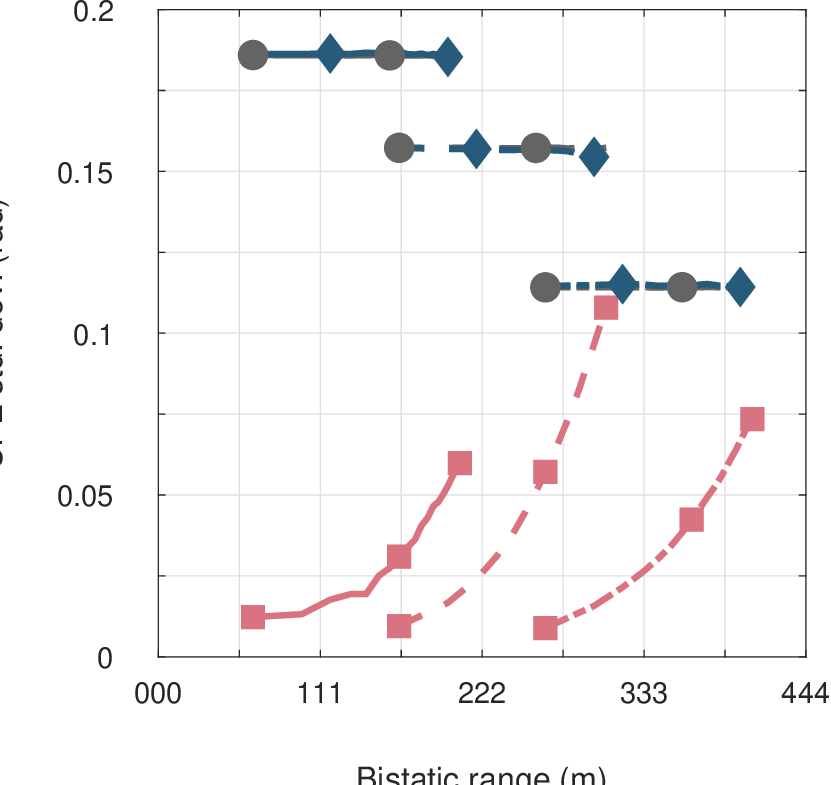}
		
		\captionsetup{justification=raggedright,labelsep=period,singlelinecheck=false}
		\caption{\ Root mean squared difference between CPE estimates of reference path and target reflection as a function of the range for \mbox{$N=256$} \mbox{({\color[rgb]{0.8471,0.4510,0.4980}\textbf{\textemdash}} and {\color[rgb]{0.8471,0.4510,0.4980}$\blacksquare$})}, \mbox{$N=2048$} \mbox{({\color[rgb]{0.8471,0.4510,0.4980}\textbf{\textendash~\textendash}} and {\color[rgb]{0.8471,0.4510,0.4980}$\blacksquare$})}, and \mbox{$N=16384$} \mbox{({\color[rgb]{0.8471,0.4510,0.4980}\textbf{\textendash\hspace{0.1cm}$\cdot$}} and {\color[rgb]{0.8471,0.4510,0.4980}$\blacksquare$})}. For comparison, the CPE estimate standard deviation is shown for the reference path and \mbox{$N=256$} \mbox{({\color[rgb]{0.3922,0.3922,0.392}\textbf{\textemdash}} and {\color[rgb]{0.3922,0.3922,0.3922}$\CIRCLE$})}, \mbox{$N=2048$} \mbox{({\color[rgb]{0.3922,0.3922,0.392}\textbf{\textendash~\textendash}} and {\color[rgb]{0.3922,0.3922,0.3922}$\CIRCLE$})}, and \mbox{$N=16384$} \mbox{({\color[rgb]{0.3922,0.3922,0.392}\textbf{\textendash\hspace{0.1cm}$\cdot$}} and {\color[rgb]{0.3922,0.3922,0.3922}$\CIRCLE$})}. In addition, the CPE estimate standard deviation of the target reflection is also shown for \mbox{$N=256$} \mbox{({\color[rgb]{0.1490,0.3569,0.4824}\textbf{\textemdash}} and {\color[rgb]{0.1490,0.3569,0.4824}$\blacklozenge$})}, \mbox{$N=2048$} \mbox{({\color[rgb]{0.1490,0.3569,0.4824}\textbf{\textendash~\textendash}} and {\color[rgb]{0.1490,0.3569,0.4824}$\blacklozenge$})}, and \mbox{$N=16384$} \mbox{({\color[rgb]{0.1490,0.3569,0.4824}\textbf{\textendash\hspace{0.1cm}$\cdot$}} and {\color[rgb]{0.1490,0.3569,0.4824}$\blacklozenge$})}. As for the results shown in Figs.~\ref{fig:sidelobe_PSD_CPEcorr} and \ref{fig:imageSIR_chebWin_CPEcorr}, a CP length of $N_\mathrm{CP}=N$ and $M=128$ OFDM symbols were considered.}\label{fig:stdDev_CPE_range}
		
	\end{figure}
	
	When performing \ac{CPE} estimation and correction in multi-target scenarios, the approaches described by \eqref{eq:CPE_est} and \eqref{eq:CPE_corr} cannot be indiscriminately applied. Instead, \ac{CIR} estimates can be obtained, e.g., based on pilot subcarriers, and the \ac{CPE} for each target can be estimated as discussed in Section~\ref{sec:sysModel}. In practical applications, however, the radar targets tend to be rather weak and not detectable without the additional processing gain of $M$ provided by Doppler shift estimation.  In addition, if the aforementioned targets are moving, their \ac{CPE} will not be distinguishable from their Doppler-shift-induced phase. A solution is then to use the dominant, reference path to estimate the \ac{CPE} and reuse this estimation for the whole \ac{OFDM} symbol, which encompasses all possible targets. While this can improve the communication performance of the \ac{OFDM}-based \ac{ISAC} system, the \ac{CPE} compensation effectiveness will be reduced for targets at higher ranges than the reference path, which will appear at $\SI{0}{\meter}$ in the bistatic radar image \cite{giroto2023_EuMW,brunner2024,giroto2024}. The reason for this is that such targets will experience a different realization of the combined transmit and receive \ac{PN} \mbox{$\theta^\mathrm{PN}_\mathrm{Tx}(t-\tau_p)-\theta^\mathrm{PN}_\mathrm{Rx}(t)$}, which will lead to different \acp{CPE}. This is illustrated in Fig.~\ref{fig:stdDev_CPE_range}, where the root mean squared difference between the \ac{CPE} estimates for the reference path and a second target is analyzed and compared to the standard deviations of the individual \ac{CPE} estimates. To obtain these results, Chebyshev windowing with $\SI{100}{dB}$ sidelobe supression was applied to separate the reference path and the radar target in the obtained \ac{CIR} estimates. In addition, only bistatic ranges between $5\%$ and $95\%$ of the maximum unambiguous range \mbox{$R^\mathrm{bist}_\mathrm{max,ua}=N\Delta R^\mathrm{bist}=N(c_0/B)$} were considered for the radar target to ensure that it is distinguishable from the reference path. Finally, the same \ac{SNR} was assumed for both the radar target at all ranges and the reference path in the obtained \acp{CIR}. Although not realistic, this ensures the same \ac{CPE} estimation accuracy for the both of them to simplify the present analysis. The presented results in Fig.~\ref{fig:stdDev_CPE_range} show that, for short \ac{OFDM} symbols, i.e., \mbox{$N=256$}, the \ac{PN} variation is fast enough to yield a non-negligibly biased \ac{CPE} estimate for the radar target after only $\SI{10}{\meter}$ range in case the estimate for the reference path is reused. For higher $N$ values, less \ac{CPE} performance degradation is observed since the induced \ac{CPE}, which is related with the term in \eqref{eq:thetaMean} that tends to zero with increasing $N$ as the transmit and receive \ac{PN} terms also do, is reduced.
	
	\subsection{Remarks on simulation results}\label{subsec:discussion}
	
	
	It is worth highlighting that the assumed \ac{PN} model shown in Fig.~\ref{fig:PN_psd_dist} presents a comparatively low \ac{PN} level compared to typical hardware. If communication or sensing is performed by a \ac{UE} or between a \ac{gNB} and a \ac{UE} instead of the \ac{gNB}-only scenarios depicted in Fig.~\ref{fig:isacArchitectures}, considerable higher \ac{PN} levels can be observed as \ac{UE} hardware is usually less advanced. This can be observed for lower \ac{FR2} bands below $\SI{30}{\giga\hertz}$ with \ac{3GPP}-compliant parameterization as discussed Section~\ref{subsubsec:3gpp}, where a combined transmit and receive \ac{PN} level of $\SI{-26.99}{dBc}$ \mbox{($\gamma=\SI{17.91}{dB}$ above the \ac{PN} level from Fig.~\ref{fig:PN_psd_dist})} was observed between two \acp{UE} in a bistatic \ac{ISAC} architecture. One example where even higher \ac{PN} level is observed is the $\SI{77}{\giga\hertz}$ setup in \cite{werbunat2024}, whose \ac{PN} \ac{PSD} is somewhat similar to what is obtained increasing the level of the original \ac{PN} model from Fig.~\ref{fig:PN_psd_dist} by $\gamma=\SI{30}{dB}$. In this case, non-negligible \ac{EVM} values are observed, which is graphically represented by the constellations in Fig.~\ref{fig:QPSKconst_psd30}.
		
	Based on the results from Sections~\ref{subsubsec:sensN} to \ref{subsubsec:3gpp}, it can be concluded that non-negligible \ac{PN}-induced sensing performance degradation in terms of mainlobe and sidelobe distortion is only observed for combined transmit and receive \ac{PN} levels above $\SI{-15}{dBc}$. At this point, \ac{PPLR}, Doppler shift \ac{PSLR}, and Doppler shift \ac{ISLR} are the first sensing performance parameters that start to degrade, e.g., as seen in Fig.~\ref{fig:sidelobe_PSD}a, d, and e. Although slightly more robust, range \ac{ISLR} can also start degrading at this \ac{PN} level depending on the \ac{OFDM} signal parameterization, e.g., as shown in \ref{fig:sidelobe_M}c. As for range \ac{PSLR}, a significant higher robustness against \ac{PN} compared to the other sensing performance parameters is observed, which non-negligible degradation being only observed for \ac{PN} levels above around $\SI{11}{dBc}$. At this point, the range processing gain $N$ \cite{giroto2021_tmtt,giroto2023_EuMW} is not anymore sufficient to sufficiently suppress the \ac{PN}-induced \ac{ICI}, resulting in the fact that the highest sidelobe may either be significantly distorted or no longer be the nearest one to the main lobe, which ideally results in a \ac{PSLR} of $\SI{-13.30}{dB}$.
	
	Still regarding \ac{PN}-induced mainlobe and sidelobe level degradation, Section~\ref{subsec:sensPerf} showed that, while the \ac{ICI} may introduce a certain degree of degradation, the dominating impairment for sensing is tends to be the \ac{CPE}, which is evidenced by the fact that Doppler shift \ac{PSLR} and \ac{ISLR} start degrading at lower combined transmit and receive \ac{PN} levels than range \ac{PSLR} and \ac{ISLR} in all investigated settings. As a consequence, target sidelobes in the Doppler shift direction tend to present irregular shape and notably higher level than in the range direction of obtained radar images. As previously mentioned, however, this effect is only observed for combined transmit and receive \ac{PN} levels from around $\SI{-15}{dBc}$ ($\gamma=\SI{29.90}{dB}$) onwards. In addition, it becomes more severe with increasing number $M$ of \ac{OFDM} symbols used for radar processing. This happens since \ac{PN} is not an additive noise, being therefore not linearly suppressed by a radar processing gain as it is the case for \ac{AWGN} \cite{giroto2021_tmtt,giroto2023_EuMW}.
	
	If \ac{SIR} is analyzed instead of mainlobe and sidelobe degradation, it can be seen that mean and minimum image \ac{SIR} are limited to around $\SI{60}{dB}$ and $\SI{30}{dB}$ at a \ac{PN} level of $\SI{-15}{dBc}$. Although no degradation of \ac{PPLR}, range \ac{PSLR} and \ac{ISLR}, and Doppler shift \ac{PSLR} and \ac{ISLR} at this \ac{PN} level is observed as previously discussed,  non-negligible sensing performance degradation is experienced especially in the Doppler shift direction, where unwanted sidelobes that cannot be suppressed by windowing drive the minimum \ac{SIR} value up. This is, e.g., illustrated by Fig.~\ref{fig:I_rD_chebwin}b, where windowing is no longer effective against Doppler shift sidelobes at a similar \ac{PN} level, i.e. $\SI{-14.90}{dBc}$. Considering \ac{CPE} estimation and correction approaches from \eqref{eq:CPE_est} \eqref{eq:CPE_corr}, respectively, significant sensing performance improvement was observed. In terms of mainlobe and sidelobe levels, only negligible degradation w.r.t. the \ac{PN}-free case was observed for combined transmit and receive \ac{PN} levels of $\SI{0}{dBc}$ or lower, which covers all possible practical scenarios. In terms of \ac{SIR}, the most relevant difference was observed for the aforementioned \ac{PN} level range, where at least $\SI{15}{dB}$ minimum \ac{SIR} improvement was observed w.r.t. the case without \ac{CPE} compensation, ensuring an \ac{SIR} of at least $\SI{30}{dB}$ for all assumed sets of \ac{OFDM} signal parameters.
	
	It is worth mentioning that, if different \ac{PN} \ac{PSD} shapes are considered, which is the case when the parameters of either the transmit, receive or both \ac{PN} contributions are different then the assumed ones in Section~\ref{sec:perfAnalysis}, the most relevant parameters to evaluate the performance of \ac{OFDM}-based \ac{ISAC} systems will still be the combined transmit and receive \ac{PN} level as well as the \ac{PLL} loop bandwidth $B_\mathrm{PLL}$, which defines the transition from \ac{CPE} to \ac{ICI} as the dominant \ac{PN}-induced effect. Related analysis were in \cite{khanzadi2014} to assess the performance of communication systems and \cite{schweizer2018}, where a brief discussion on the effects of \ac{PN} on stepped-carrier \ac{OFDM} radar was presented.

	\section{Conclusion}\label{sec:conclusion}
	
	This article presented a sensing performance analysis of \ac{OFDM}-based \ac{ISAC} under influence of \ac{PN}. In this context, an \ac{OFDM}-based \ac{ISAC} system model was formulated and the \ac{PN} induced impairments to it mathematically described. It discussed that \ac{PN} causes \ac{CPE} and \ac{ICI}. The first effect results in a distinct phase rotation to each \ac{OFDM} symbol, while the latter effect already causes degradation within each \ac{OFDM} symbol due to interference among subcarriers. More specifically, the performed analysis showed that monostatic sensing at lower ranges benefits from \ac{PN} reduction, as transmit and receive \ac{PN} are correlated and therefore add destructively to a certain extent. In bistatic sensing, \ac{PN} at the transmitter and receiver sides is uncorrelated and their combination equals to the sum of uncorrelated Gaussian processes, resulting in a $\SI{3}{dB}$ \ac{PN} level increase regardless of the target range. It was also shown that higher-order modulations are more affected by \ac{PN}, as they are more sensitive to \ac{CPE} and \ac{ICI} due to their dense constellations that also do not have a constant envelope.
	
	After radar signal processing on the receive \ac{OFDM} frame with the two aforementioned \ac{PN}-induce impairments, power loss at the main lobe of radar target reflections is observed in the range-Doppler shift radar image. Furthermore, an increased sidelobe level in both range and Doppler shift directions is experienced, mainly due to \ac{ICI} and \ac{CPE}, respectively. These effects can severely degrade the quality of obtained radar images, possibly preventing the detection of actual targets or resulting in ghost targets. The latter issue is especially in the Doppler shift direction, since a distorted phase pattern will be fed to the \ac{DFT}-based Doppler shift estimation. This results in a rather high sidelobe increase, which forms a stripe in the radar image that cannot be suppressed by windowing.It was also observed in the presented analysis that the an increasing number of subcarrier and increasing modulation order lead to more severe \ac{PN}-induced sensing performance degradation, as the aforementioned \ac{ICI} and \ac{CPE} are intensified. In addition, an increasing number of \ac{OFDM} symbols for radar processing leads to peak power loss and sidelobe level increase instead of additional processing gain as expected against \ac{AWGN}, which is the case since \ac{PN} is not an additive noise. Finally, it was shown that the mainlobe and sidelobe level degradation is only negligibly affected by the Doppler shift of targets.
	
	The discussed \ac{PN}-induced mainlobe and sidelobe level distortions, however, are only severe for rather high combined transmit and receive \ac{PN} levels. Even in cases where little to no degradation of \ac{PPLR}, range \ac{PSLR} and \ac{ISLR}, and Doppler shift \ac{PSLR} and \ac{ISLR} is observed, the \ac{SIR} between a target reflection peak and the interference floor in the radar image is reduced due to the influence of \ac{PN}. This effect is, however, tolerable until around $\SI{-15}{dBc}$, where the changes in the mainlobe and sidelobe levels are still negligible. More severe sensing performance degradation is therefore only expected in cases where hardware with less stringent specifications is used. For the specific case of \ac{OFDM}-based \ac{ISAC} systems in the lower \ac{FR2} bands that are parameterized in compliance with \ac{3GPP} specifications and use \ac{gNB} or \ac{UE} hardware with similar \ac{PN} to what is specified by the \ac{3GPP} TR 38.803 \cite{3GPPTR38803}, no severe sensing performance degradation is induced by \ac{PN}, regardless of whether monostatic or bistatic architectures are adopted. Finally, it was observed that significant sensing performance enhancement is achievable with \ac{CPE} estimation and correction. Although this does not fully compensates for the \ac{PN}-induced effects, since \ac{ICI} is not addressed, the main issue of unwanted sidelobes in the Doppler shift direction can be effectively compensated for expected combined transmit and receive \ac{PN} levels in practice.

	
	%
	
	\section*{Acknowledgment}
	
	L. Giroto de Oliveira would like to thank Dr. Mateus de Lima Filomeno from the Federal University of Juiz de Fora, Brazil, Dr. Silvio Mandelli and Marcus Henninger from Nokia Bell Labs, Stuttgart, Germany, as well as Tsung-Ching Tsai from the Karlsruhe Institute of Technology, Germany, for the valuable discussions.
	
	\bibliographystyle{IEEEtran}
	\bibliography{main}

\begin{thebibliography}{10}
\providecommand{\url}[1]{#1}
\csname url@samestyle\endcsname
\providecommand{\newblock}{\relax}
\providecommand{\bibinfo}[2]{#2}
\providecommand{\BIBentrySTDinterwordspacing}{\spaceskip=0pt\relax}
\providecommand{\BIBentryALTinterwordstretchfactor}{4}
\providecommand{\BIBentryALTinterwordspacing}{\spaceskip=\fontdimen2\font plus
\BIBentryALTinterwordstretchfactor\fontdimen3\font minus
  \fontdimen4\font\relax}
\providecommand{\BIBforeignlanguage}[2]{{%
\expandafter\ifx\csname l@#1\endcsname\relax
\typeout{** WARNING: IEEEtran.bst: No hyphenation pattern has been}%
\typeout{** loaded for the language `#1'. Using the pattern for}%
\typeout{** the default language instead.}%
\else
\language=\csname l@#1\endcsname
\fi
#2}}
\providecommand{\BIBdecl}{\relax}
\BIBdecl

\bibitem{chafii2023}
M.~Chafii, L.~Bariah, S.~Muhaidat, and M.~Debbah, ``Twelve scientific
  challenges for {6G}: Rethinking the foundations of communications theory,''
  \emph{IEEE Commun. Surv. Tut.}, vol.~25, no.~2, pp. 868--904, Second Quarter
  2023.

\bibitem{liu2022}
F.~{Liu et al.}, ``Integrated sensing and communications: Towards
  dual-functional wireless networks for {6G} and beyond,'' \emph{IEEE J. Sel.
  Areas Commun.}, vol.~40, no.~6, pp. 1728--1767, Jun. 2022.

\bibitem{viswanathan2020}
H.~Viswanathan and P.~E. Mogensen, ``Communications in the {6G} era,''
  \emph{IEEE Access}, vol.~8, pp. 57\,063--57\,074, Mar. 2020.

\bibitem{rajatheva2020}
N.~{Rajatheva et al.}, ``White paper on broadband connectivity in {6G},''
  \emph{arXiv preprint arXiv:2004.14247 [eess.SP]}, Apr. 2020.

\bibitem{kadelka2023}
A.~Kadelka, G.~Zimmermann, J.~Plach\'y, and O.~Holschke, ``A {CSP}'s view on
  opportunities and challenges of integrated communications and sensing,'' in
  \emph{2023 IEEE 3rd Int. Symp. Joint Commun. Sens.}, Mar. 2023, pp. 1--6.

\bibitem{shatov2024}
V.~{Shatov et al.}, ``Joint radar and communications: Architectures, use cases,
  aspects of radio access, signal processing, and hardware,'' \emph{IEEE
  Access}, pp. 1--1, 2024.

\bibitem{alkhateeb2023}
A.~Alkhateeb, S.~Jiang, and G.~Charan, ``Real-time digital twins: Vision and
  research directions for {6G} and beyond,'' \emph{IEEE Commun. Mag.}, vol.~61,
  no.~11, pp. 128--134, Nov. 2023.

\bibitem{imran2024}
S.~Imran, G.~Charan, and A.~Alkhateeb, ``Environment semantic communication:
  Enabling distributed sensing aided networks,'' \emph{arXiv preprint
  arXiv:2402.14766 [cs.IT]}, Feb. 2024.

\bibitem{thomae2021}
R.~Thom\"a, T.~Dallmann, S.~Jovanoska, P.~Knott, and A.~Schmeink, ``Joint
  communication and radar sensing: An overview,'' in \emph{2021 15th European
  Conf. Antennas Propag.}, Mar., pp. 1--5.

\bibitem{wild2023}
T.~Wild, A.~Grudnitsky, S.~Mandelli, M.~Henninger, J.~Guan, and F.~Schaich,
  ``{6G} integrated sensing and communication: From vision to realization,'' in
  \emph{2023 20th Eur. Radar Conf.}, Sept. 2023, pp. 355--358.

\bibitem{roberts2024}
I.~P. Roberts, Y.~Zhang, T.~Osman, and A.~Alkhateeb, ``Real-world evaluation of
  full-duplex millimeter wave communication systems,'' \emph{IEEE Trans.
  Wireless Commun. (Early Access)}, pp. 1--1, Mar. 2024.

\bibitem{barneto2019}
C.~B. {Barneto et al.}, ``Full-duplex {OFDM} radar with {LTE} and {5G} {NR}
  waveforms: Challenges, solutions, and measurements,'' \emph{IEEE Trans.
  Microw. Theory Tech.}, vol.~67, no.~10, pp. 4042--4054, Oct. 2019.

\bibitem{barneto2021}
C.~B. {Barneto}, S.~D. {Liyanaarachchi}, M.~{Heino}, T.~{Riihonen}, and
  M.~{Valkama}, ``Full duplex radio/radar technology: The enabler for advanced
  joint communication and sensing,'' \emph{IEEE Wireless Commun.}, vol.~28,
  no.~1, pp. 82--88, Feb. 2021.

\bibitem{thomae2019}
R.~S. {Thom\"a et al.}, ``Cooperative passive coherent location: A promising
  {5G} service to support road safety,'' \emph{IEEE Commun. Mag.}, vol.~57,
  no.~9, pp. 86--92, Sept. 2019.

\bibitem{kanhere2021_multistatic}
O.~Kanhere, S.~Goyal, M.~Beluri, and T.~S. Rappaport, ``Target localization
  using bistatic and multistatic radar with {5G} {NR} waveform,'' in \emph{2021
  IEEE 93rd VVeh. Technol. Conf.}, Apr. 2021, pp. 1--7.

\bibitem{mollen2023}
C.~Moll\'en, G.~Fodor, R.~Baldemair, J.~Huschke, and J.~Vinogradova, ``Joint
  multistatic sensing of transmitter and target in {OFDM}-based {JCAS}
  system,'' in \emph{2023 Joint Eur. Conf. Netw. Commun. 6G Summit}, Jun. 2023,
  pp. 144--149.

\bibitem{thomae2023}
R.~Thomä and T.~Dallmann, ``Distributed {ISAC} systems – multisensor radio
  access and coordination,'' in \emph{2023 20th Eur. Radar Conf.}, Sept. 2023,
  pp. 351--354.

\bibitem{yajnanarayana2024}
V.~Yajnanarayana and P.~Geuer, ``Bi-static sensing in {OFDM} wireless systems
  for indoor scenarios,'' \emph{arXiv preprint arXiv:2403.04201 [cs.IT]}, Mar.
  2024.

\bibitem{geng2020}
Z.~{Geng}, ``Evolution of netted radar systems,'' \emph{IEEE Access}, vol.~8,
  pp. 124\,961--124\,977, Jul. 2020.

\bibitem{giroto2023_EuMW}
L.~{Giroto de Oliveira et al.}, ``Bistatic {OFDM}-based joint
  radar-communication: Synchronization, data communication and sensing,'' in
  \emph{2023 20th Eur. Radar Conf.}, Sept. 2023, pp. 359--362.

\bibitem{pegoraro2024}
J.~{Pegoraro et al.}, ``{JUMP}: Joint communication and sensing with
  unsynchronized transceivers made practical,'' \emph{IEEE Trans. Wireless
  Commun. (Early Access)}, pp. 1--16, Feb. 2024.

\bibitem{brunner2024}
D.~{Brunner et al.}, ``Bistatic {OFDM}-based {ISAC} with over-the-air
  synchronization: System concept and performance analysis,'' \emph{arXiv
  preprint arXiv:2405.04962 [eess.SP]}, May 2024.

\bibitem{omri2019}
A.~Omri, M.~Shaqfeh, A.~Ali, and H.~Alnuweiri, ``Synchronization procedure in
  {5G} {NR} systems,'' \emph{IEEE Access}, vol.~7, pp. 41\,286--41\,295, Mar.
  2019.

\bibitem{giroto2024}
L.~{Giroto de Oliveira et al.}, ``Pilot-based {SFO} estimation for {OFDM}-based
  bistatic integrated sensing and communication,'' \emph{arXiv preprint
  arXiv:2407.07567 [eess.SP]}, Jul. 2024.

\bibitem{bookSFO}
L.~Smaini, \emph{{RF} Analog Impairments Modeling for Communication Systems
  Simulation: Application to {OFDM}-based Transceivers}.\hskip 1em plus 0.5em
  minus 0.4em\relax John Wiley \& Sons, Ltd, 2012.

\bibitem{khanzadi2014}
M.~R. Khanzadi, D.~Kuylenstierna, A.~Panahi, T.~Eriksson, and H.~Zirath,
  ``Calculation of the performance of communication systems from measured
  oscillator phase noise,'' \emph{IEEE Trans. Circuits Syst. I, Reg. Papers},
  vol.~61, no.~5, pp. 1553--1565, May 2014.

\bibitem{henninger2023}
M.~Henninger, S.~Mandelli, A.~Grudnitsky, T.~Wild, and S.~ten Brink, ``{CRAP}:
  Clutter removal with acquisitions under phase noise,'' in \emph{2023 2nd Int.
  Conf. 6G Netw.}, Oct. 2023, pp. 1--8.

\bibitem{henninger2024}
M.~Henninger, S.~Mandelli, A.~Grudnitsky, and S.~ten Brink, ``{CRAP} part {II}:
  Clutter removal with continuous acquisitions under phase noise,'' \emph{arXiv
  preprint arXiv:2402.11939 [eess.SP]}, Feb. 2024.

\bibitem{ayhan2016}
S.~Ayhan, S.~Scherr, A.~Bhutani, B.~Fischbach, M.~Pauli, and T.~Zwick, ``Impact
  of frequency ramp nonlinearity, phase noise, and {SNR} on {FMCW} radar
  accuracy,'' \emph{IEEE Trans. Microw. Theory Tech.}, vol.~64, no.~10, pp.
  3290--3301, Oct. 2016.

\bibitem{kou2024}
J.~Kou, M.~Bauduin, A.~Bourdoux, and S.~Pollin, ``Distributed {PMCW} radar
  network in presence of phase noise,'' \emph{arXiv preprint arXiv:2405.09680
  [eess.SP]}, May 2024.

\bibitem{werbunat2024}
D.~{Werbunat et al.}, ``On the synchronization of uncoupled multistatic {PMCW}
  radars,'' \emph{IEEE Trans. Microw. Theory Tech. (Early Access)}, pp. 1--13,
  Feb. 2024.

\bibitem{schweizer2018}
B.~Schweizer, D.~Schindler, C.~Knill, J.~Hasch, and C.~Waldschmidt, ``On
  hardware implementations of stepped-carrier {OFDM} radars,'' in \emph{2018
  IEEE/MTT-S Int. Microw. Symp.}, Jun. 2018, pp. 891--894.

\bibitem{aguilar2024}
J.~Aguilar, D.~Werbunat, V.~Janoudi, C.~Bonfert, and C.~Waldschmidt,
  ``Uncoupled digital radars creating a coherent sensor network,'' \emph{IEEE
  J. Microw.}, vol.~5, no.~3, pp. 459--472, Jun. 2024.

\bibitem{keskin2023}
M.~F. Keskin, H.~Wymeersch, and V.~Koivunen, ``Monostatic sensing with {OFDM}
  under phase noise: From mitigation to exploitation,'' \emph{IEEE Trans.
  Signal Process.}, vol.~71, pp. 1363--1378, Apr. 2023.

\bibitem{keskin2023_2}
M.~F. Keskin, C.~Marcus, O.~Eriksson, H.~Wymeersch, and V.~Koivunen, ``On the
  impact of phase noise on monostatic sensing in {OFDM} {ISAC} systems,'' in
  \emph{2023 IEEE Radar Conference (RadarConf23)}, May 2023, pp. 1--6.

\bibitem{koivunen2024}
V.~Koivunen, M.~F. Keskin, H.~Wymeersch, M.~Valkama, and N.~Gonz\'alez-Prelcic,
  ``Multicarrier {ISAC}: Advances in waveform design, signal processing and
  learning under non-idealities,'' \emph{arXiv preprint arXiv:2406.18476
  [eess.SP]}, Jun. 2024.

\bibitem{walt2023}
P.~W. van~der Walt and W.~Steyn, ``Characterizing phase noise in
  self-referencing radar,'' \emph{IEEE Aerosp. Electron. Syst. Mag.}, vol.~38,
  no.~12, pp. 4--13, Dec. 2023.

\bibitem{armada2001}
A.~Garcia~Armada, ``Understanding the effects of phase noise in orthogonal
  frequency division multiplexing ({OFDM}),'' \emph{IEEE Trans. Broadcast},
  vol.~47, no.~2, pp. 153--159, Jun. 2001.

\bibitem{giroto2021_tmtt}
L.~{Giroto de Oliveira}, B.~{Nuss}, M.~B. {Alabd}, A.~{Diewald}, M.~{Pauli},
  and T.~{Zwick}, ``Joint radar-communication systems: Modulation schemes and
  system design,'' \emph{IEEE Trans. Microw. Theory Tech.}, vol.~70, no.~3, pp.
  1521--1551, Mar. 2022.

\bibitem{dhar2017}
D.~Dhar, P.~van Zeijl, D.~Milosevic, H.~Gao, and A.~van Roermund, ``Modeling
  and analysis of the effects of {PLL} phase noise on {FMCW} radar
  performance,'' in \emph{2017 IEEE Int. Symp. Circuits Syst.}, May 2017, pp.
  1--4.

\bibitem{keysight2024}
{Keysight Technologies}. \textit{E8257D PSG Microwave Analog Signal Generator}
  (2024). {Accessed}: Jun. 21, 2024. {[Online]}. {Available}:.
  \url{https://www.keysight.com/us/en/assets/7018-01211/data-sheets/5989-0698.pdf}.

\bibitem{3GPPTS38211}
3GPP, ``{3rd Generation Partnership Project; Technical Specification Group
  Radio Access Network; NR; Physical channels and modulation},'' 3rd Generation
  Partnership Project (3GPP), Tech. Rep. TS 38.211 V16.2.0 (2020-07), 2020.

\bibitem{hussin2012}
S.~Hussin, K.~Puntsri, and R.~Noe, ``Efficiency enhancement of {RF}-pilot-based
  phase noise compensation for coherent optical {OFDM} systems,'' in \emph{17th
  Int. OFDM Workshop 2012}, Aug. 2012, pp. 1--5.

\bibitem{casas2002}
R.~Casas, S.~Biracree, and A.~Youtz, ``Time domain phase noise correction for
  {OFDM} signals,'' \emph{IEEE Trans. Broadcast.}, vol.~48, no.~3, pp.
  230--236, Sept. 2002.

\bibitem{zou2007}
Q.~Zou, A.~Tarighat, and A.~H. Sayed, ``Compensation of phase noise in {OFDM}
  wireless systems,'' \emph{IEEE Trans. Signal Process.}, vol.~55, no.~11, pp.
  5407--5424, Nov. 2007.

\bibitem{syrjala2009}
V.~Syrjala, M.~Valkama, N.~N. Tchamov, and J.~Rinne, ``Phase noise modelling
  and mitigation techniques in {OFDM} communications systems,'' in \emph{2009
  Wireless Telecommun. Symp.}, Apr. 2009, pp. 1--7.

\bibitem{suyama2009}
S.~Suyama, H.~Suzuki, K.~Fukawa, and J.~Izumi, ``Iterative receiver employing
  phase noise compensation and channel estimation for millimeter-wave {OFDM}
  systems,'' \emph{IEEE J. Sel. Areas Commun.}, vol.~27, no.~8, pp. 1358--1366,
  Oct. 2009.

\bibitem{syrjala2010}
V.~Syrj\"al\"a and M.~Valkama, ``Analysis and mitigation of phase noise and
  sampling jitter in {OFDM} radio receivers,'' \emph{Int. J. Microw. Wireless
  Technol.}, vol.~2, pp. 193 -- 202, Apr. 2010.

\bibitem{wang2016}
R.~Wang, H.~Mehrpouyan, M.~Tao, and Y.~Hua, ``Channel estimation, carrier
  recovery, and data detection in the presence of phase noise in {OFDM} relay
  systems,'' \emph{IEEE Trans. Wireless Commun.}, vol.~15, no.~2, pp.
  1186--1205, Feb. 2016.

\bibitem{leshem2017}
A.~Leshem and M.~Yemini, ``Phase noise compensation for {OFDM} systems,''
  \emph{IEEE Trans. Signal Process.}, vol.~65, no.~21, pp. 5675--5686, 2017.

\bibitem{oh2020}
J.~Oh and T.-K. Kim, ``Phase noise effect on millimeter-wave pre-{5G}
  systems,'' \emph{IEEE Access}, vol.~8, pp. 187\,902--187\,913, Oct. 2020.

\bibitem{armada1998}
A.~Armada and M.~Calvo, ``Phase noise and sub-carrier spacing effects on the
  performance of an {OFDM} communication system,'' \emph{IEEE Commun. Lett.},
  vol.~2, no.~1, pp. 11--13, Jan. 1998.

\bibitem{robertson1995}
P.~Robertson and S.~Kaiser, ``Analysis of the effects of phase-noise in
  orthogonal frequency division multiplex ({OFDM}) systems,'' in
  \emph{Proceedings IEEE Int. Conf. Commun.}, vol.~3, Jun. 1995, pp. 1652--1657
  vol.3.

\bibitem{qi2018}
Y.~Qi, M.~Hunukumbure, H.~Nam, H.~Yoo, and S.~Amuru, ``On the phase tracking
  reference signal ({PT-RS}) design for {5G} new radio ({NR}),'' in \emph{2018
  IEEE 88th Veh. Technol. Conf.}, Aug. 2018, pp. 1--5.

\bibitem{zzhang2023}
Z.~{Zhang et al.}, ``A general channel model for integrated sensing and
  communication scenarios,'' \emph{IEEE Commun. Mag.}, vol.~61, no.~5, pp.
  68--74, May 2023.

\bibitem{jzhang2024}
J.~{Zhang et al.}, ``Integrated sensing and communication channel:
  Measurements, characteristics, and modeling,'' \emph{IEEE Commun. Mag.},
  vol.~62, no.~6, pp. 98--104, Jun. 2024.

\bibitem{li2024}
Y.~{Li et al.}, ``User detection in {RIS}-based {mmWave} {JCAS}: Concept and
  demonstration,'' \emph{IEEE Trans. Wireless Commun. (Early Access)}, pp.
  1--16, Feb. 2024.

\bibitem{ismail2023}
A.~R.~M. Ismail, M.~Guenach, A.~Bourdoux, and H.~Steendam, ``The effect of
  phase noise in {OCDM},'' in \emph{2023 IEEE Global Commun. Conf.}, Dec. 2023,
  pp. 3240--3245.

\bibitem{lellouch2016}
G.~Lellouch, A.~K. Mishra, and M.~Inggs, ``Design of {OFDM} radar pulses using
  genetic algorithm based techniques,'' \emph{IEEE Trans. Aerosp. Electron.
  Syst.}, vol.~52, no.~4, pp. 1953--1966, Aug. 2016.

\bibitem{liao2024}
Z.~{Liao et al.}, ``Pulse shaping for random {ISAC} signals: The ambiguity
  function between symbols matters,'' \emph{arXiv preprint arXiv:2407.15530
  [eess.SP]}, Jul. 2024.

\bibitem{overdevest2020}
J.~{Overdevest}, F.~{Jansen}, F.~{Uysal}, and A.~{Yarovoy}, ``Doppler influence
  on waveform orthogonality in 79 {GHz} {MIMO} phase-coded automotive radar,''
  \emph{IEEE Trans. Veh. Technol.}, vol.~69, no.~1, pp. 16--25, Jan. 2020.

\bibitem{3GPPTR38803}
3GPP, ``{3rd Generation Partnership Project; Technical Specification Group
  Radio Access Network; Study on new radio access technology: Radio Frequency
  ({RF}) and co-existence aspects (Release 14)},'' 3rd Generation Partnership
  Project (3GPP), Tech. Rep. TR 38.803 V14.3.0 (2022-03), 2022.

\bibitem{mandelli2023survey}
S.~Mandelli, M.~Henninger, M.~Bauhofer, and T.~Wild, ``Survey on integrated
  sensing and communication performance modeling and use cases feasibility,''
  in \emph{2023 2nd Int. Conf. 6G Netw.}, Oct. 2023, pp. 1--8.

\bibitem{tosi2024}
P.~Tosi, M.~Henninger, L.~{Giroto de Oliveira}, and S.~Mandelli, ``Feasibility
  of non-line-of-sight integrated sensing and communication at {mmWave},''
  \emph{arXiv preprint arXiv:2406.12828 [eess.SP]}, Jun. 2024.

\end{thebibliography}
	
\end{document}